\documentclass[fleqn,usenatbib]{mnras}

\usepackage[T1]{fontenc}

\DeclareRobustCommand{\VAN}[3]{#2}
\let\VANthebibliography\thebibliography
\def\thebibliography{\DeclareRobustCommand{\VAN}[3]{##3}\VANthebibliography}

\usepackage{xspace}
\usepackage{graphicx}	
\usepackage{amsmath}	
\usepackage{amssymb}	
\usepackage{newtxtext,newtxmath}

\newcommand{\atlas}{$\mathrm{ATLAS}^{\mathrm{3D}}$\xspace}
\newcommand{\vf}{$v_{\mathrm{rms}}$\xspace}

\newcommand{\g}{$\gamma = -2.22 \pm 0.05$\xspace}
\newcommand{\gearly}{$\gamma = -2.23 \pm 0.04$\xspace}
\newcommand{\ag}{$\gamma = -2.25 \pm 0.02$\xspace}
\newcommand{\ff}{$\gamma = -2.01 \pm 0.04$\xspace}

\newcommand{\re}{\ensuremath{\mathrm{R}_{\mathrm{e}}}\xspace}
\newcommand{\Msun}{\ensuremath{\mathrm{M}_\odot}\xspace}
\newcommand{\kms}{\ensuremath{\mathrm{km\,s}^{-1}}\xspace}
\newcommand{\rmass}{\ensuremath{\mathrm{R}^{3\mathrm{D}}_{1/2}}\xspace}
\newcommand{\comovingbox}{\ensuremath{\mathrm{Mpc}\,\mathrm{h}^{-1})^3}\xspace}
\newcommand{\comovingmass}{\ensuremath{{\mathrm{M}_\odot}\,\mathrm{h}^{-1}}\xspace}

\title[Environmental impact on galaxy mass distributions]{The MAGPI Survey: Impact of environment on the total internal mass distribution of galaxies in the last 5 Gyr}

\author[Caro Derkenne et al.]{
Caro Derkenne,$^{1,2}$\thanks{E-mail:caro.derkenne@hdr.mq.edu.au}
Richard M. McDermid,$^{1,2}$
Adriano Poci,$^{3}$
J. Trevor Mendel,$^{2,4}$
Francesco D'Eugenio,$^{5,6}$
\newauthor
Seyoung Jeon,$^{7}$
Rhea-Silvia Remus,$^{8}$
Sabine Bellstedt,$^{9}$ 
Andrew J. Battisti,$^{2,4}$
Joss Bland-Hawthorn,$^{2,10}$
\newauthor
Anna Ferr\'{e}-Mateu,$^{11,12}$ 
Caroline Foster,$^{2,13}$
K. E. Harborne,$^{2,9}$
Claudia D.P. Lagos,$^{9,14}$ 
Yingjie Peng,$^{15,16}$
\newauthor
Piyush Sharda,$^{2,4,17}$
Gauri Sharma,$^{18,19,20}$
Sarah Sweet,$^{2,21}$
Kim-Vy H. Tran,$^{2,13}$
Lucas M. Valenzuela,$^{8}$
\newauthor
Sam Vaughan,$^{1,2,22}$
Emily Wisnioski,$^{2,4}$
and Sukyoung K. Yi$^{7}$ \\
$^{1}$ Research Centre for Astronomy, Astrophysics, and Astrophotonics, School of Mathematical and Physical Sciences, Macquarie University,
NSW 2109, Australia\\
$^{2}$ ARC Centre of Excellence for All Sky Astrophysics in 3 Dimensions (ASTRO 3D), Australia \\
$^{3}$ Centre for Extragalactic Astronomy, University of Durham, Stockton Road, Durham DH1 3LE, United Kingdom,\\
$^{4}$ Research School of Astronomy and Astrophysics, Australian National University, Canberra, ACT 2611, Australia\\
$^{5}$ Kavli Institute for Cosmology, University of Cambridge, Madingley Road, Cambridge, CB3 0HA, United Kingdom\\
$^{6}$ Cavendish Laboratory - Astrophysics Group, University of Cambridge, 19 JJ Thompson Avenue, Cambridge, CB3 0HE, United Kingdom\\
$^{7}$ Department of Astronomy and Yonsei University Observatory, Yonsei University, Seoul 03722, Republic of Korea\\
$^{8}$ Universi{\"a}ts-Sternwarte, Fakul{\"a}t f{\"u}r Physik, Ludwig-Maximilians-Universit{\"a}t M{\"u}nchen, Scheinerstr. 1, 81679 M{\"u}nchen, Germany\\
$^{9}$ International Centre for Radio Astronomy Research, The University of Western Australia, 35 Stirling Highway, Crawley, WA 6009, Australia\\
$^{10}$ Sydney Institute for Astronomy, School of Physics, Building A28, The University of Sydney, NSW 2006, Australia\\
\emph{\normalsize Remaining affiliations are listed at the end of the paper}
}

\date{Accepted XXX. Received YYY; in original form ZZZ}

\pubyear{2022}

\begin{document}
\label{firstpage}
\pagerange{\pageref{firstpage}--\pageref{lastpage}}
\maketitle

\begin{abstract}
We investigate the impact of environment on the internal mass distribution of galaxies using the Middle Ages Galaxy Properties with Integral field  spectroscopy (MAGPI) survey. We use 2D resolved stellar kinematics to construct Jeans dynamical models for galaxies at  mean redshift $z \sim 0.3$, corresponding to a lookback time of $3-4$ Gyr. The internal mass distribution for each galaxy is parameterised by the combined mass density slope $\gamma$ (baryons $+$ dark matter), which is the logarithmic change of density with radius. We use a MAGPI sample of 28 galaxies from low-to-mid density environments and compare to density slopes derived from galaxies in the high density Frontier Fields clusters in the redshift range $0.29 <z < 0.55$, corresponding to a lookback time of $\sim 5$ Gyr. We find a median density slope of $\gamma = -2.22 \pm 0.05$ for the MAGPI sample, which is significantly steeper than the Frontier Fields median slope  ($\gamma = -2.00 \pm 0.04$), implying the cluster galaxies are less centrally concentrated in their mass distribution than MAGPI galaxies. We also compare to the distribution of density slopes from galaxies in Atlas3D at $z \sim 0$, because the sample probes a similar environmental range as MAGPI. The Atlas3D median total slope is $\gamma = -2.25 \pm 0.02$, consistent with the MAGPI median. Our results indicate environment plays a role in the internal mass distribution of galaxies, with no evolution of the slope in the last 3-4 Gyr. These results are in agreement with the predictions of cosmological simulations.

\end{abstract}




\section{Introduction}

The impact of environment on the galaxies they host is an open question in astronomy. The MAGPI \footnote{Based on observations obtained at the Very Large Telescope (VLT) of the European Southern Observatory (ESO), Paranal, Chile (ESO program ID 1104.B-0526).} (Middle Ages Galaxy Properties with Integral Field Spectroscopy) has been specifically designed to answer this question. MAGPI is a MUSE/VLT Large Program survey, aimed at spatially mapping the ionised gas and stellar properties of galaxies in a key transition period in cosmic time. The survey targets 60 massive galaxies around redshift $0.3$, and spans field to group environments. In addition to the primary targets, the observed fields contain a wealth of secondary objects across a range of redshifts. In this work, we use this unique data set to begin to untangle what role the large-scale structure surrounding a galaxy plays in shaping its internal mass distribution. 

The mass assembly of galaxies is dominated by the collapse and accretion of dark matter, in combination with the accretion of stars and gas at later times, mostly associated with infalling dark matter. The total mass (baryonic and dark matter) distribution can be described as a density profile of the form $\rho(r) \propto r^{\gamma}$, for which $\gamma$ ($<$ 0) is the total density slope, and indicates how steeply the mass density falls away with radius. 

Dynamical modelling has been applied in the local Universe to measure the total density slopes of galaxies, finding a clustering of values around, or just steeper than, $\gamma = -2$ \citep{thomas_2011_dynamical,cappellari_2015_slopes,serra_2016_linear,poci,bellstedt_sluggs_2018,li_manga_2019}. These dynamical modelling studies use either the highly general Schwarzschild orbit superposition technique \citep{schwarzschild_1979_numerical} or the simpler Jeans approach \citep{jeans_1922_motions}. The measurement of total density slopes in the Frontier Fields clusters by \citet{derkenne} pushed the Jeans dynamical modelling technique up to redshift $0.55$, and found no evidence for the evolution of the slope when comparing to the methodologically consistent study of \citet{poci} in the local Universe. However, indirect observations of dark matter using the rotation curves of star forming galaxies indicate dark matter haloes  were denser at earlier times, suggesting the total profile was correspondingly steeper as well \citep{sharma_2022_observational}.

The technique of gravitational lensing has been applied to observationally determine total density slopes up to redshift $\sim 0.8$. These studies indicate that density slopes are more shallow (less negative) at earlier times than in the local Universe, from $\sim -1.7$ at $z \sim 0.6$ to slightly steeper than  $-2$ locally \citep{ruff_sl2s_2011, bolton_boss_2012,rui_2018_strong}. Gravitational lensing requires the most massive of galaxies to act as lenses to the even more distant Universe, perhaps leading to a bias in those samples. Furthermore, dynamical modelling and gravitational lensing techniques make use of different assumptions, and it is unclear what impact these methodological systematics have on the resulting density slope measurements.

Cosmological simulations could provide an effective way to predict the total density slopes and their evolution with redshift. However, only relatively recent simulation suites have been able to achieve this in practice, as a result of their sufficiently high particle resolution. Magneticum is one such suite of simulations.
The Magneticum simulations are a set of hydrodynamical simulations with high enough resolution to probe scales within the half-mass radius of galaxies \citep{dolag_2015_magneticum}. These simulations indicate the average density slope of early-type galaxies was steep in the early Universe, of value  $\sim -3$ at $z \sim 3$ \citep{remus_co-evolution_2017}. At these early times, galaxy assembly is dominated by gas accretion from cosmic filaments and gas-rich mergers. Both are dissipative processes which result in compact galaxies with high density central regions. At later times dry mergers dominate, which are particularly effective at increasing the size of the galaxy without a significant increase in mass \citep{naab_2009_minor,hilz_2013_minor,remus_2013_dark}, due to their efficient redistribution of mass and angular momentum towards the outskirts of galaxies \citep{lagos_2018_connection}.
Consequently, the total density slope is driven from steep to shallower values of $\gamma \sim -2$ in the local Universe, although gas-rich mergers can steepen it again, although such mergers become rare at lower redshifts as the cold gas fraction of galaxies decreases \citep{remus_2013_dark}. Density slopes of $\gamma \approx -2$ are called `isothermal' as they correspond to the density slope of a singular isothermal sphere.

There is no consensus on the degree of the evolution of total density slopes across redshift from simulations. The magneto-hydrodynamical cosmological simulation set IllustrisTNG \citep{pillepich_2018_simulating}, shows little evolution of the total density slope below $z \sim 1$ \citep{wang_early_2019}. By comparison, Magneticum total density slopes changed from $\gamma \sim -2.3$ to $\gamma sim - 2.0$   between $z=1$ and $z=0$. These differences in prediction arise from the specific recipes used in each simulation set, such as gas cooling, stellar winds, and active galactic nuclei (AGN) feedback.

The evolution discussed above does not consider the host environment of a galaxy, which might also impact the internal total mass distribution. In high density cluster environments there are numerous processes which can impact galaxy evolution that cannot act on an isolated galaxy in the field. First, ram-pressure stripping acts to strip away the hot gas of galaxies as they move through the intergalactic medium \citep{gunn_1972_infall,boselli_2022_ram}. Second, galaxies in clusters have high velocity dispersions and tend to interact frequently at high speed; these encounters, so-called “fly-bys” and “harassments”, can tidally distort galaxies in both morphology and kinematics \citep{moore_1996_harassment,mihos_2003_interactions,Bialas_2015_harassment,scott_2018_abell}. Lastly, dense environments act to truncate the dark matter haloes of galaxies that have crossed the cluster core compared to galaxies in the field \citep{limousin_2007_truncation,limousin_2009_probing}. 

Considering only dark matter, a lensing study of 12 early-type galaxies found the density profile of dark haloes may depend on environment, with haloes in high density environments experiencing expansion due to satellite accretion, and thereby becoming more shallow than dark haloes in less dense environments \citep{oldham_2018_dark}. Total density slopes have been shown to depend on dark matter fractions both in observations and simulations \citep{sonn_sl2s_2013,remus_co-evolution_2017}. Galaxies with lower dark matter fractions tend to have correspondingly steeper density slopes due to the dominating stellar component within the measurement region. As dark halo properties can vary with environment, we might therefore expect the total density slope to vary as well.

Gas-poor, dry mergers are another key mechanism which can influence the evolution of a galaxy, and are thought to drive the total density slope towards isothermal values \citep{remus_2013_dark}. Using a sample of spectroscopically observed galaxies in the  zCOSMOS redshift survey \citep{lilly_2007_zcosmos}, \citet{ravel_2011_zcomos} revealed a correlation between local density and merger rates, with major mergers occurring preferentially in high density regions. \citet{watson_2019_galaxy} found enhanced merger rates for galaxies in clusters compared to field galaxies at redshift $\sim2$. This result has also been found by \citet{Jian_2012_environmental}  in the Millenium  simulations \citep{springel_2005_simulations}: they find a strong dependence between merger rates and local overdensities, with galaxies in overdense regions experiencing merger rates up to a factor of twenty greater than those in underdense regions, modulo the semi-analytic model used. Further evidence for increased merger rates in dense environments is the morphology-density relation. Elliptical galaxies are preferentially found in cluster environments, with their transformation from disc to elliptical morphologies caused by merger events \citep{deeley_2017_galaxy}. From this we might hypothesise that galaxies in groups and high density environments have had, on average, more mergers in their history than isolated galaxies of comparable mass, and should therefore have total density slopes closer to isothermal values. In this sense, isothermal total density slopes are indicative low-energy preferential mass distribution \citep{remus_2013_dark}.

Finally, we can look at the dependence of the size-mass distribution of galaxies with environment, as galaxies that are more compact are likely to have steeper total density slopes \citep{sonn_sl2s_2013,poci,derkenne}. \citet{maltby_2010_environmental} used a photometric sample of $\sim 1200$ galaxies across lenticular, spiral, and elliptical morphologies at redshift $0.2$, and found no statistically significant dependence of a galaxy's position on the size-(stellar) mass plane with environment, with the exception of some spirals, indicating that internal physical processes dominate over environment in driving size-mass evolution, in agreement with the results of \citet{grutzbauch_2011_how} at redshifts $0.1 - 0.4$. At the extreme end, it is noted that the most massive galaxies are found in correspondingly dense environment \citet{calvi_2013_impact}. In contrast to the above, \citet{cebrian_2014_effect} show galaxies in cluster environments tend to be slightly more compact than their field counterparts.

To begin to answer the complex question of the relation between total mass density slopes and environment we present Jeans dynamical modelling results and total density slopes of 30 galaxies from the MAGPI survey. Out of these 30, we present a total density slope analysis for 28 galaxies. We use 2D, resolved stellar kinematic maps to constrain the total gravitational potential  with Jeans anisotropic models.  This sample of galaxies is drawn from a mix of field and group environments, less extreme and less dense than the Frontier Fields clusters studied by \citet{derkenne}, and roughly comparable in environment to the local sample of \atlas galaxies modelled by \citet{poci}. We compare to these two other studies in particular as they both used an identical modelling technique and definition of the total potential to what we implement in this work, and their inclusion allows us to span environments: field galaxies, groups, and clusters.

By using a combined sample from studies that use a consistent methodology we can determine what impact the broad categorisation of environment has on the total density slope, and investigate whether there has been any evolution in the total density slope between the MAGPI and Frontier Fields samples in the `middle ages' compared to the \atlas sample in the local Universe. When the full MAGPI sample is observed we intend to use finer environmental metrics for analysis purposes. We also compare our observational results to the simulation predictions from the Magneticum, IllustrisTNG, and Horizon-AGN simulations \citep{dubois_2014_dancing}. 

Section \ref{sec:data} describes the MAGPI survey data, sample selection, and data-quality cuts made in this work. Section \ref{sec:data} also outlines all other data sets used for comparison in this work. Section \ref{sec:method} describes the measurement of the stellar kinematics, stellar potentials, Jeans models definitions and the calculation of the total density slope. The results are presented in Section \ref{sec:results}, with a discussion following in Section \ref{sec:discussion}. Conclusions are given in Section \ref{sec:conclusions}. 

Throughout this article, we use a flat $\Lambda$CDM cosmology with $H_0 = 70\ \kms \mathrm{Mpc}^{-1}$ and $\Omega_m = 0.3$. All scales are converted (angular and physical units) using the angular diameter distance given by the object's redshift and the above cosmology. At the nominal redshift of the MAGPI survey, $ z \sim 0.3$, this results in $\sim 4.45$ kpc per arcsecond. For comparison, a galaxy only 40 Mpc distant (local, and approximately the distance of galaxies in the \atlas comparison sample) has a  scale factor of $\sim 0.19$ kpc per arcsecond. 

\section{Data}
\label{sec:data}
In this section we give an overview of the MAGPI data we use to derive total density slopes, and the two main comparison data sets of \atlas and Frontier Fields. We follow with a description other literature data sets and cosmological simulations we compare to.

\subsection{MAGPI}
\label{sec:data_magpi}

This work uses the first tranche of observed data from the MAGPI\footnote{\url{https://magpisurvey.org}} survey (the 16 fields observed before August 2021 of the 56 total fields to be observed). MAGPI is a VLT/MUSE Large Program still gathering observations at the time of writing (PIs: Foster, Harborne, Lagos, Mendel, and Wisnioski). The survey targets 60 primary galaxies at $z \sim 0.3$ with stellar masses estimated at $\mathrm{M}_{\star}>7 \times 10 ^{10} \Msun$. Primary targets were selected from the GAMA survey, chosen so that the final MAGPI sample spans environments from isolated galaxies to galaxies in groups. The fields we present here are representative of the full sample. Two archival cluster data sets act as high-density environment supplements: clusters Abell 2744 at $z=0.308$ and Abell 370 at $z=0.375$ (program ID 096.A-0710, PI: Bauer and program IDs 095.A-0181 and 096.A-0496, PI: Richard, respectively). 
The survey aims and description are presented in full by \citet{MAGPI}. Briefly, the survey aims to map stellar and ionised gas components of galaxies in the relatively unobserved `middle ages' of the Universe, with comparable relative spatial resolution to local Universe studies like the Sydney-Australian-Astronomical-Observatory Multi-object Integral-Field Spectrograph (SAMI) survey \citep{croom_2012_sydney}, and Mapping Nearby Galaxies at Apache Point Observatory (MaNGA) \citep{bundy_2015_manga} survey. These observations will provide the key to unlocking the role of environment and assessing the impacts of merging, metal mixing, and energy sources in galaxies at this time. The survey includes foreground and background objects, extending the redshift baseline the survey can probe.

The program uses MUSE in the wide-field mode with the nominal wavelength range, resulting in spectra in the range $4650 - 9300\,$\AA\ in steps of $1.25\,$\AA\ per pixel. A ground-layer adaptive optics (GLAO) system is used to reduce the impact of seeing on the observations, resulting in a gap in wavelength coverage from $ 5780$ – $6050\,$\AA\ due to the GALACSI system sodium laser notch filter \citep{hartke_2020_MUSE}. The point spread function (PSF) full-width half-maximum (FWHM) is indicated for each field used in this work in Table \ref{tab:PSF_table}. The MAGPI fields cover a $1 \times 1$ arcminute field of view with $0.2$ arcsecond per pixel spatial sampling. Each field is observed across six observing blocks of $2 \times 1230$ s exposures, resulting in a total on-source integration time of 4.4 hours per field (246 hours on-source total). 

\begin{figure*}
	\includegraphics[width=2\columnwidth]{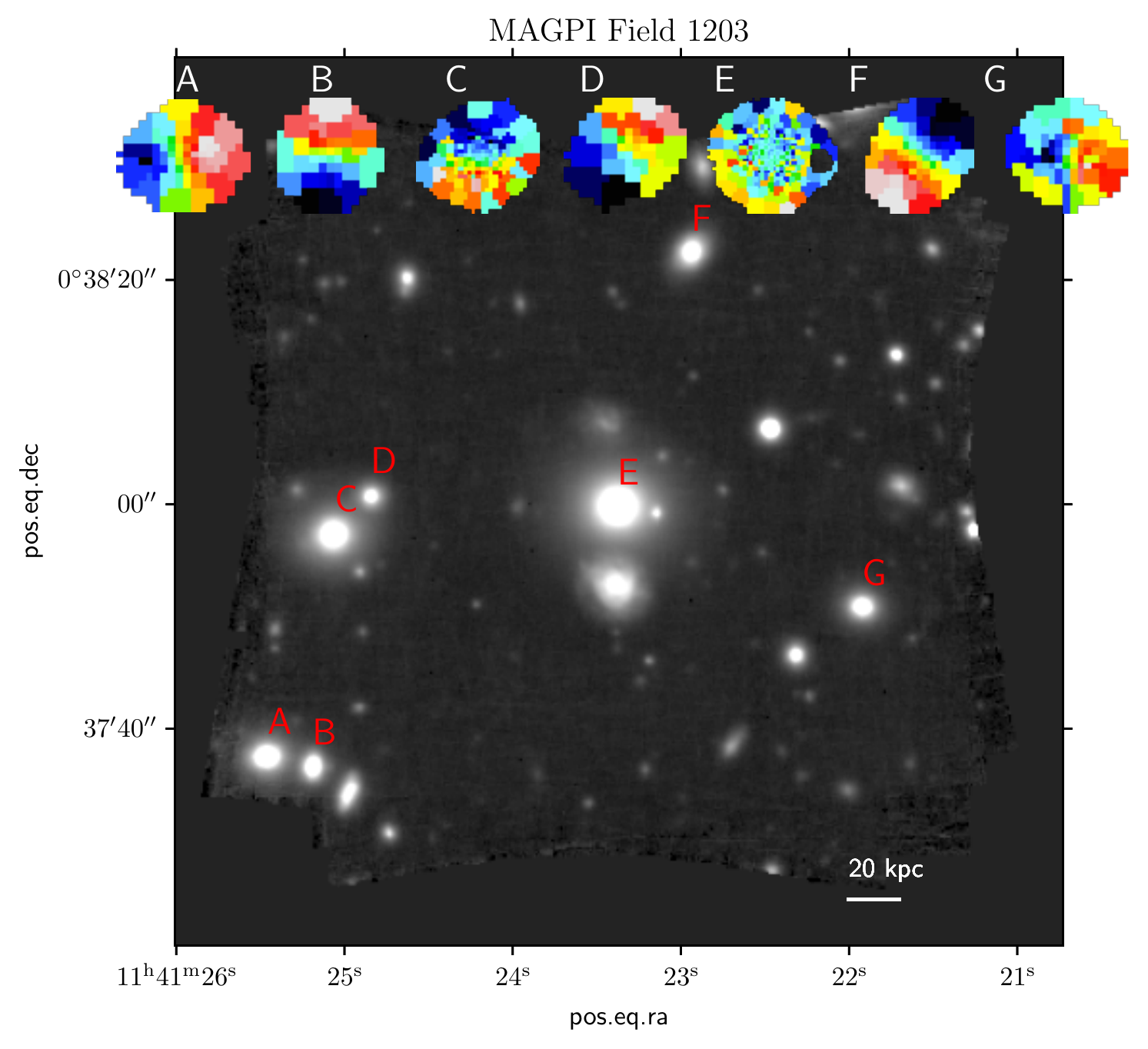}
	\caption{MAGPI field 1203 as an equivalent \textit{r}-band image extracted from the MUSE data cube by summing over a broad spectral band in the wavelength direction. Field 1203 is the richest field used in this work in terms of number of galaxies that meet the data quality requirements for measuring a total density slope. A 20 kpc scale at $z=0.3$ is inset. The letter labels indicate which inset velocity field match which object in the field. These fields are indicative only, and are not to scale spatially. Object E is the central target. The MAGPI IDs are as follows: A = 1203040085, B = 1203060081, C = 1203070184, D = 1203087201, E = 1203196196, F = 1203230310, G = 1203305151.}
    \label{fig:field_1203}
\end{figure*}

The data reduction process will be described in detail in an upcoming paper (Mendel et. al., in prep.), but here we provide a brief overview. Reduction was based on the MUSE reduction pipeline \citep{weilbacher_2020_data}, with sky-subtraction completed using Zurich Atmosphere Purge sky-subtraction software \citep{soto_2016_zap}. All fields have segmentation maps and estimates of galaxy structural parameters, such as the effective radius created using {\sc{ProFound}} \citep{robotham_2018_profound}. Each source identified within a field is post-processed to be situated at the centre of a `minicube' based on the main MUSE cube, which is sized to encompass the maximum extent of the ``dilated" segmentation map determined by {\sc{ProFound}}. An example of the richness of the MAGPI fields is shown in Figure \ref{fig:field_1203}. 

At present the highest resolution and deepest imaging data for MAGPI galaxies are the MUSE observations themselves. Mock images are created from the MUSE cube for each field in the SDSS $g_{\mathrm{mod}}$, $r$, $i$, and $z_{\mathrm{mod}}$ bands based on the mean flux density within the filter region. The $g$ and $z$ bands are only partially covered by MUSE, and are so labelled `mod(ified)'. 

We investigated the impact of creating Multi-Gaussian Expansions (MGEs) of the stellar light based on MUSE cubes with detailed, end-to-end simulations (see Appendix \ref{sec:simulations}). The potential issue is that poorly resolved galaxies will have a modelled light distribution which is less centrally peaked than it should be, resulting in a bias towards more shallow total density profiles. Our simulations show poorly resolved galaxies with multiple effective radii of kinematic coverage bias slopes towards steeper values, whereas resolved galaxies with less than an effective radius of kinematic coverage bias slopes to more shallow values.  We determine there is no inherent bias in the recovered parameters using MUSE-based MGEs so long as modelled galaxies have elliptical effective radii (\re) greater than or equal to the PSF FWHM, and stellar kinematics extend up to at least $1.5$ \re.  We applied these cuts to the sample, giving a sample of 30/72 MAGPI galaxies from 14/16 fields. Further cuts based on analysis and not data quality, outlined in Section \ref{sec:calculation}, adjust the final sample used to calculate the median total density slope value as 28 MAGPI galaxies. 

\begin{figure*}
	\includegraphics[width=2\columnwidth]{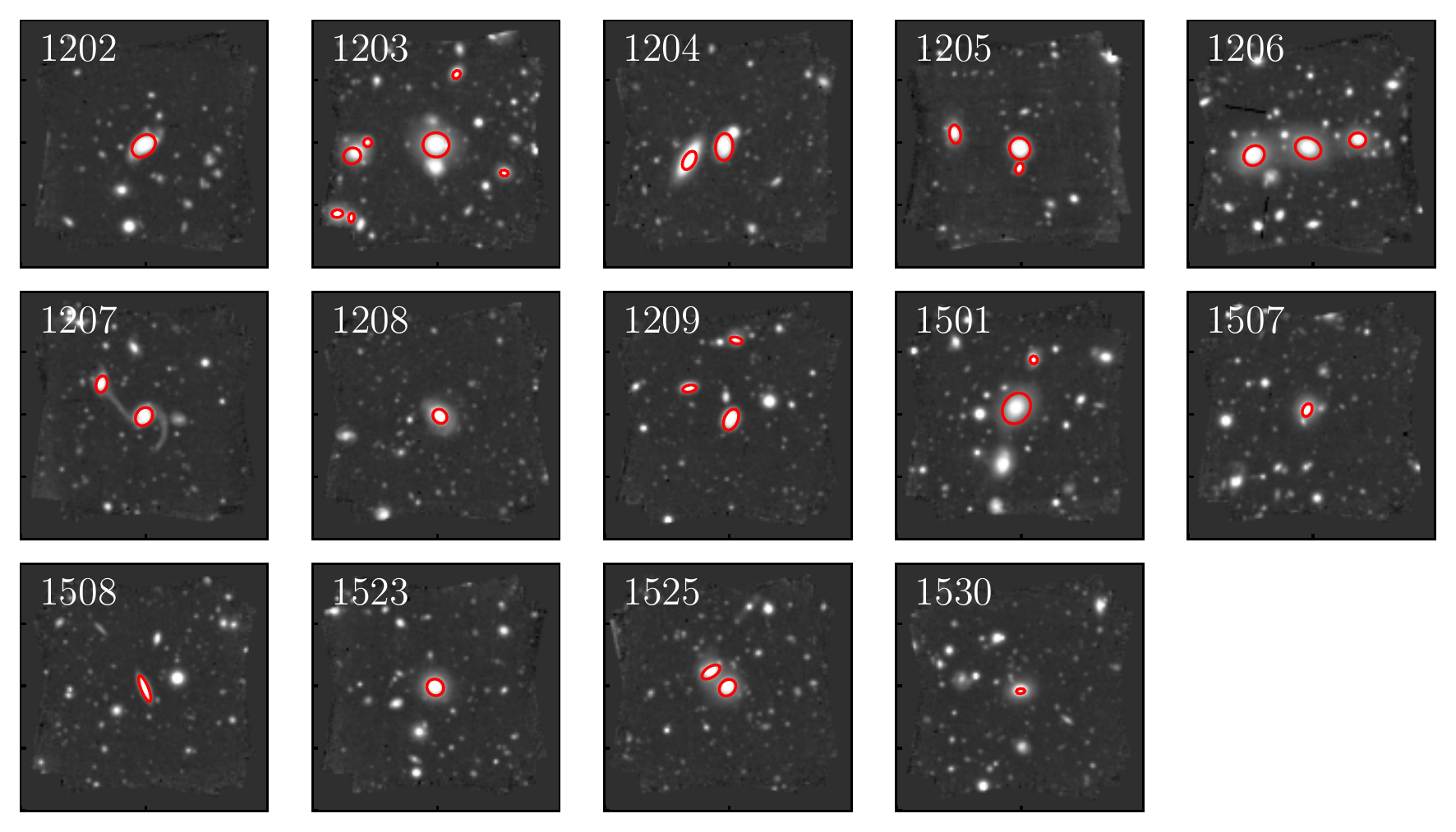}
    \label{fig:all_fields}
    \caption{All the MAGPI SDSS $r$-band images of fields used in this study. This is a subset (14/16) of the available MAGPI fields at the time of writing due to data quality cuts on the sample (see Section \ref{sec:data}). Inset on each image is the MAGPI field name. Red ellipses show all the MAGPI objects used in this work; the ellipses have major-axis equal to twice the major-axis from the PSF-deconvolved MGEs, for visibility.}
\end{figure*}

Figure \ref{fig:all_fields} shows the thumbnails of all MAGPI fields analysed to create the final sample of galaxies used in this work. 

\begin{table}
	\centering
	\caption{Column 1 lists the MAGPI fields used in this work, with galaxies selected based on data-quality cuts outlined in Appendix \ref{sec:simulations}. Column 2 lists the PSF FWHM in arcseconds, as measured from an MGE of a stack of point-sources in each field in the SDSS $r$-band, described in Section \ref{sec:mges}.}
	\label{tab:PSF_table}
	\begin{tabular}{lc} 
		\hline
		Field & PSF FWHM [arcsec] \\\hline
		1202 & 0.70  \\
        1203 & 0.63  \\
        1204 & 0.78  \\
        1205 & 0.67  \\
        1206 & 0.71  \\
        1207 & 0.62  \\
        1208 & 0.74  \\
        1209 & 0.59  \\
        1501 & 0.65  \\
        1507 & 0.59  \\
        1508 & 0.57  \\
        1523 & 0.59  \\
        1525 & 0.63  \\
        1530 & 0.69  \\
		\hline
	\end{tabular}
\end{table}

\subsection{Main comparison data sets: \texorpdfstring{\atlas}{atlas} and Frontier Fields}
We compare the MAGPI total density slopes to two main observational samples. The first is the Frontier Fields \citep{lotz_2017_frontier} sample of galaxies, modelled in \citet{derkenne}. These  modelled galaxies are drawn from the five Frontier Fields clusters that have optical integral field unit (IFU) data; Abell 2744, Abell 370, Abell S1063, MACS J0416.1-2403 and MACS J1149.5+223. These clusters were chosen as part of the Frontier Fields program for the likelihood of having a high number of gravitational lensing systems, and so necessarily represent very dense galaxy environments. The clusters have MUSE/VLT data cubes and \textit{Hubble Space Telescope} (\textit{HST}) data, and span $0.29 < z<0.55$. The sample as modelled in \citet{derkenne} is 90 early-type galaxies.

The second comparison sample are the 260 \atlas galaxies \citep{cappellari_2011_atlas}, a volume-limited sample of galaxies in the local Universe. Objects were selected to be closer than 42 Mpc with a  magnitude limit of $M_K < -21.5$ mag. IFU data was collected by the SAURON integral-field spectrograph mounted on the William Herschel Telescope on La Palma. Optical data was gathered using the  Wide-Field Camera at the 2.5m Isaac Newton Telescope, if imaging did not already exist as part of the Sloan Digital Sky Survey (SDSS). \citet{poci} used Jeans dynamical models to constrain the total gravitational potential for 258 of these galaxies (Model 1 in that work). In this work we only include \atlas galaxies for analysis that have a data quality flag $\geq 1$ as defined in \citet{cappellari_2013_benchmark} (187/258 galaxies). The dynamical models of \citet{derkenne} and \citet{poci} use the same formulation of the total gravitational potential.

\subsection{SLUGGS and Coma}
\citet{thomas_2011_dynamical} constructed Schwarzschild dynamical models of 17 galaxies in the Coma cluster, a massive cluster in the local Universe ($z = 0.0231$), in order to constrain their luminous and dark matter distributions. The models assumed separate luminous (from an MGE) and dark matter (a Navarro-Frenk-White halo) components  to form the total density profile, whereas in this work we use a single profile to describe the total density profile. Models were constructed using the orbit superposition technique of Schwarzschild modelling \citep{schwarzschild_1979_numerical}. The dynamical models of \citet{thomas_2011_dynamical} provide a valuable data set of total density profiles measured in a massive cluster environment in the local Universe. We use the total density profiles as constrained by the models presented in \citet{thomas_2011_dynamical}, but re-calculate the total density slope within the radial range $r = [\re/10, 2\,\re]$ to match that used for MAGPI, Frontier Fields, and \atlas galaxies (see Section \ref{sec:calculation}).

Another local sample of galaxies with constrained total density profiles from dynamical models are the SLUGGS (SAGES Legacy Unifying Globulars and GalaxieS Survey) galaxies, as presented by \citet{bellstedt_sluggs_2018}. Globular cluster kinematics at large radii were combined with stellar kinematics from \atlas data at small radii for 22 galaxies. Many modelling choices for the SLUGGS sample overlap with the modelling approach in this work; a double power-law potential was used to describe the total gravitational potential with a fixed transition between a free inner slope and constant outer density slope at 20 kpc (see Section \ref{sec:models} for the Jeans model descriptions used in this work). However, a hyperparameter was used to independently weight the globular cluster data and stellar kinematics in order to fit the combined \vf fields. It is as of yet unclear what systematics this introduces compared to a method that uses stellar IFU data alone. For comparisons we use the most likely parameters from the {\sc{emcee}} posterior distributions from the models as given by \citet{bellstedt_sluggs_2018}. As with the Coma data, we re-calculate the SLUGGS total density slopes within the radial range of $r = [\re/10, 2\,\re]$.

\subsection{Simulations: Horizon-AGN, Magneticum, and IllustrisTNG}

The first simulation comparison sample we use comes from the Horizon-AGN simulation. The Horizon-AGN simulation uses the adaptive mesh refinement code RAMSES \citep{teyssier_2002_cosmological} and has a  co-moving volume of $(100$ \comovingbox with  a dark matter particle resolution of $m_{\mathrm{DM}} = 8 \times 10^{7}$ \Msun. The simulations incorporates gas cooling, stellar winds from {\sc{starburst99}} \citep{leitherer_1999_starburst}, and AGN feedback following \citet{dubois_2012_self}.

We present Horizon-AGN  simulation total density slopes calculated in the radial range $r = [\re/10, 2\,\re]$ with a single power-law fit, to closely match the method used to measure total density slopes for MAGPI galaxies from the total density profile. The Horizon-AGN sample presented here is comprised of galaxies from 14 simulation snapshots spanning $0.018 < z < 0.305$, with stellar masses between $10^{11}-10^{12}$ \Msun. Because of gravitational softening within the simulations, no region within a physical scale of 2 kpc is used to fit the total density slope, although we do not expect this choice to drive the observed trends. The median value of 1 \re for the Horizon-AGN sample is 1 kpc, which is smaller than the 2 kpc bound imposed. However, the single power-law density slope is not particularly sensitive to these innermost regions of simulated galaxies. 

The second simulation data set we use for comparisons are the  published total density slopes from the Magneticum Pathfinder simulations as calculated by \citet{remus_co-evolution_2017}. For that work, the cosmological box volume  is $(48$ \comovingbox, with a dark matter particle resolution of $m_{\mathrm{DM}} = 3.6 \times 10^{7}$ \comovingmass. To avoid the effects of softening and resolution, total density slopes are calculated on the radial range of $0.4-4$ \rmass, where \rmass indicates the 3D half-mass radius. As an additional criteria, if $0.4$ \re is smaller than twice the gravitational softening length, then twice the softening length is adopted as the inner radial bound. This change of bound only occurs for very few galaxies.  The simulations are run with smoothed particle hydrodynamics in the form of the {\sc{gadget3}} code \citep{hirschmann_2014_cosmological}, and include gas cooling, passive magnetic fields, and AGN feedback \citep{springel_2003_cosmogloical,springel_2005_modelling,fabjan_2010_simulating}. The studied galaxies have stellar masses  greater than $\mathrm{M}_{\star} = 5 \times 10^{10}$ \Msun.

Finally, we also compare to slopes calculated by \citet{wang_early_2019} using galaxies in the IllustrisTNG simulations. The particular cosmological box volume used in that work is $( 110.7$ \comovingbox with a dark matter particle resolution of $m_{\mathrm{DM}} = 7.5 \times 10^{6}$ \Msun. IllustrisTNG uses the adaptive moving-mesh code {\sc{arepo}} \citep{springel_2010_galilean}. Stellar and AGN feedback follows the prescriptions of \citet{pillepich_2018_simulating} and \citet{weinberger_2018_supermassive}. The published slopes of \citet{wang_early_2019} are calculated using the range $0.4-4$ \rmass, the inner bound again set to avoid gravitational softening effects. The sample includes early-type galaxies with stellar masses concentrated within an aperture of 30 kpc  between $\mathrm{M}_{\star} = 10^{10.7}$ \Msun and  $\mathrm{M}_{\star} = 10^{11.9}$ \Msun.

We stress that the implementations of gas physics and feedback are different in these simulations, and these differences will necessarily impact the internal mass distribution of simulated galaxies. For example, the exact implementation of gas cooling in the simulations can lead to differences in the mass distribution,  as cooling can cause star formation at later times and increase galaxy central densities.

\subsection{Other literature data sets}
We collate several other literature data sets against which we compare our results. First, \citep{li_manga_2019} constructed Jeans models for a  sample of 2110 nearby galaxies from the MaNGA survey. Second are density slopes measured using galaxies from the Sloan Lens ACS (SLACS) survey \citep{bolton_2006_sloan}, published in \citet{auger_sloan_2010} and \citet{barnabe_two_2011}, and based on gravitational lensing techniques. Thirdly, we compare to density slopes presented in \citet{bolton_boss_2012}, which combined strong gravitational lensing with a dynamical analysis to measure total density slopes, utilising data from SLACS and the BOSS Emission Line Lensing Survey (BELLS). Fourth are the density slopes from the  Strong Lenses in the Legacy Survey (SL2S), as published by \citet{ruff_sl2s_2011} and \citet{sonn_sl2s_2013}. Finally, we compare to density slopes derived from a combined analysis of BELLS and BELLS GALaxy-Ly$\alpha$ EmitteR sYstems (GALLERY) Survey, and SL2S galaxies, as published in \citet{rui_2018_strong}, again using a strong gravitational lensing technique.

\section{Methods}
\label{sec:method}

\subsection{Multi-Gaussian Expansion of stellar light}
\label{sec:mges}
The stellar light distribution is parameterised as a series of 2D Gaussians, to be used as input to the Jeans dynamical models as the luminous tracer. An assumed viewing angle de-projects the observed surface brightness to a 3D luminosity density \citep{emsellem_1994_multi}. Here we use the Python package {\sc{mgefit}} to fit the stellar light \citep{MGE}. MGEs were fit to SDSS $r$-band,  chosen for image depth. 

Sources other than the target galaxy were masked using the {\sc{ProFound}} \citep{robotham_2018_profound} segmentation maps. The initial fit area was cut as a square of sides $20$ \re (from MAGPI data {\sc{ProFound}} outputs, as described in 
\citealp{MAGPI}) to ensure the fit was not artificially contracted, as this thumbnail size includes sky-dominated boundaries. The threshold level of the fit was determined by the median of the set of pixels associated with no source according to the segmentation maps. 

A regularised fit was performed on the $r$-band mock images, meaning the roundest possible solution was found with the least number of Gaussian components that did not perturb the absolute deviation of the model excessively far from the best-fit model (here we allow for a $3$ per cent deviation, empirically determined). Regularisation allows for the largest possible range of galaxy inclinations, still consistent with the data, to be tested in the subsequent Jeans models.

MGE models of the PSF  were made using stacked point sources in each of the MAGPI fields. The MGE surface brightness model of each galaxy was then analytically convolved with the corresponding PSF and the fit optimised against the observed galaxy surface brightness \citep{MGE}. This optimised, underlying MGE surface brightness model with no PSF convolution was used as the luminous tracer in the Jeans models described in Section \ref{sec:models}.

\begin{figure*}
	\includegraphics[width=2\columnwidth]{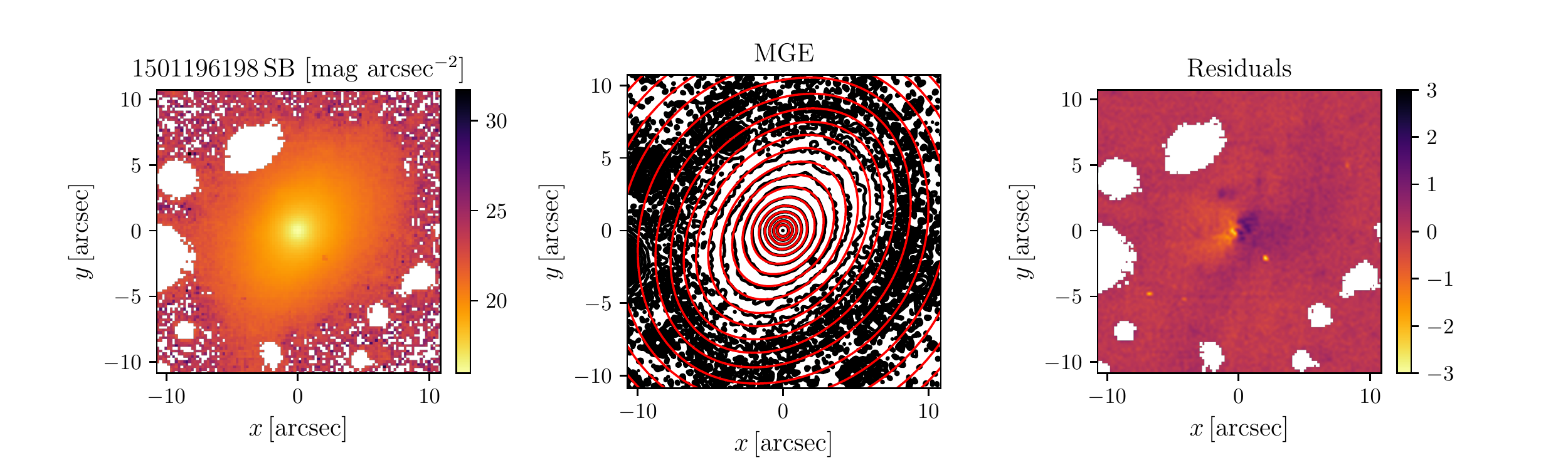}
	\caption{Left: The SDSS \textit{r}-band surface brightness of MAGPI object 1501196198, in the STMAG magnitude system \citep{stone_1996_spectrophotometry}. The white regions show masked pixels due to neighbouring galaxies. These pixels were excluded from the subsequent fit. Middle: The PSF-convolved MGE model in red, over plotted on the black galaxy isophotes, in steps of 0.5 magnitudes per square-arcsecond. The MGE contours mostly cover the image contours. Right: The residuals (data - MGE model) divided by the noise of the flux image.}
    \label{fig:mge_example}
\end{figure*}

This work uses Jeans models with a total potential which sets the scaling of the gravitational potential, meaning the scaling of the luminous component is irrelevant; only the relative scaling, widths and shapes of the luminous Gaussians impact results. For each galaxy, $N$ 2D Gaussian components are used to describe the light, consisting of the total image counts enclosed by the $N$\textsuperscript{th} Gaussian in the image units, its sigma-width in pixels ($\sigma$), and the observed axial ratio of the Gaussian ($q$). To prepare the MGEs as inputs to the Jeans models the Gaussian enclosed counts ($C$) in image units were re-normalised  to a peak surface brightness ($C_0$) for each of the $N$ components:
\begin{equation}
    C_0 = \frac{C}{2\uppi\sigma^2q}.
\end{equation}
The Gaussian dispersions were converted from pixel to observed units by multiplying by the image pixel scale. An example MGE model for a MAGPI galaxy is shown in Figure \ref{fig:mge_example}.

At this point, galaxies with very poorly fitting MGEs (as judged by visual inspection) were excluded from the sample. This exclusion is mainly restricted to galaxies with high elliptical shapes but small effective radii where the PSF-deconvolution could not be accurately performed and the resulting model bore little resemblance to the galaxy's light distribution. This means we exclude some small, highly flattened systems. Our final sample spans up to 0.73 in ellipticity, with a mean ellipticity of 0.26; this distribution is comparable to the \atlas survey \citep{emsellem_2011_census}.

The effective radius was derived as the circularised arcsecond extent which contains half the measured light of the galaxy. These radii are used in all subsequent analysis, and are provided in Table \ref{tab:results_table}.

\subsection{2D stellar kinematics}
\label{sec:kinematics} 
Full-spectrum fitting using the Python package {\sc pPXF} was used to derive the stellar kinematic fields \citep{ppxf_1,ppxf_2}. In this work, we set the higher order Hermite coefficients (h3 and h4) to zero which ensures the line-of-sight velocity distribution is described by a Gaussian. From the velocity and velocity dispersion fields, the \vf field is constructed as \vf$=\sqrt{v^2 + \sigma^2}$. 

We used the Indo-US template library \citep{valdes_2005_indous} as it has a sufficiently high spectral resolution of $1.35\,$\AA$\,$FWHM so that the stellar templates could be convolved with a Gaussian kernel to match the mean spectral resolution within the spectral window used (for MUSE this is $\sim 2.5\,$\AA, \citealp{mentz_2016_abundance}).

The minicube of each MAGPI source was thresholded to a signal-to-noise of 1.5 per pixel in the fitted spectral region, which is approximately $3500 - 6500\,$\AA\ in rest-frame but depends on the redshift of each galaxy. Each galaxy was Voronoi binned  to a signal-to-noise of 10 per spectral pixel per bin using the Python package {\sc{vorbin}} \citep{cappellari_2003_voronoi}. The binning ensures a high enough signal-to-noise across the field to recover kinematics without excessively truncating the radial coverage. It is also exactly matched to the binning process used by \citet{derkenne}, which ensures the resulting dynamical models can be fairly compared against that work.

We restricted the range of Indo-US templates for each spectrum fit by first using the MGE model parameters (effective radius, ellipticity) to create an elliptical 1 \re co-added spectrum. This central spectrum was fitted with the entire Indo-US template library, after which the template library for the spectral fit of each individual bin was limited to the top-weighted 20 template spectra of the central fit. In detail, a fifth-order multiplicative polynomial was used in the fit, with no additive polynomials.

\begin{figure*}
	\includegraphics[width=2\columnwidth]{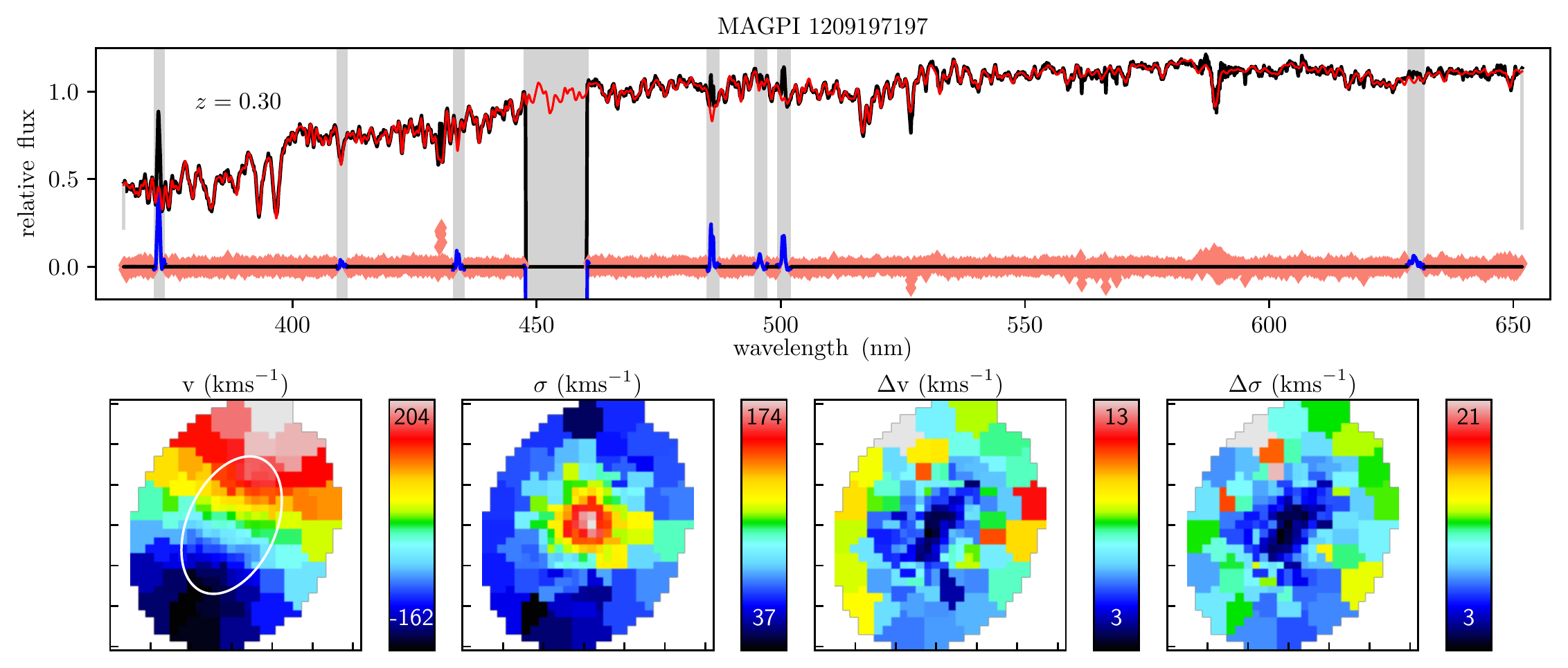}
	\caption{Top row: The full-spectrum fit of MAGPI galaxy 1209197197. The spectrum is shown in black, and is the co-added spectra within an elliptical aperture with semi-major axis equal to 1 \re. The best-fit to the central spectrum is shown in red, with excluded regions shown in grey (including the notch filter from the  GALACSI laser band). Blue indicates masked ionised gas emission. The residuals from the fit are shown in pink. The spectral fit was made using {\sc pPXF} \citep{ppxf_1,ppxf_2}. Bottom row: The extracted velocity, velocity dispersion, 1-sigma uncertainty on the velocity, and 1-sigma uncertainty on the velocity dispersion. A white ellipse on the velocity panel shows 1 \re. The x- and y-axis ticks are 1 arcsecond intervals.}
    \label{fig:ppxf}
\end{figure*}

Uncertainties on the fitted velocity and velocity dispersions were estimated by randomly shuffling the residuals between the observed spectrum and the best-fit spectrum and adding them back onto the observed spectrum in a Monte-Carlo process across 100 iterations. Resulting uncertainties were reduced by $\sqrt{2}$ to account for the doubling of the noise this process involved. An observed spectrum in the rest-frame, the best-fit determined by {\sc pPXF}, and the kinematic fields with associated errors are shown in Figure \ref{fig:ppxf}. At this point in the analysis, galaxies with contaminated kinematics due to a projected or actual neighbour were removed upon visual inspection (2 objects). 

\subsection{Jeans Modelling}
\label{sec:models}

We use Jeans anisotropic models (JAM) created using the Python package {\sc{jampy}} to constrain the total gravitational potential \citep{cappellari_measuring_2008}. Following the {\sc{jampy}} method, a galaxy is a large system of stars, the velocities and positions of which can be described using a distribution function. If we assume the system is in a steady state with a smooth gravitational potential, then the distribution function must obey the collisionless Boltzmann equation. However, this equation has infinite solutions, which necessitates further assumptions. To achieve a unique solution, the gravitational potential is assumed to be axisymmetric, and in the implementation of {\sc{jampy}} used here it is assumed that the velocity ellipsoid is aligned with the cylindrical coordinate system. The velocity ellipsoid is the 3D distribution of velocity dispersions in the radial, azimuthal, and $z$ directions, where $z$ is along the axis of symmetry. Aligning the velocity ellipsoid with cylindrical coordinates is observationally motivated by the fact that anisotropy in galaxies can be well characterised as a flattening of the velocity ellipsoid in the $z$ direction \citep{cappellari_2007_sauron}. In JAM, the global anisotropy term is defined as:

\begin{equation}
  \beta_z \equiv 1 - \left(\frac{\sigma_z}{\sigma_R}\right)^2,
\end{equation}
\label{anisotropy_equation}
where $R$ denotes the radial direction. The inclusion of this terms allows deviations from perfectly isotropic orbital structures.

These assumptions yield the general axisymmetric, anisotropic Jeans equations, given in equations 8 and 9 in \citet{cappellari_measuring_2008}. An assumed inclination, anisotropy, and gravitational potential is used to predict the root mean square velocity, defined as $v_{\text{rms}} = \sqrt{v^2 + \sigma^2}$. The observed velocity and velocity dispersions fields from Section \ref{sec:kinematics} were combined into a single measurement per Voronoi bin. The weighting of each bin in the models was determined by use of the error vector

\begin{equation} 
  \delta v_{\text{rms}} = \frac{\sqrt{(v  \delta v)^2  + (\sigma   \delta \sigma)^2}}{v_{\text{rms}}}.
\end{equation}
\label{eq:vrms_uncertainty}

The JAM model predictions are analytically convolved with the circular PSF of each field (see Table \ref{tab:PSF_table}) before comparison to the observed \vf field. In this sense, the optimised \vf field is PSF deconvolved, and the constrained total density profile should be likewise independent of the PSF. The model quality is judged against the observed \vf field using a $\chi^2$ statistic, taking each Voronoi bin as a degree of freedom.

In our models the total gravitational potential (baryons and dark matter) is described as a spherical double power-law of the form
\begin{equation}
  \rho(r) = \rho_s \left(\frac{r}{r_s}\right)^{\gamma'} \left(\frac{1}{2} + \frac{1}{2}\frac{r}{r_s}\right)^{-\gamma'-3}, 
  \label{eq:rho_tot}
\end{equation}

where $\rho$ is the total density as a function of galactocentric radius, $\gamma'$ is the \textit{inner} density slope, $r_{\rm{s}}$ is the `break' radius at which the potential changes from a free inner slope to a fixed logarithmic outer slope of $-3$, fixed at 20 kpc, and $\rho_{\mathrm{s}}$ is the density of the total profile at radius $r_{\mathrm{s}}$. This profile is a `Nuker' profile, a type of generalised Navarro-Frenk-White (NFW) profile \citep{lauer_1995_nuker,navarro_1997_nfw}. This profile is input to the models as a 1D MGE. The total density slope $\gamma$ is calculated from this profile once the parameters have been fit to the observed data, discussed in Section \ref{sec:calculation}, and is the defined as

\begin{equation}
    \gamma = \frac{d\log(\rho)}{d\log(r)}.
\label{eq:define_gamma}
\end{equation}

The formulation of the total potential given in Equation \ref{eq:rho_tot} allows for a luminous component in the inner regions of the galaxy (as traced by the luminous MGE described in Section \ref{sec:mges}) situated within a dark matter halo. We implement a spherical potential and use a fixed break radius of 20 kpc to be consistent with the dynamical models of \citet{poci} and \cite{derkenne}. We explore the impact of this choice of break radius in Appendix \ref{sec:app_break_radius}.

Furthermore, \citet{poci} found that considering a non-spherical halo did not change the total density slope values or result in improved models. \citet{bellstedt_sluggs_2018} found that a variable break radius could not be well constrained by the data, despite the large radial coverage of SLUGGS \citep{brodie_2014_sluggs} data. When the break radius was left free, \citet{bellstedt_sluggs_2018} found that the derived total density slopes were consistent (within error) with those found using a 20 kpc fixed break radius. 

In our definition of the total potential there are four free parameters:
\begin{enumerate}
    \item The inner density slope $\gamma'$, bounded between $-3.5$ and $0.5$.
    \item The log density at the break radius $\rho_{\mathrm{s}}$, bounded by $-4$ and $0$ with $\rho$ in units of \Msun$\mathrm{pc}^{-3}$. 
    \item The inclination of the galaxy $i$, used to de-project the observed luminous surface density. The maximum inclination allowed is 90 degrees (edge on). The minimum inclination is set as $i= \cos^{-1}q$, where $q$ is the smallest observed axial ratio from the luminous MGE fit.
    \item The global anisotropy $\beta_z$, which accounts for the flattening of the velocity ellipsoid in the z-direction, bounded by $-0.5$ and $0.5$.
\end{enumerate}

\begin{figure*}
	\includegraphics[width=2\columnwidth]{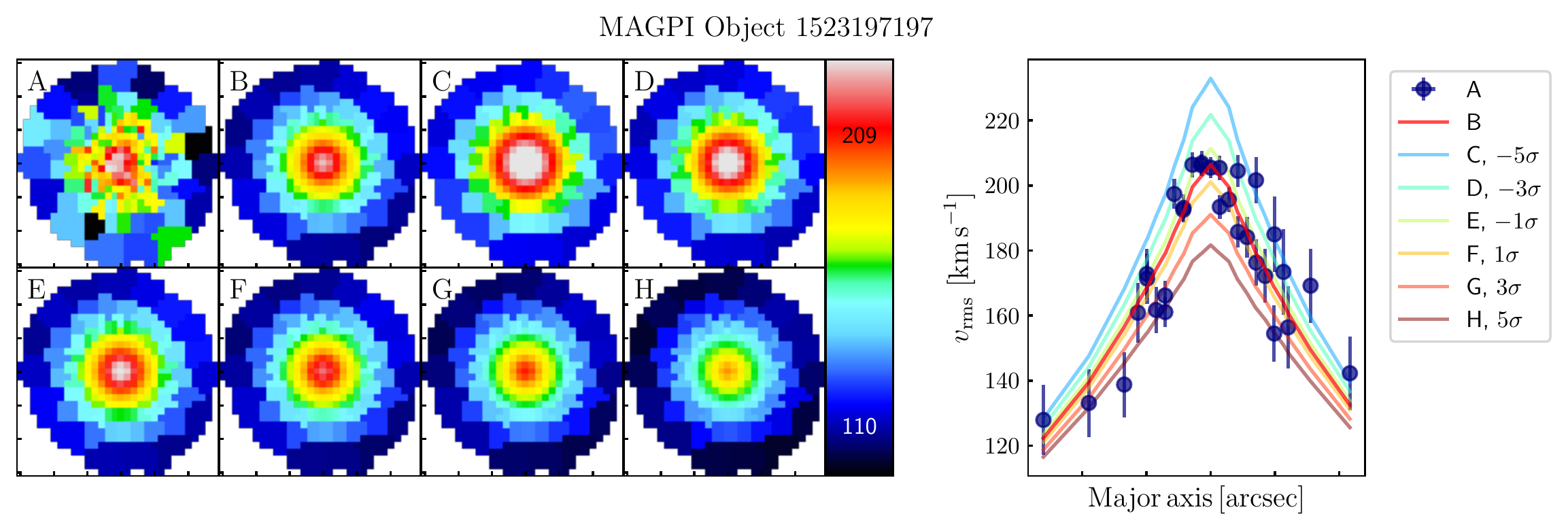}
	\caption{Top row: Panel A: the observed \vf field for MAGPI galaxy 1523197197. Panel B: the median JAM model from the posterior distributions of the {\textsc{emcee}} process. Panels C-H: The resulting \vf field at $\pm 1,3,5\sigma$ in $\gamma'$ around the median solution to illustrate the changes in structure and scale of the \vf field.  The x- and y-axis ticks are 1 arcsecond intervals. The total density slope and $1\sigma$ uncertainty for this object is $\gamma = -2.28^{+0.01}_{-0.01}$. Right: A \vf major axis cut of the models and observed data, corresponding to the fields shown in Panel A. The \vf errors are shown for the observed data (A).}
    \label{fig:random_draw}
\end{figure*}

To estimate the parameter posterior distributions the Python package {\sc{emcee}} \citep{foreman_2013_emcee,hogg_2018_emcee} was used, with hundreds of thousands of independent model evaluations made to estimate the posterior distribution for each free parameter. We used 30 independent walkers with a maximum of 10,000 steps, although chains could terminate in fewer steps if the deemed converged by auto-correlation time estimates. The $\chi^2$ statistic of the observed and modelled \vf fields were used to estimate the likelihood of any particular model. Flat (uniform) priors were assumed, that is, the likelihood of a particular model was informed \textit{only} by the $\chi^2$ statistic. Figure \ref{fig:random_draw} shows a set of Jeans models around the median model from the posterior distribution to visualise the {\sc{emcee}} process and associated uncertainties.

\begin{figure}
	\includegraphics[width=1\columnwidth]{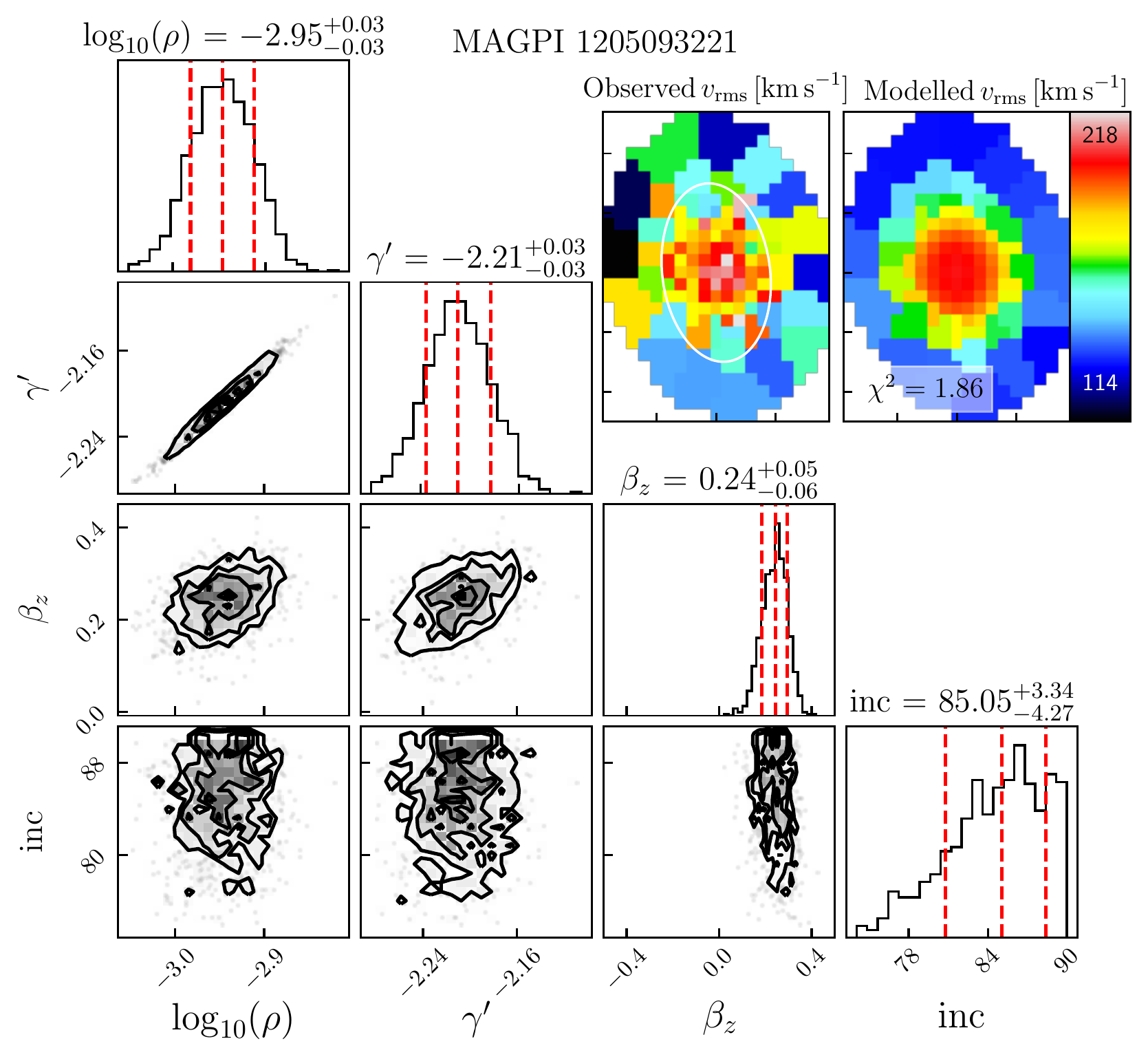}
	\caption{An example corner plot from the {\textsc{emcee}} process for MAGPI object 205093221, with the 16\textsuperscript{th}, 50\textsuperscript{th}, and 84\textsuperscript{th} percentiles of each posterior distribution shown as red dashed lines. Diagonal panels show estimates of the posterior probability distributions, marginalised over one parameter. Off-diagonal panels show the estimated posterior distribution marginalised over every pair of parameter in two dimensions. In general there is some covariance between $\rho_\mathrm{s}$ and $\gamma'$, as evidenced by the ellipsoidal two-dimensional estimated posterior distribution for those parameters. This covariance is due to both parameters affecting the enclosed mass of the model. Inset, Left: The observed \vf field with 1 \re shown as a white ellipse. Inset, Right: The \vf model corresponding to the median of the posterior distributions. The reduced $\chi^2$ value is also given. The x- and y-axis ticks are 1 arcsecond intervals.}
    \label{fig:corner}
\end{figure}

Figure \ref{fig:corner} shows the estimated marginalised posterior distributions for each of the four free parameters in the Jeans models, with the \vf field constructed from the median of the posterior distributions inset. None of the galaxies in our sample have best-fitting values of the inner total density slope on the boundaries of the original prior. This indicates that the constraints for the parameter chains are data-driven, not prior-driven.

Although the fixed break radius of 20 kpc is outside the kinematic coverage for all but two of the galaxies in the MAGPI sample, the density at this radius is still well constrained. This is due to the fact that the density at the break radius affects the enclosed mass, and therefore the level of the \vf field, for all radial positions interior to the break radius where kinematic data does exist.

\subsection{Calculation of the total slope \texorpdfstring{$\gamma$}{gamma}}
\label{sec:calculation}

A double power-law is used in this work, which includes a smooth transition between inner and outer regimes at a fixed transition radius of 20 kpc. The slope (Equation \ref{eq:define_gamma}) itself is measured as a single power-law fit across some radial range. The specific radial range over which an average total density slope is measured will therefore have an impact on the result. We argue it is crucial to compare the MAGPI sample and main comparison samples, \atlas and Frontier Fields, over a consistent physical radial range.

Due to the differing redshifts of the samples, it is impossible to compare all samples across a common radial range without including radial regions that are either within the PSF or beyond the kinematic coverage of the data. The PSF region for MAGPI and Frontier Fields objects is comparable in size to the galaxy effective radii, whereas for \atlas the PSF is less than 10 per cent of the median effective radius for that sample.

To make the comparison of total density slopes across studies fair, we re-compute total density slopes for the \atlas and Frontier Fields samples using identical radial bounds as for MAGPI galaxies. Across samples and for all galaxies we define the radial bounds as $r = [\re/10, 2\,\re]$. The total density slope is then calculated as a 1D power-law fit to the analytic total profile within these radial bounds. 

For Frontier Fields, there are 68/90 galaxies with kinematics at least beyond $2$ \re, however due to the PSF of $\sim 0.6 \mathrm{\,\,arcsec\,FWHM}$ across the fields the total slope calculation is extrapolated to radii smaller than the PSF.

For \atlas, there are only 18/258 galaxies with \vf fields up to $2$ \re \citep{emsellem_2011_census}, and so for most objects we measure the total density slope by extrapolating beyond the region of kinematics used to constrain the total potential. That is, it is assumed that the total potential constrained by kinematics up to $1$ \re\, is the same potential that would be constrained if the object had kinematics extending to $2$ \re. The end-to-end simulations presented in Appendix \ref{sec:simulations} imply this is the case since there is no change on the input versus output inner density slope as the kinematic coverage is increased (see far right columns of Figure \ref{fig:muse_image_simulations_bias}). 

Of the 30 galaxies in MAGPI judged as viable for measurement of the total slope, 20 have kinematics that extend to 2 \re, and 10 have kinematics that extend somewhere between 1.5 and 2 \re. For these 10 objects the total slope measurement is extrapolated beyond the region of kinematics, and for all 30 objects the total slope measurement is extrapolated to within the PSF. 

Finally, there is a break in the relation between velocity dispersion and the total density slope \citep{poci,li_manga_2019}. Galaxies with a central velocity dispersion below $\sim 100$ \kms have a strong trend with total density slope, but above $\sim 100$ \kms there is only a mild correlation driven by morphological type. 
To remove the velocity dispersion as confounding factor, we further select galaxies across the samples that have a central velocity dispersion greater than $100$ \kms. This cut mainly affects the \atlas sample. This final sample cut leaves 28 galaxies from MAGPI, 150 galaxies from \atlas, and 64 galaxies from the Frontier Fields clusters.

Total masses were calculated as twice the (spherical) mass integrated from the best-fit potential within 1 \re, as in \citet{cappellari_2013_benchmark}. This calculation depends on the effective radius, density at the break radius, $\rho_{\mathrm{s}}$, and inner density slope, $\gamma'$. Errors on the total masses were determined using the standard deviation of 100 masses calculated from random samples of the {\sc{emcee}} chains. Errors on the total slope were calculated in the same way, by drawing 100 random samples of the estimated posterior distribution to construct the total density profile, and measuring the single power-law slope from that.  All total masses and slopes are presented in Appendix \ref{sec:results_app}, Table \ref{tab:results_table}. 

We have re-calculated the total density slopes for Coma, SLUGGS, and Horizon-AGN, to match the radial range used for the MAGPI, Frontier Fields, and \atlas sample, being $r = [\re/10, 2\,\re]$. While this does not alleviate issues of methodological biases, we believe it does aid comparison between studies. A summary of the radial ranges and slopes for all studies are given in Appendix \ref{sec:results_app}, Table \ref{tab:comparison_gamma}.

\section{Results}
\label{sec:results}
\begin{figure*}
	\includegraphics[width=2\columnwidth]{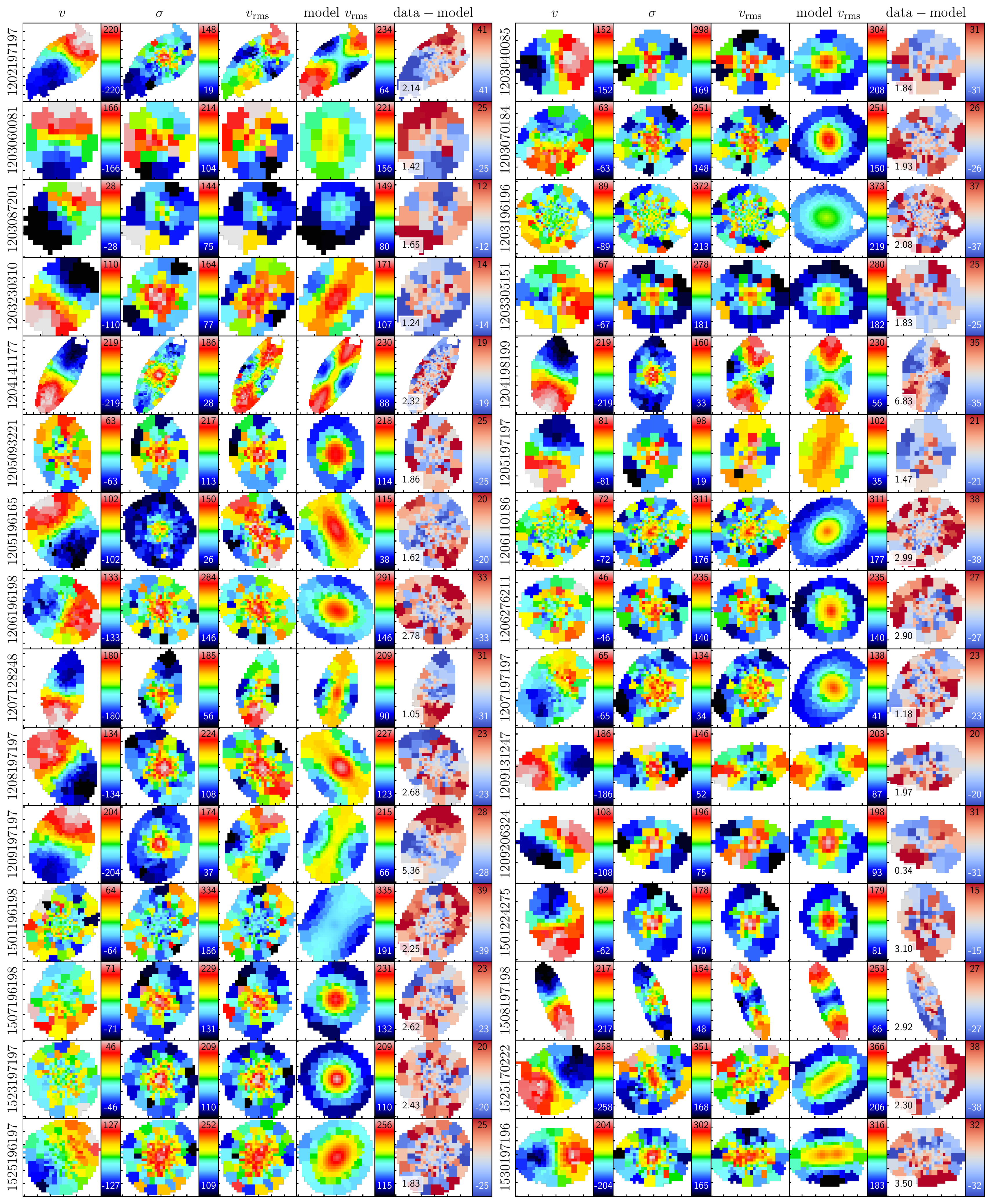}
    \caption{The velocity, velocity dispersion, observed \vf = $\sqrt{v^2 +\sigma^2}$ field, JAM modelled \vf field, and residual field for each galaxy used in this work. The units for all panels are \kms. The modelled \vf field shares the same colour scale as the observed \vf field. The reduced $\chi^2$ value is inset on the residual pane for each galaxy. The x- and y-axis ticks are 1 arcsecond intervals. The MAGPI ID of each galaxy is given on the side of each row.}
    \label{fig:fields_1}
\end{figure*}

Figure \ref{fig:fields_1} shows all the kinematic fields and Jeans models for the 30 MAGPI galaxies  with 1 \re greater than the PSF FWHM and with kinematic coverage to at least $1.5$ \re. As mentioned above, two of these galaxies (1205197197 and 1205196165) have a velocity dispersion too low to be included in the final analysis of total density slopes (that is, they are not used to compute median value or correlations). We also include a foreground object at $z \sim 0.1$ and background object at $z \sim 0.5$, as both met the data quality requirements.

Table \ref{tab:results_table} includes the IDs, right ascensions and declinations, redshifts, central velocity dispersions, total mass, circularised effective radius, and total density slope with associated uncertainty for these 30 galaxies. Due to the data quality cuts imposed on the sample the formal uncertainties on the total density slope are small, and comparable to local Universe studies like that of \citet{bellstedt_sluggs_2018}. In the following sections we assess the kinematic fits, compare the MAGPI total density slopes to the Frontier Fields and \atlas sample, compare to other dynamical and lensing works, compare to the predictions of simulations. We also investigate any correlations between the total density slope and other parameters. 

The median density slope for the 28 MAGPI galaxies is \g (standard error on the median), with a sample standard deviation of $0.22$. If all visually classified spiral galaxies are excluded (8 objects), the results are consistent within the uncertainties with a median of \gearly and a sample standard deviation of $0.15$. In all of the following, the results do not depend on whether these spiral galaxies are excluded and so we leave them in the final sample.

\subsection{Assessment of kinematic fits}
\label{sec:kinematic_fit_assessment}
Most of the MAGPI galaxies (24/30) have reduced $\chi^2$ values less than three. MAGPI galaxies 1207197197 and 1204198199 have the highest reduced $\chi^2$ values of $\sim 6$, both with clear structure in their residuals maps in Figure \ref{fig:fields_1}. MAGPI galaxy 1207197197 in particular has a photometric twist which was not captured in our constant position angle MGE model. Similarly, 1207128248 has particularly high residuals, although this object has a reduced $\chi^2$ value of roughly one due to its high kinematic uncertainties, incorporated into the {\sc{emcee}} determination of estimated parameters for the potential. Although spirals are generally well represented with a disk potential, they may also have bars which are not accounted for in the Jeans axisymmetric framework. MAGPI object 1204198199 may have a bar which drives the higher discrepancy between the model and observed \vf fields. In some cases the spiral systems have \vf models that match the observations particularly well, such as object 1204135171, at $z \sim 0.1$.

We split our sample into galaxies with $\chi^2$ values below or equal to two ($N=12$) and those with $\chi^2$ values greater than two ($N=16$), to test whether our conclusions are driven by the poorer kinematic fits. There is a non-significant difference between the two samples using a Kruskal-Wallis test statistic (p-value = 0.05), and the derived medians are consistent within uncertainties ($\gamma = -2.30 \pm 0.14$ for reduced $\chi^2$ values less or equal to 2, and $\gamma=-2.17 \pm 0.1$ for reduced $\chi^2$ values greater than 2.)

\subsection{Comparison to Frontier Fields and \texorpdfstring{\atlas}{atlas}}
\label{sec:homogenous-comparisons}

\begin{figure}
	\includegraphics[width=1\columnwidth]{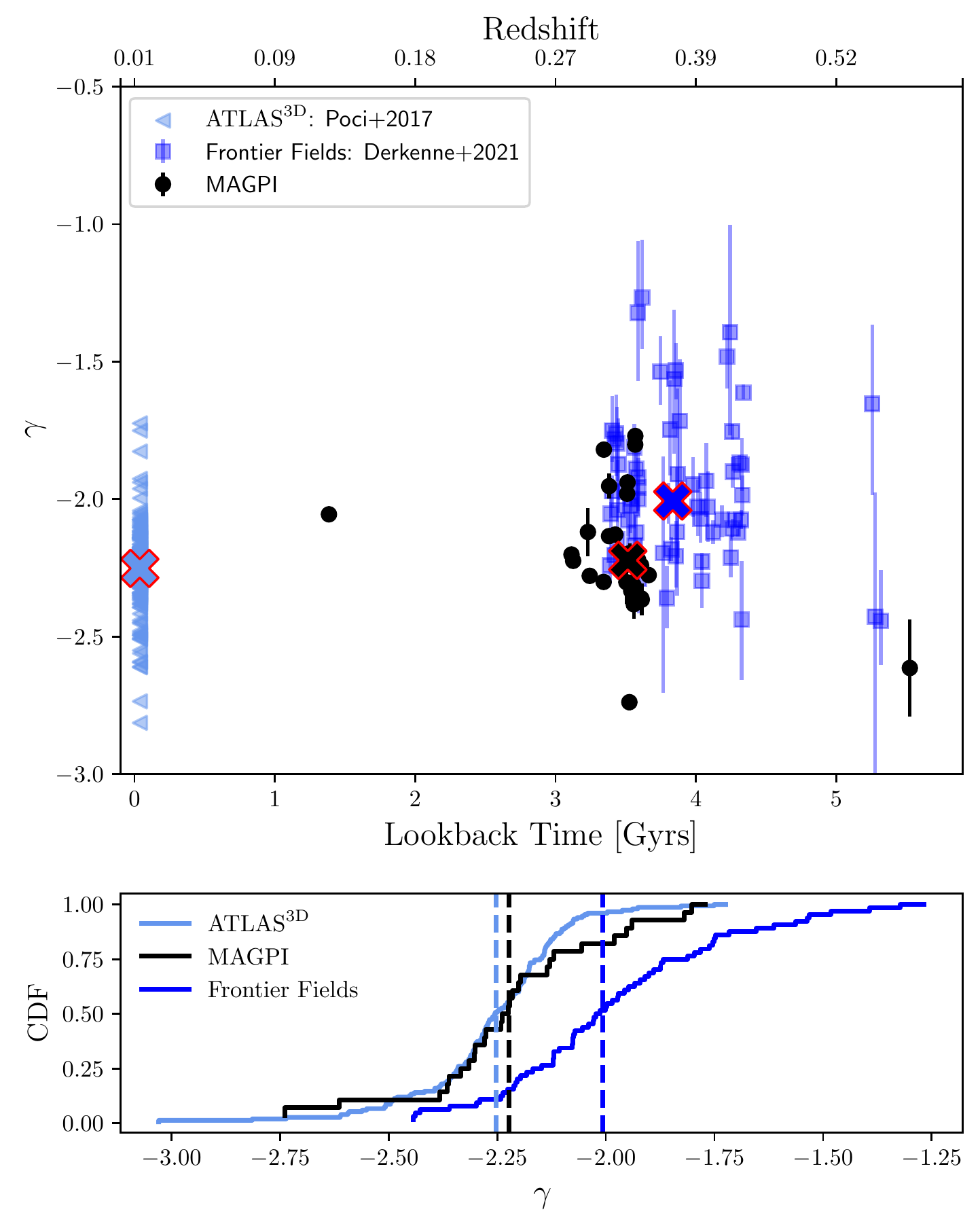}
	\caption{Top: The total density slope as a function of redshift/lookback time for the MAGPI sample and two main comparison samples, Frontier Fields and \atlas.  The MAGPI values are shown in black, Frontier Fields as navy blue squares, and \atlas as light blue triangles. Filled crosses show the median values. Bottom: The cumulative distribution function (CDF) of the MAGPI, Frontier Fields, and \atlas samples. Vertical dashed lines indicate the sample medians.}
    \label{fig:main_results}
\end{figure}

Figure \ref{fig:main_results} shows the distribution of MAGPI total density slopes as a function of redshift, with the slopes for the \atlas sample and Frontier Fields sample re-calculated with our new radial constraints.

The median density slope for MAGPI galaxies of \g compares well to that of the \atlas galaxies in the local Universe, with \ag. The re-analysed total density slopes of the Frontier Fields galaxies have a median of \ff, which is significantly shallower than for the MAGPI galaxies. This median value is different to the published value in \citet{derkenne} of ${\gamma = -2.11 \pm 0.03}$ due to the additional data quality cuts made in this work, and the re-measurement of the slope between $0.1-2$ \re. The bottom panel of Figure \ref{fig:main_results} shows the cumulative distribution function of three samples, with the Frontier Fields sample shifted towards shallower values compared to the overlapping MAGPI and \atlas samples in this space.

Two statistical tests were carried out to assess the differences in the population of total density slopes calculated from MAGPI, \atlas, and the Frontier Fields sample. A Kruskal-Wallis test was used to test the null hypothesis of whether the populations share a median value, and an Anderson-Darling test was used to evaluate the null hypothesis that the observed samples originated from the same underlying distribution (considering the shape of the distribution as well). We found a significant difference between the MAGPI and Frontier Fields total density slope distributions, but no significant difference between the MAGPI and \atlas total density slope distributions.

\begin{figure}
	\includegraphics[width=\columnwidth]{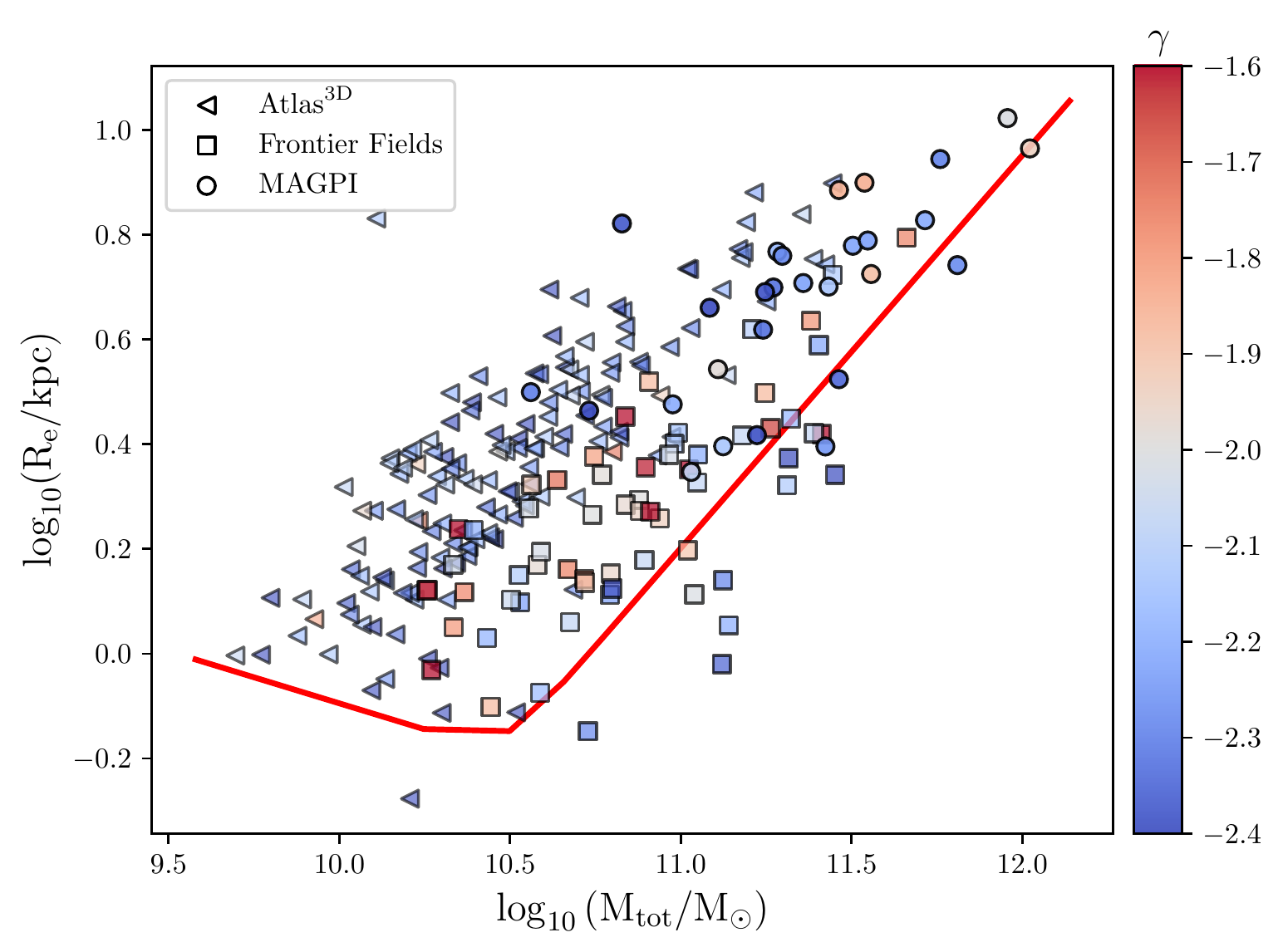}
	\caption{The \textit{total} mass-size plane for the comparison samples with central velocity dispersion $ > 100$ \kms. The x-axis shows total dynamical masses, defined as twice the (spherical) mass integrated from the best-fit potential within 1 \re, as in \citet{cappellari_2013_benchmark}. Since the total masses are in part constructed from the total density slopes, the Frontier Fields galaxies by construction have higher masses for their sizes due to their on average shallower total density slopes. The y-axis shows circularised effective radii, with radii for Frontier Fields galaxies from \citet{derkenne}, and radii for \atlas galaxies from \citet{cappellari_2011_atlas}. The MAGPI circularised effective radii are shown in Table \ref{tab:results_table}. The red solid line shows the `Zone of Exclusion' in the \atlas survey mass-size plane from \citet{cappellari_2013_mass}.}
    \label{fig:mass-size}
\end{figure}

The MAGPI galaxies studied are, on average, larger and more massive than the galaxies studied in the Frontier Fields work (see Figure \ref{fig:mass-size}). Could these sample differences drive the observed differences in total density slopes? If the three comparison samples (MAGPI, \atlas, and Frontier Fields) are cut to a common size range of $3-8$ kpc there is still a significant difference between the MAGPI and Frontier Fields sample, and still no significant difference in the observed populations of total density slopes from MAGPI and \atlas. Similarly, if a common mass cut is made comparing only galaxies with a spherical total mass between $10^{10.5} - 10^{11.5}\, \mathrm{M}_{\odot}$ the results still hold. We  note however that the samples become considerably smaller which makes the statistical comparison less sound. For all statistical tests see Table \ref{tab:p_values}.

We conclude the \atlas and MAGPI total density slopes share a common distribution, but that the MAGPI and Frontier Fields slope distributions are significantly different. We stress that the MAGPI models and Frontier Fields Jeans models and parameter estimations were constructed in an almost identical manner, and argue that the difference in median total density slope of the two populations has a physical origin. As the Frontier Fields total density slopes represent galaxies in extreme, dense environments, we interpret this difference in slope distribution as being due to host environment. The consistency of the total slope distributions between the MAGPI and \atlas samples then indicates no evolution in the total slope across $3-4$ Gyr of cosmic time. 

The Frontier Fields galaxies are generally more compact than MAGPI galaxies, yet exhibit shallower total density slopes. That the Frontier Fields galaxies have shallower total density slopes for a more compact luminous component indicates they could have higher dark matter fractions than the MAGPI sample.

\begin{table*}
	\centering
	\caption{Column 1 lists the samples compared against each other in a 2-way test. Mass cut means all samples are restricted to include galaxies within a cross-sample common mass range of $10^{10.5} - 10^{11.5}\, \Msun$. Radius-cut means all samples are restricted to only include galaxies within the common radial range $3 - 8$ kpc - see Figure \ref{fig:mass-size} for the distributions of the three samples in the mass-size plane. KW=Kruskal-Wallis, AD=Anderson-Darling. The Python implementation of the Anderson-Darling test ({\sc{scipy.stats}}) has p-values clipped between 0.25 and 0.001. The median total slope is shown with the median standard error. In all tests the MAGPI and \atlas sample medians are consistent, but the MAGPI and Frontier Fields sample medians are inconsistent.}
	\label{tab:p_values}
	\begin{tabular}{lccccc} 
		\hline
		Sample & KW p-value & AD p-value & MAGPI $\gamma$ & Frontier Fields $\gamma$ & \atlas $\gamma$ \\\hline
MAGPI, Frontier Fields &$9.5 \times 10^{-5}$ &  $<0.001$ & $-2.22 \pm 0.05$ & \ff & $-$\\
MAGPI, \atlas & $0.28$ & $0.20$ & $-2.22 \pm 0.05$ & $-$ & $-2.25 \pm 0.02$ \\
MAGPI, Frontier Fields (mass cut) &$4 \times 10^{-4}$& $<0.001$ &$-2.24 \pm 0.06$ & $-2.03 \pm 0.04$ & $-$\\
MAGPI, \atlas (mass cut) &$0.94$& $>0.25$ & $-2.24 \pm 0.06$ & $-$ & $-2.26 \pm 0.02$\\
MAGPI, Frontier Fields (radius cut) &$0.002$&$0.002$ & $-2.23 \pm 0.06$ & $-1.87 \pm 0.09$ &$-$\\
MAGPI, \atlas (radius cut) &$0.77 $&$>0.25$ & $-2.23 \pm 0.06$ & $-$ & $-2.26 \pm  0.02$\\
		\hline
	\end{tabular}
\end{table*}

\begin{figure}
	\includegraphics[width=1\columnwidth]{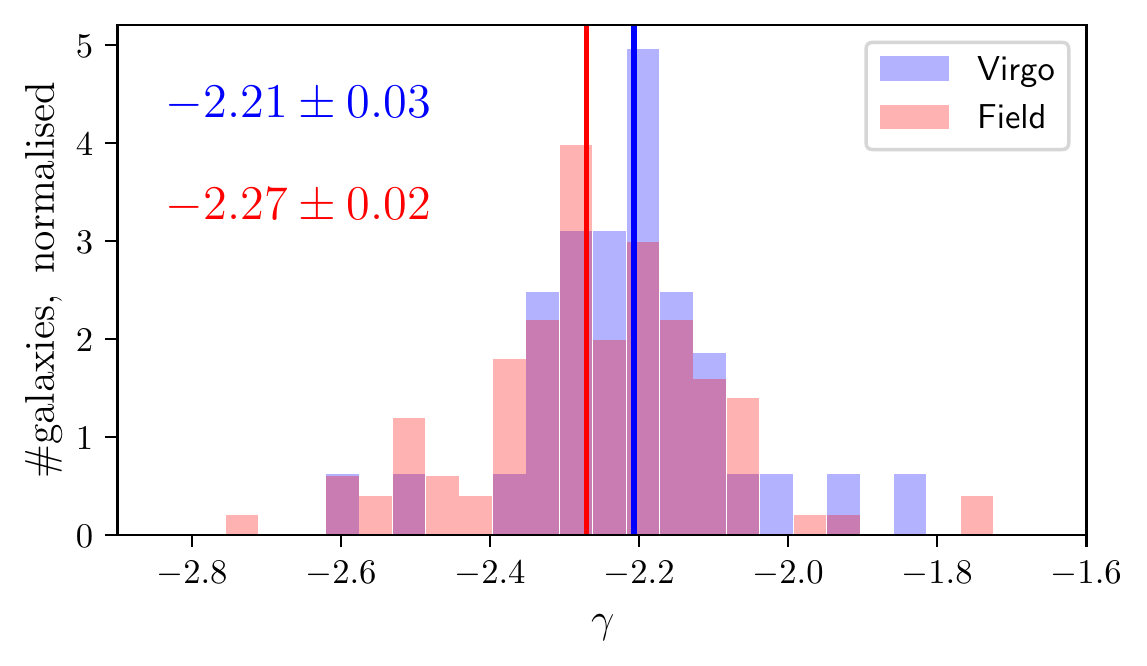}
	\caption{The distribution of total density slopes for \atlas galaxies within the Virgo cluster and from the field. There is evidence of the Virgo total slopes being mildly more shallow than for \atlas galaxies not in the Virgo cluster.}
    \label{fig:virgo_not_virgo}
\end{figure}

If the difference in total slope distributions between the MAGPI and Frontier Fields sample is due to host environment, we can also explore the impact on environment within the \atlas sample itself. The \atlas sample is drawn from a mixture of field galaxies and those from the local Virgo cluster, with 37/150 of the \atlas galaxies analysed in this work drawn from Virgo. Figure \ref{fig:virgo_not_virgo} shows the distribution of \atlas total density slopes from the field and those from within the Virgo cluster. There is evidence of a mild offset between the total slopes of galaxies in Virgo and those from the field, although the Kruskal-Wallis and Anderson-Darling statistical tests are unable to discriminate between them. The Virgo slopes are slightly more shallow, which agrees with the observed difference between MAGPI and Frontier Fields slopes. We explore why we see a pronounced difference between MAGPI and Frontier Fields slopes but not between the field and Virgo \atlas samples in Section \ref{sec:discussion}.

\subsection{Comparison to other dynamical and lensing works}
\label{sec:nonhomogenous-comparisons}

\begin{figure*}
	\includegraphics[width=2\columnwidth]{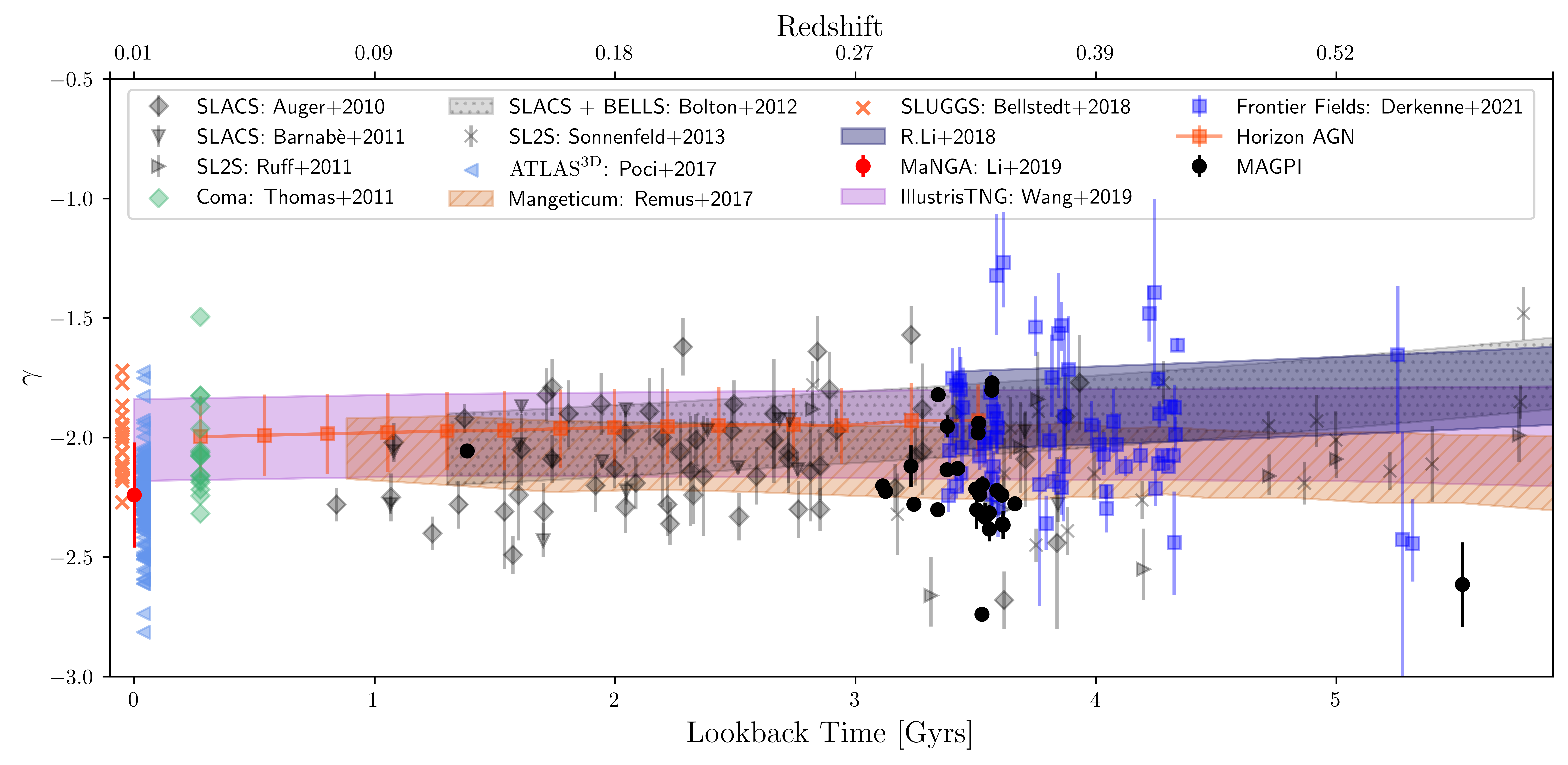}
	\caption{Top: The total density slope as a function of redshift/lookback time. The MAGPI values are shown in black, Frontier Fields as navy blue squares, and \atlas as light blue triangles. Data points for Coma are shown as green diamonds using models from \citet{thomas_2011_dynamical}. At redshift zero are also shown total density slopes from \citet{bellstedt_sluggs_2018} (orange crosses) and from the MaNGA survey as an average (red circle) with standard deviation \citep{li_manga_2019}. These points have been slightly offset in the $x$-direction for visibility. Dark grey symbols show total density slopes measured using gravitational lensing techniques \citep{auger_sloan_2010,barnabe_two_2011,ruff_sl2s_2011,sonn_sl2s_2013}. The grey dotted band shows the fitted relation, of width 1-sigma, between total density slopes and redshift from \citet{bolton_boss_2012}, of the form $\gamma(z) = (-2.11 \pm 0.02) + z(0.6 \pm 0.15)$. The blue band shows a relation from \citet{rui_2018_strong} of the form $d\langle\gamma\rangle/dz = 0.309^{+0.160}_{-0.166}$. The orange striped band shows the average total density slopes from the Magneticum simulations with a 1-sigma width \citep{remus_co-evolution_2017}. The purple band shows the average total density slopes from the IllustrisTNG with a 1-sigma width \citep{wang_early_2019}. The dark orange squares show the Horizon-AGN total density slopes, computed as a single power-law slope in the radial range of $0.1 - 2$ \re, and 1-sigma width.}
    \label{fig:literature_results}
\end{figure*}

In this section we compare to other literature studies with different methodologies and systematics. We show these literature results, as well as the predictions of the Horizon-AGN, IllustrisTNG, and Magneticum simulations, on Figure \ref{fig:literature_results}. To fully understand how the methodological choices made in each study impact the resulting total density slopes it would be necessary to use complementary methods on the same sample, which is beyond the scope of this work.

The total slopes of galaxies in the Coma cluster \citep{thomas_2011_dynamical}, shown on Figure \ref{fig:literature_results} in light green, are are similarly shallow to the Frontier Fields galaxies. The median total density slopes for 17 galaxies in the Coma cluster is $-2.00 \pm 0.06$, which is consistent with the Frontier Fields median of \ff. However, we note the generalised orbit-superposition technique of Schwarzschild modelling was used to constrain the total potential for that work, rather than axisymmetric Jeans modelling used for Frontier Fields galaxies. The consistency suggests the internal mass distribution of a galaxy is partly a function of its environment, with no evolution in the total slope across equivalent environments in the past 5 Gyr. 

The median total density for MaNGA galaxies in the local Universe are shown on Figure \ref{fig:literature_results} as a red circle \citep{li_manga_2019}. For that work, a mass-weighted inner density slope definition  was used within $1$ \re, and a free scale radius (as opposed to the 20 kpc fixed scale radius used in this work). The MaNGA total density slope is equivalently steep to the MAGPI and \atlas samples, with median value $\gamma = -2.22 \pm 0.006$. 

The SLUGGS data are shown as orange crosses on Figure \ref{fig:literature_results}. The same break radius and total potential definition were used in that work in combination with Jeans modelling to constrain the total density slope. These galaxies are a subset of the \atlas sample, where \atlas stellar IFU data was coupled with globular cluster data at large radii. The median total density slope for this data is $-2.06 \pm 0.04$, which is much shallower than the MAGPI or \atlas medians, and not consistent with our proposed scenario of environmental impact on the internal mass distribution. The shallower SLUGGS slopes could be due to the combination of globular cluster data with stellar IFU data, or could be due to the extended radial coverage of the SLUGGS data.

We also show published total density slopes from several gravitational lensing works, specifically those of \citet{auger_sloan_2010}, \citet{barnabe_two_2011},  \citet{ruff_sl2s_2011}, \citet{bolton_boss_2012}, \citet{sonn_sl2s_2013}, and \citet{rui_2018_strong}, covering a combined redshift range of $0.06 < z < 1$. \citet{rui_2018_strong} note a significant dependence on the radial range used to calculate the slope, in that for a fixed galaxy the slope becomes steeper for an increasing radial range. Although likely due to different methodological reasons than those noted in this work, the bias works the same way as in this work. In general the Einstein radius is used to constrain the gravitational potential in lensing  studies, which is typically on a scale equivalent to the effective radius. We present MAGPI, Frontier Fields and \atlas slopes calculated on a similar radial range to the lensing works in Appendix \ref{sec:app_break_radius}, Table \ref{tab:comparison_gamma}. Regardless of the exact range used, the lensing slopes are more shallow than the MAGPI ones for similar redshifts. 

Furthermore, lensing works with large redshift baselines show an evolution of the total slope from shallow at early times to steeper in the local Universe \citep{ruff_sl2s_2011,bolton_boss_2012,rui_2018_strong}. This is tension with our finding that, across a smaller redshift baseline, that there has been no evolution in the total density slope between MAGPI and \atlas galaxies. Even if there is a methodological offset when comparing the techniques, this discrepancy regarding the evolution remains. Galaxies that act as gravitational lenses are necessarily massive and dense, but the evolution with redshift is found to hold even after considering other parameters such as stellar mass and radius, and radial range used to constrain the slope \citep{bolton_boss_2012,rui_2018_strong}. We note the distribution of total density slopes has large intrinsic scatter, and so increasing sample sizes at higher redshifts, with complementary methods, will be crucial to determining the slope evolution.

\subsection{Comparison to simulations}
\label{sec:sim-comparisons}

In Figure \ref{fig:literature_results} we show a comparison to three simulations; Magneticum, IllustrisTNG, and Horizon-AGN.
The IllustrisTNG total density slopes, and the Horizon-AGN total density slopes in the redshift range $0.018 < z < 0.305$ are consistent within the uncertainties, and show little to no evolution in this span of cosmic time. The Horizon-AGN slopes do become marginally steeper for $ z < 0.3$, from $\gamma \sim -1.93$ to  $\gamma \sim -1.99$, but this is still consistent with no evolution in this period. This agrees with the lack of evolution seen between the MAGPI and \atlas slopes, and broadly consistent with the Magneticum results, as a wider redshift baseline is needed before the predictions become significantly different.

\begin{figure}
	\includegraphics[width=\columnwidth]{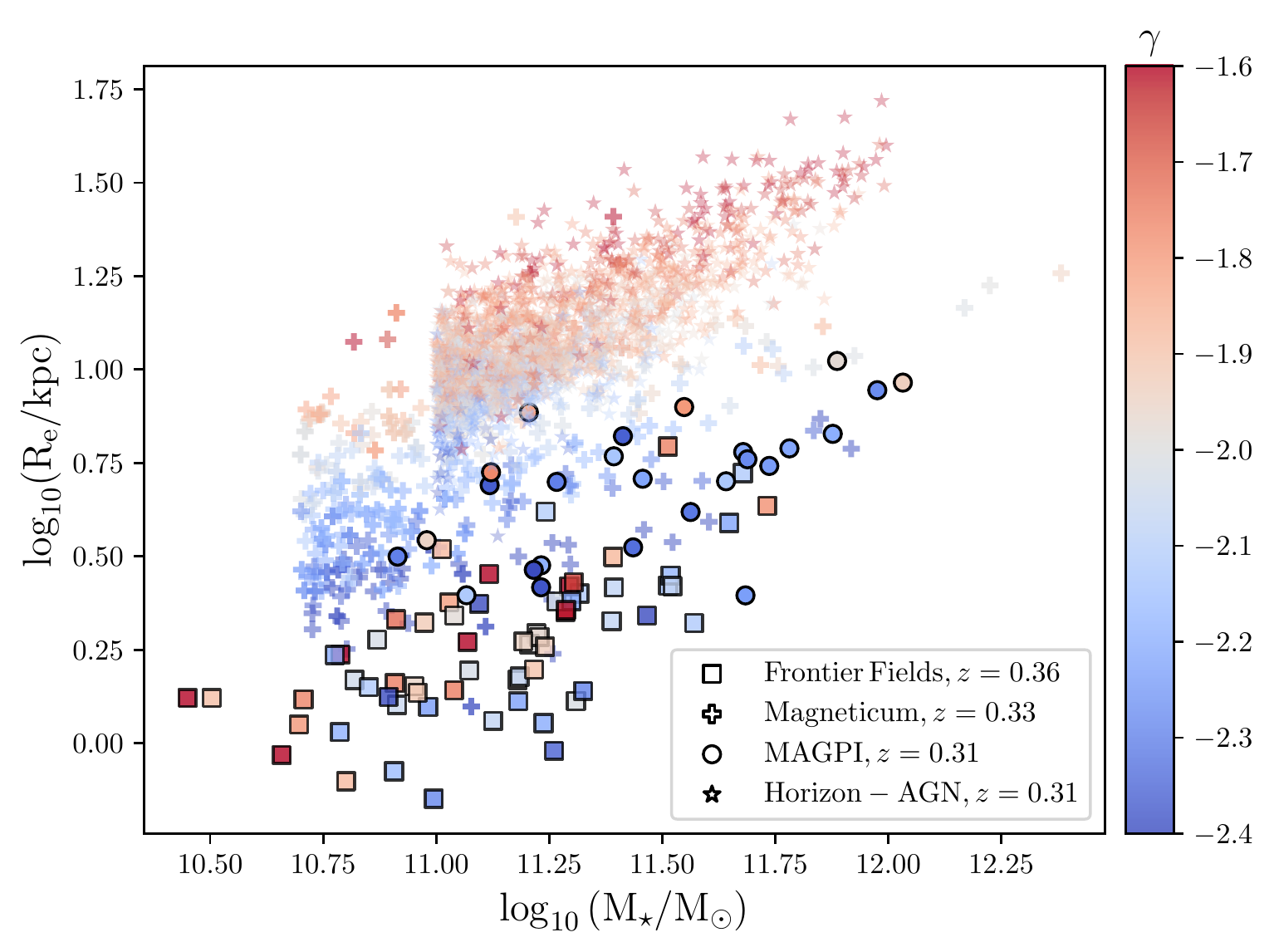}
    \caption{The size-mass plane for Magneticum, Horizon-AGN, MAGPI, and Frontier Fields galaxies at $z \sim 0.3$; for the MAGPI and Frontier Fields samples this is the mean redshift, and for Magneticum and Horizon-AGN this corresponds to a single snapshot. For Magneticum, the 3D half-mass radius, \rmass, is plotted. The simulated galaxies are more extended for equivalent stellar masses than MAGPI galaxies.}
    \label{fig:size-mass_sims}
\end{figure}

Around the nominal MAGPI redshift of 0.3, the simulation slopes are on average more shallow than the MAGPI ones. In the local Universe the median total density slopes from all simulations considered here are more shallow than the \atlas median, comparing \ag to roughly $\gamma = -2.0$, but are similar to the Frontier Fields median. In Figure \ref{fig:size-mass_sims} we show the size-mass plane for Magneticum, Horizon-AGN, Frontier Fields, and MAGPI galaxies at $z \sim 0.3$. At fixed stellar mass, the simulated galaxies are typically more extended in radius than either the MAGPI or Frontier Fields samples. A small set of Magneticum galaxies approximate the size-mass distribution of MAGPI galaxies, and have more similar total density slopes. In general, however, the simulated and observed samples do not overlap. There is also a visible correlation particularly along the radius-axis between simulated galaxies and total density slope, in that galaxies with larger radii tend to have shallower total density slopes at fixed mass. We explore these correlations in Section \ref{sec:correlations}.

We therefore stress that the simulated and observed galaxies do not overlap well in the mass-size plane, and this difference could be intertwined with the offset in total density slope medians between the different samples, seen in Figure \ref{fig:literature_results}.

\begin{figure}
	\includegraphics[width=\columnwidth]{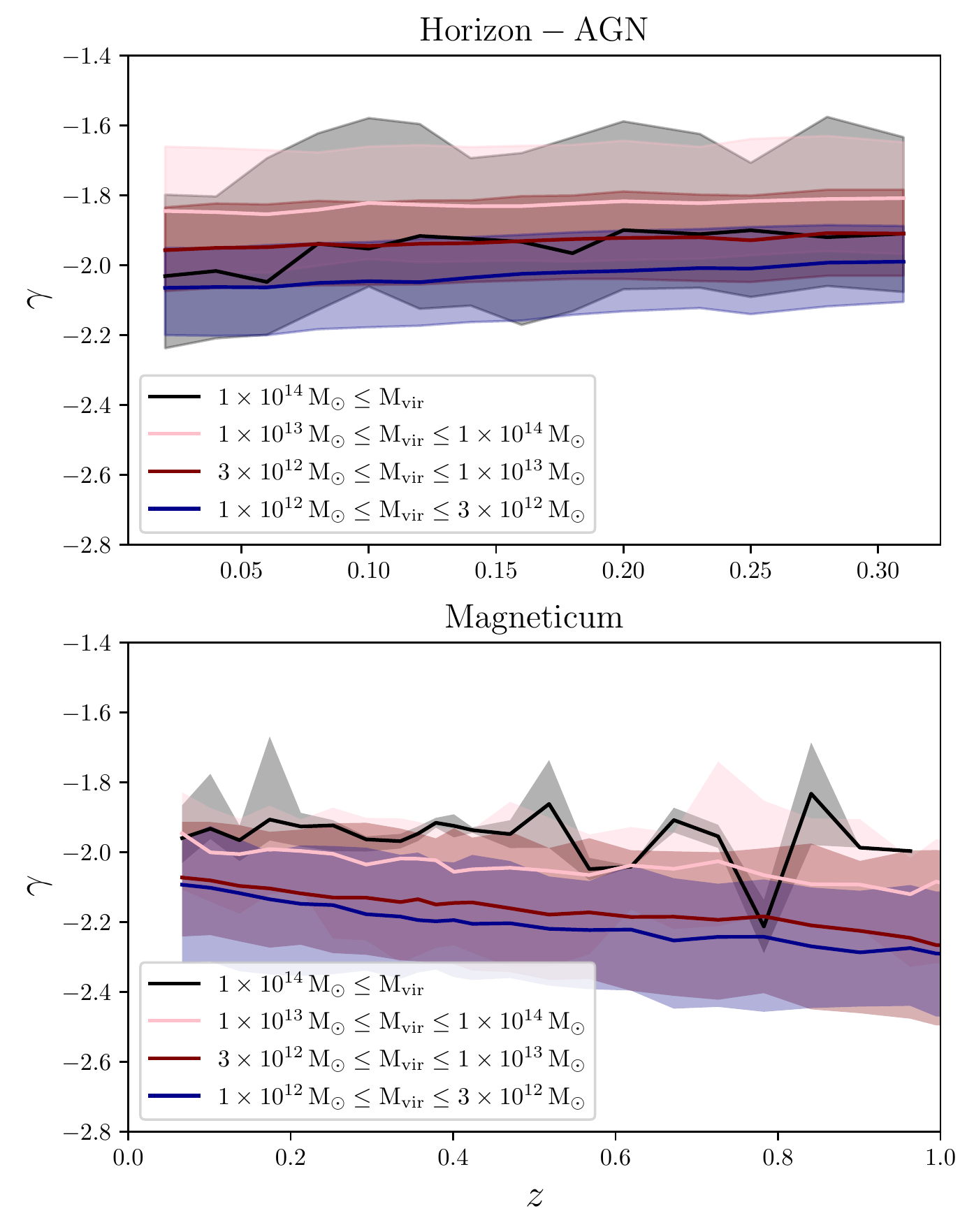}
    \caption{Top: Horizon-AGN galaxies divided into host halo mass bins: The blue line shows small galaxies with virial mass of the main host halo between 1 and $3 \times 10^{12}$\Msun, the maroon line shows massive galaxies with host halo mass between $3 \times 10^{12}$\Msun and $1 \times 10^{13}$\Msun, the pink line shows galaxies in groups with host halo mass between $1 \times 10^{13}$\Msun and $1 \times 10^{13}$\Msun and $1 \times 10^{14}$\Msun, and finally the black line shows galaxies in cluster environments, with a host halo mass greater than $10^{14}$ \Msun. The shading represents the standard deviation. Bottom row: Magneticum galaxies from \citet{remus_co-evolution_2017} with the same cuts as the Horizon-AGN galaxies. The cluster masses in Magneticum are not as massive as the Frontier Fields environments, but a division between total slope and host mass is still seen. As with the different in total density slope between MAGPI and Frontier Fields galaxies, Magneticum galaxies in denser environments (clusters) have on average shallower total density slopes. For Horizon-AGN galaxies the trend is the same for groups, massive galaxies, and small galaxies, except clusters no longer have the most shallow slopes.}
    \label{fig:magneticum_hagn}
\end{figure}

We used the Magneticum and Horizon-AGN galaxies to test whether the separation of total slopes with environments also exists in the simulations, shown in Figure \ref{fig:magneticum_hagn}. To do this, galaxies were binned by virial mass of the main host galaxy. We used these divisions as a proxy for environment, where the bins represent: small galaxies, in a halo mass bin between 1 and $3 \times 10^{12}$ \Msun; massive galaxies in a halo between $3 \times 10^{12}$ \Msun and $1 \times 10^{13}$ \Msun; galaxies in groups with halo mass between {$1 \times 10^{13}$\ \Msun} and {$1 \times 10^{14}$\ \Msun}; and galaxies in cluster environments, with a host halo mass greater than {$10^{14}$\ \Msun}.

The `cluster' bin is still less massive than the Frontier Fields clusters of order {$10^{15}$\ \Msun}.  Nonetheless, a trend is clearly visible in the Magneticum simulation where galaxies belonging to more massive haloes have shallower total density slopes than those in less massive haloes. Partly this is due to the trends in the Magneticum simulation between stellar mass and total density slope (see Section \ref{sec:correlations}), and stellar mass and halo mass are themselves correlated. However, a linear model for Magneticum total slopes has more predictive power and a lower Akaike information criterion (AIC) when both stellar mass and halo mass are used as explanatory variables rather than stellar mass alone ($-350$ compared to $-224$). Therefore we conclude there exists a dependence between total density slope and host halo mass. This dependence in the Magneticum simulations supports our finding of environmental difference in internal mass distribution between MAGPI and Frontier Fields galaxies. From Figure \ref{fig:magneticum_hagn} we also see that the Magneticum total density slopes reach MAGPI-like values at $z \sim 0.8$, which could be related to the size-mass offset in samples discussed above.

For the Horizon-AGN galaxies there is again a separation in total density slopes between the small galaxy, massive galaxy, and group bins, which follows the trend observed with Magneticum galaxies. However, the cluster bin have slopes comparable to small or massive galaxies, and it is the group galaxies that have the most shallow slopes on average. As with Magneticum galaxies, a linear model including both stellar mass and dark matter halo mass has more predictive power than stellar mass alone using the AIC ($-2444$ compared to $-2433$). The disagreement regarding the environmental trend for the cluster could stem from differences in the sample selection criteria. For instance, in Horizon-AGN, both the brightest cluster galaxy and relatively small satellites within the cluster region are excluded from the sample stellar mass range of $10^{11}-10^{12}$ \Msun.

\subsection{Correlations with the total density slope}
\label{sec:correlations}
We investigate whether there exist any significant correlations between the total density slope and other intrinsic galaxy parameters, such as central velocity dispersion, effective radius, total mass, stellar mass, and stellar mass surface density. For the \atlas galaxies we take circularised effective radii from \citet{cappellari_2011_atlas} and the central velocity dispersion from \citet{cappellari_2013_benchmark}. For Frontier Fields galaxies the circularised effective radii and central velocity dispersions are from \citet{derkenne}. Total masses were calculated in the same way for all three samples, defined as twice the mass integrated given the median parametrisation of the potential within a sphere of one effective radius \citep{cappellari_2013_benchmark}. 

Stellar masses for MAGPI were calculated using the spectral energy distribution fitting software {\sc{prospect}} with a Chabrier \citep{chabrier_2002_galactic} initial mass function. The spectral energy distribution was fit using pixel-matched imaging in the \textit{ugriZYJHKs} bands from the GAMA survey \citep{bellstedt_2020_gama}, with photometry derived by the MUSE-based {\sc{profound}} segmentation maps for each galaxy. For the \atlas sample, stellar masses were calculated as the total luminosity from the MGEs of \citet{scott_2013_atlas} multiplied by the SDSS $r$-band mass-to-light ratio from \citet{cappellari_2013_mass}, giving a total stellar mass. For the Frontier Fields sample, $V$-band stellar mass-to-light ratios were calculated from a MILES \citep{vazdekis_2010_evolutionary} template library {\sc{pPXF}} spectral fit with a saltpeter \citep{saltpeter_1955_luminosity} initial mass function. These were converted to a Chabrier initial mass function by dividing by 1.53, as in \citet{driver_2011_galaxy}. The total stellar mass was then calculated as the total luminosity from the MGEs multiplied by this mass-to-light ratio. 

The stellar surface mass density is given as 
\begin{equation}
\Sigma_{\star} = \frac{\mathrm{M}_{\star}}{2\uppi\re^2}.
\label{sellarmasssurface}
\end{equation}

\begin{figure*}
	\includegraphics[width=2\columnwidth]{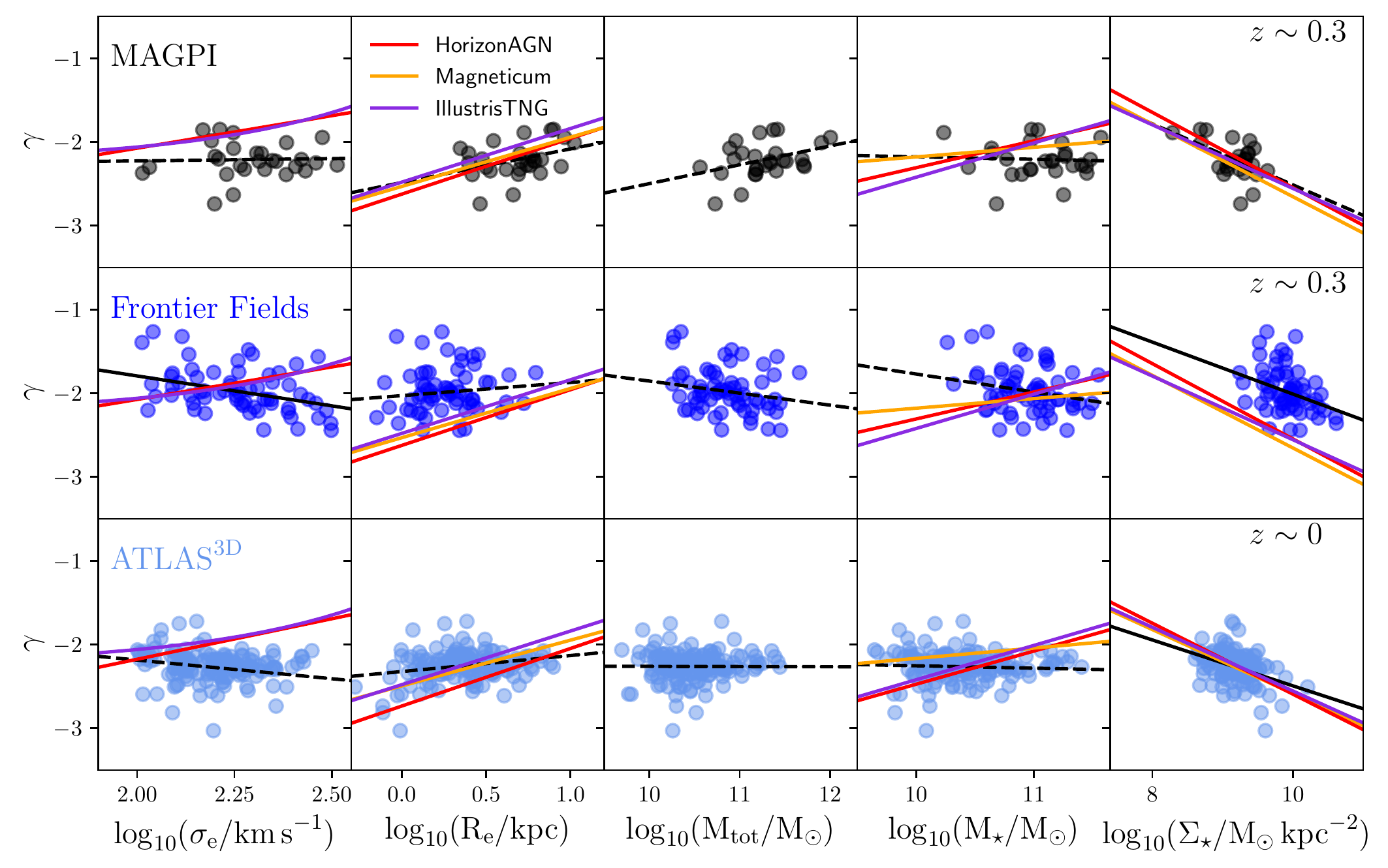}
	\caption{Linear correlations between the total density slope and other galaxy parameters: central velocity dispersion, effective radius, total mass, stellar mass, and stellar mass surface density. For the MAGPI, Frontier Fields and \atlas samples, panels with a black dashed line indicates a non-significant fit with a p-value greater than 0.01. Solid black lines show correlations with a p-value smaller than 0.01, fitted using the Python package {\sc{ltsfit}} \citep{cappellari_2013_benchmark} and of the form $y = a + b(x-x_0)$, where $x_0$ is the median of the $x$-values. The IllustrisTNG correlations are taken directly from \citet{wang_2020_early} at $z =0$, and the central velocity dispersion is $\sigma_{\mathrm{e}/2}$ instead of $\sigma_{\mathrm{e}}$. Horizon-AGN correlations are shown for galaxies from the simulation snapshot at $0.305$ on the MAGPI and Frontier Fields rows, and for galaxies from the redshift $0.018$ snapshot for the \atlas row. Likewise, Magneticum correlations are shown for galaxies including snapshots between $0.25 < z< 0.35$ on the MAGPI and Frontier Fields rows, and for galaxies in snapshots with $z < 0.1$ for the \atlas row. All significant relations are given in Table \ref{tab:correlations}.}
    \label{fig:correlations}
\end{figure*}

The relations are plotted in Figure \ref{fig:correlations}, and all significant relations are given in Table \ref{tab:correlations}. The MAGPI galaxies show no significant correlation with any other parameter, possibly due to the small sample size. For the velocity dispersion, this could also be due to the greater mix of morphological type, as early-type galaxies have a slight negative trend between slope and velocity dispersion, whereas spiral galaxies do not \citep{li_manga_2019}. The only other parameter the observational sample correlates with is the stellar mass surface density, which is intuitive as galaxies that are more compact have steeper slopes.

We also compared to the trends seen in simulations, using the sample of Horizon-AGN and Magneticum galaxies, as well as the linear correlations published in \citet{wang_2020_early}. Interestingly, the simulations all predict that galaxies with greater velocity dispersions have shallower total density slopes, although for the observational sample we see the opposite. This discrepancy continues with the stellar mass, as simulations predict the more massive the stellar content of a galaxy the more shallow its total slope, whereas the observational samples show no trend to a mildly negative one.  For the \atlas and MAGPI samples the trends between total slope and stellar mass surface density (although not significant for MAGPI) are quite close to the ones seen for Magneticum, Horizon-AGN, and IllustrisTNG galaxies. The Frontier Fields galaxies show the same trend (the more compact a galaxy is the steeper its total slope), however, the galaxies are offset from the simulated relations. The correlation between stellar surface density and total density slopes is also seen in the measurements from gravitational lensing \citep{sonn_sl2s_2013}. 

A potential reason for the difference in observed trends between the simulation and observational works is the manner in which the slopes are measured. As the simulations have access to the dark matter, star, and gas particles directly, the total density slope is measured by fitting a power-law to the co-added simulation particles in spherical, concentric shells. What systematic differences this introduces when comparing the slopes to observational studies can be addressed in future work by constructing mock IFU observations of simulation data and re-measuring the slopes.

\begin{table*}
	\centering
	\caption{A summary of the linear correlations between the total slope and parameters shown in Figure \ref{fig:correlations}. Only significant correlations with a Spearman p-value of less than 0.01 are shown. Correlations with total mass are not given in this table, as no trends were significant for the MAGPI, \atlas, and Frontier Fields samples. We fit each data set using the Python package {\sc{ltsfit}} \citep{cappellari_2013_benchmark}, and given relations for the total slope are of the form  $y = a + b(x-x_0)$, where $x_0$ is median of the $x$ values. For the IllustrisTNG relations we take the correlations directly from \citet{wang_2020_early}, with the gradient shown as $m$. The stellar surface mass density $\Sigma_{\star}$ is defined as $\mathrm{M}_{\star}/2\uppi\re$. For IllustrisTNG, the correlation with velocity dispersion uses a central aperture dispersion $\sigma_{\mathrm{e}/2}$.}
	
	\label{tab:correlations}
	\begin{tabular}{lcccc} 
		\hline
Sample & $\log_{10}(\sigma_{\mathrm{e}}/\kms)$ & $\log_{10}(\re/\mathrm{kpc})$  & $\log_{10}(\mathrm{M}_{\star}/\Msun)$ & $\log_{10}(\Sigma_{\star}/\Msun\mathrm{\,kpc}^{-2})$ \\

		\hline
MAGPI & - & - &  - & -  \\
Frontier Fields &$ -1.98 -0.71(x -2.27)$  & - &- &$-1.99 -0.31(x-9.62)$ \\
Atlas3D &-  & -  &- &$-2.26 -0.27(x-9.16)$ \\ 
Magneticum, $z < 0.1$ & - & $-2.04 +0.55(x -0.83)$ &$-2.04 +0.12(x-11.02)$& $-2.05 -0.38(x -8.59)$\\ 
Magneticum, $0.25 < z < 0.35$ & - & $-2.06 +0.59(x -0.79)$ & $-2.06 +0.12(x -11.02)$ & $-2.07 -0.43(x-8.65)$ \\ 
Horizon-AGN, $z=0.018$ & $-2.00 +0.97(x-2.18)$ &$-2.00 +0.69 (x-1.07)$ &  $-2.01 +0.40(x-11.19)$ & $-1.99 -0.42(x-8.58)$ \\
Horizon-AGN, $z=0.305$ & $-1.93 +0.78(x-2.18)$ & $-1.93 +0.67(x-1.04)$ &  $-1.93 +0.32(x-11.17)$ & $-1.92-0.45(x-8.62)$ \\
IllustrisTNG, $z=0$ & $m = - 0.00019$ & $m = 0.64$ & $m = 0.41$ & $m = -0.45$ \\
\hline
	\end{tabular}
\end{table*}

\section{Discussion}
\label{sec:discussion}
In our results we have shown there is a significant difference in the distribution of total density slopes of MAGPI and Frontier Fields galaxies, with Frontier Fields galaxies having on average total density profiles with shallower slopes. This difference suggests the mass distribution of Frontier Fields galaxies are, on average, less centrally concentrated (the mass density falls off less rapidly with radius) than for MAGPI galaxies. The differences observed in the total mass density slopes of  the populations hold if the samples are cut to overlapping radii and mass ranges, and when cut by velocity dispersion to remove any potential trends with that parameter. The detailed methodology of  deriving the total density slopes for MAGPI and Frontier Fields objects are near identical, and any dependence of total density slope with radial range is removed by measuring all slopes on the same radial range.

\subsection{The Frontier Fields, Virgo, and Coma environments}

A remaining difference between the samples is environment. The Frontier Fields galaxies are drawn from dense cluster environments, whereas MAGPI targets are drawn from a range of environments that includes field galaxies and galaxies in groups (see \citealp{MAGPI} for a summary, in particular their Figure 4. Note we do not include any Frontier Field objects as part of the MAGPI sample in this work, although Frontier Fields clusters Abell 2744 and Abell 370 are supplementary to the MAGPI survey). The Frontier Fields clusters were selected as ideal candidates for lensing systems \citep{lotz_2017_frontier}, and all are massive and X-ray luminous. The clusters span {$\sim 1 \times 10^{45}\mathrm{erg\,s}^{-1}$} as a bolometric X-ray luminosity for MACSJ0416.1-2403 \citep{mann_2012_xray}, to {$\sim 3 \times 10^{45}\mathrm{erg\,s}^{-1}$} at 2-10 keV for Abell 2744 \citep{allen_1998_resolving}. 

The Frontier Fields clusters all show significant substructure, interpreted as evidence of ongoing formation and mergers \citep{mann_2012_xray,zitrin_2009_discovery,gruen_2013_weak,richard_2010_abell}, with Abell 2744 in particular having up to eight different mass concentrations \citep{jauzac_2016_extraordinary}. In fact, the structure of Abell 2744 is so extreme it has been used to question whether such a cluster is even possible within the $\Lambda$CDM model, as no equivalent clusters could be found within the the Millenium XXL simulation volume \citep{schwinn_2017_abell}, although such a structure has been identified in the Magneticum simulation suite with the detailed comparison of cluster and simulation masses depending on projection  \citep{kimmig_2022_hateful}. Due to the complexity of the clusters, gravitational lensing has been effective in estimating total masses. \citet{merten_2011_creation} place the total mass of Abell 2744 at {$\mathrm{M}(r < 1.3 \mathrm{\,Mpc}) = (1.8 \pm0.4) \times 10^{15}$ \Msun} with {$h=0.7$} using a mass model created from weak lensing, with mass estimates of the other clusters consistently falling above {$\mathrm{M} \sim 10^{15}$ \Msun} \citep{grillo_2015_clash,zheng_2012_magnified,Williamson_2011_sunyaev,lagattuta_2022_pilot}. 

There is little observed difference in total density slope between \atlas galaxies drawn from Virgo and those from the field. Virgo is less X-ray luminous than the Frontier Fields clusters, classified as having low X-ray luminosity ({$< 10^{44}\,\mathrm{\,erg\,s}^{-1}$}) and low central galaxy density \citep{jones_1984_structure}. Mass estimates also fall well below those of the Frontier clusters, with \citet{simionescu_2017_witnessing} placing the virial mass of Virgo at {$\mathrm{M}_{200} = (1.05 \pm 0.02) \times 10^{14}$ \Msun for $r_{200} = (974.1 \pm 5.7)\, \mathrm{\,kpc}$}. To stress the difference, a recent strong lensing analysis of Abell 2744 puts the total mass at {$M = (1.77 \pm 0.07) \times 10^{14} $ \Msun} at a radius of just 200 kpc \citep{bergamini_2023_new}. 

In contrast to Virgo, Coma is more equivalent in its mass to the Frontier Fields clusters. A recent Bayesian deep-learning analysis of the mass of Coma by \citet{ho_2022_dynamical} gives a mass of {$\mathrm{M}_{200} = 10^{15.10 \pm 0.15}\,h^{-1}$ \Msun} within {$(1.78 \pm 0.03)\,h^{-1}\,\mathrm{Mpc}$} of the cluster centre (see their figure 3 for this result compared to historical mass estimates). The X-ray luminosity of the cluster is also more than an order of magnitude greater than that of Virgo, at {$\sim 8 \times 10^{44}\,\mathrm{erg\,s}^{-1}$} in the band $0.1-2.4$ keV \citep{reiprich_2002_mass}. 

The nature of the Frontier Fields program  means studied galaxies are situated well within the core regions of the clusters because of the limited field of view  of MUSE \citep{lotz_2017_frontier}. Due to the different observing strategy of the \atlas survey, the galaxies drawn from Virgo are not limited to the very core regions. This detail of survey design could also impact how comparable the two environments are, as the local density changes with distance from the cluster core.  \citet{li_manga_2019} found that satellites within haloes have a density slope that is steeper (more negative) than centrals by about $\sim 0.1$. Although none of the galaxies modelled in \citet{derkenne} are the `brightest cluster galaxies' in the eight substructures identified by \citep{jauzac_2016_extraordinary}, we might be comparing the very dense regions of the Frontier Fields clusters to a variety of local densities in Virgo. In future work we aim to address this question quantitatively with MAGPI data, as the full sample will probe a large range of local environments, and we will be able to apply consistent metrics. 

\subsection{Drivers of the total slope}

From the Magneticum simulations it is known that progressive dry mergers drive total density slopes towards isothermal values \citep{remus_2013_dark}. Once a galaxy has an isothermal total density profile, subsequent dry mergers do not perturb the slope away again.  Dry mergers tend to increase the stellar mass at large radii, resulting in shallower total density slopes.  Minor mergers in particular have increased efficiency with regard to size-growth compared to major mergers \citep{naab_2009_minor}. Major merger rates for galaxies below $z =0.5$ are uncommon, placed at an occurrence of just $0.058 \pm 0.009$ per Gyr \citep{lopez_2015_alhambra}. Minor mergers are found to be $\sim 3$ times more likely that major mergers at redshift $z= 0.7$ with little evidence the merger rate evolves with redshift below $z =1$, making minor mergers also infrequent across the redshift baseline we consider \citep{lotz_2011_major}. That we see no evidence for evolution in the total density slope is therefore consistent with the low occurrence of mergers at later times. The evolution observed by gravitational lensing is potentially a factor of those works targeting massive galaxies, which are more likely to have undergone mergers and therefore have closer to isothermal slopes.

While galaxies in cluster environments move with high velocities and are unlikely to undergo mergers, galaxy clusters themselves are assembled from the in-fall of groups, as seen with the many substructures that make up the Frontier Fields clusters, and especially Abell 2744 \citep{jauzac_2016_extraordinary}. Galaxies in groups would experience higher merger rates at earlier times as local density and merger rates are coupled \citep{Jian_2012_environmental,shankar_2013_size,watson_2019_galaxy}. This is seen in the cold dark matter haloes as well, where haloes in high density environments undergo most of their mass aggregation at earlier times compared to haloes is less dense environments \citep{maulbetsch_2007_dependence}. \citet{papovich_2012_candels} studied a sample of galaxies in a proto-cluster at redshift $z = 1.62$, and found quiescent galaxies in clusters are larger than those in the field at that redshift. This implies, like the dark matter haloes, accelerated growth at higher redshifts for galaxies in cluster environments. Galaxies in clusters might therefore have undergone more mergers in their history than isolated fields galaxies, explaining why we see a difference in the internal mass distributions of galaxies in the MAGPI survey compared to those drawn from the Frontier Fields environments. 

Ram-pressure stripping and harassment can also occur in the cluster environments, which in extreme cases can result in the unusual `jellyfish' galaxies with extended gas streams and clumpy star formation marking their violent fall into the cluster environment \citep{owers_2012_shocking,poggianti_2017_gasp}. As stars are preferentially stripped from a galaxy potential only after the majority of the dark matter \citep{smith_2016_preferential}, this suggests galaxies in a cluster potential with visible distortions of their stellar matter are also significantly transformed in their total mass distribution. This result is evidence of the cluster environment itself transforming the properties of a galaxy.

\section{Summary and Conclusions}
\label{sec:conclusions}
The MAGPI survey is a MUSE large-program survey observing massive galaxies in the `middle-ages' of the Universe. With over two hundred hours of observing time on the VLT this survey will enable the spatial mapping of ionised gas and stars in galaxies 3-4 Gyr ago, providing a detailed understanding of galaxy evolution spanning field to cluster environments. Foreground and background objects in the fields supplement this time baseline further. 

In this work we present an early study of the impact of environment in shaping the internal mass distributions of galaxies, with the intention of providing a quantitative relationship between environmental metrics and internal mass distribution when the full sample is observed. We present total mass density slopes for 28 galaxies in the MAGPI survey. To construct the density slopes, 2D resolved stellar kinematic maps measured from MUSE data are used in combination with multi-Gaussian models of the stellar light to construct Jeans dynamical models, with the total potential (baryons and dark matter)  described by a generalised NFW profile. 

We compare the MAGPI total density slope distribution primarily to two other works. First, the total density slopes for \atlas galaxies as measured using a consistent methodology by \citet{poci}. These galaxies are in the local Universe ($< 40 $ Mpc) and are from field environments and the Virgo cluster, a dynamically young cluster of moderate mass. Second, we compare to the density slopes from Frontier Fields galaxies as measured by \citet{derkenne}. These galaxies are all from extreme and dense cluster environments in the redshift range $0.29 < z < 0.55$. Combining these methodologically consistent studies allows us to trace the internal mass distribution of galaxies from field to cluster environments across 5 Gyr of cosmic time in a self-consistent manner.

We determined that the radial range across which the total slope is measured is critical when using a generalised NFW total potential with fixed break radius. A mismatch between radial ranges can cause the measured slope to change by $\sim 0.1$, which erases the signature in density slope from host environment or other parameters. For this reason, we re-analyse total mass density slopes from \atlas and Frontier Fields galaxies using the same radial range as MAGPI galaxies. We measure total mass density slopes in the radial range  $r = [\re/10, 2$ \re$]$. 

The main results are the following:
\begin{enumerate}
    \item We found a median total mass density slope of \g (standard error) for the MAGPI 28 galaxies with central velocity dispersion greater than $100$ \kms. These results are unchanged if we consider only early-type morphology galaxies, with median slope \gearly. 
    \item  We found a median slope of \ff for 68 Frontier Fields galaxies at median redshift 0.348, and a median slope of \ag for 150 galaxies from the \atlas sample in the local Universe.
    \item We found no difference between the distribution and median slope of MAGPI and \atlas galaxies, even after cutting the samples to common size and mass ranges. The similarity of the samples indicates the total density slope for galaxies in intermediate density environments has not evolved across the past $3-4$ Gyr of cosmic time. The MAGPI and \atlas samples are treated as roughly equivalent in terms of their host environments in comparison to the very dense Frontier Fields cluster environments. This lack of evolution is broadly consistent with the hydrodynamical cosmological simulation results of Magneticum, IllustrisTNG, and Horizon-AGN. We do not see evidence of the density slope evolutionary trends derived from gravitational lensing studies.
    \item A statistically significant difference is found in the distribution of total density slopes between the MAGPI and Frontier Fields sample (\g compared to \ff, respectively). This difference remains statistically significant when cutting the samples to common radius and mass ranges.
    \item The difference in slope distribution suggest that environmental factors play a role in the internal mass distributions of galaxies.  Dense, cluster environments are more likely to host galaxies with shallow mass distributions than field environments, potentially due to the way clusters assemble their mass from galaxy groups. The separation of total density slopes with environment (as indicated by host halo mass) is also predicted by the Magneticum simulations. The dependence of slopes with environment in the Horizon-AGN simulations is not as clear.
\end{enumerate}

For this work we have used different galaxy samples to broadly categorise galaxies as residing in field, group, or cluster environments. When all MAGPI fields are observed we can expect to more than triple the sample size used here, which will allow for a quantitative test between environmental metrics (such as group number) and total density slope. The establishment of the impact of environment with total density slope in the `Middle Ages' can be compared against local Universe surveys that span a wide range of environments, such as the SAMI survey or the recently commissioned Hector\footnote{\url{https://hector.survey.org.au/}} survey. The launch of the James Webb Space Telescope also makes it possible to push this type of resolved stellar kinematic methodology further back in cosmic time, where the predictions of the different simulations (and the measurements of gravitational lensing) significantly differ.

\section*{Acknowledgements}
We thank the anonymous reviewer for their constructive feedback.
We wish to thank the ESO staff, and in particular the staff at Paranal Observatory, for carrying out the MAGPI observations. MAGPI targets were selected from GAMA. GAMA is a joint European-Australasian project based around a spectroscopic campaign using the Anglo-Australian Telescope.
GAMA is funded by the STFC (UK), the ARC (Australia), the AAO, and the participating institutions.
GAMA photometry is based on observations made with ESO Telescopes at the La Silla Paranal
Observatory under programme ID 179.A-2004, ID 177.A-3016. CL, JTM and CF are the recipients of an ARC Discovery Project DP210101945. CF is the recipient of an Australian Research Council Future Fellowship (project number FT210100168) funded by the Australian Government. CL, JTM, and EW
acknowledge support by the Australian Research Council Centre of Excellence for All Sky
Astrophysics in 3 Dimensions (ASTRO 3D), through project number CE170100013.
FDE acknowledges funding through the ERC Advanced grant 695671 ``QUENCH'' and support by the Science and Technology Facilities Council (STFC). AFM has received support from the Spanish Ministry grants CEX2019-000920-S, PID2021-123313NA-I00 and 4RYC2021-031099-I. PS is supported by the Australian Government Research Training Program (RTP) Scholarship. PS also acknowledges support in the form of Oort Fellowship at Leiden Observatory, and the International Astronomical Union-Gruber Foundation Fellowship. LMV acknowledges support by the COMPLEX project from the European Research Council (ERC) under the European Union’s Horizon 2020 research and innovation program grant agreement ERC-2019-AdG 882679. S.K.Y. acknowledges support from the Korean National Research Foundation (2020R1A2C3003769, 2022R1A6A1A03053472). GS thanks SARAO PDF grant (UID 97882). Y.P. acknowledges National Science Foundation of China (NSFC) Grant No. 12125301, 12192220, 12192222, and the science research grants from the China Manned Space Project with NO. CMS-CSST-2021-A07. 
This research was conducted on the land of the Wallumattagal clan of the Dharug Nation.
\section*{Data Availability}
\label{sec:data_avail}
The MAGPI spectral data underlying this work are available from the ESO Science Archive Facility ({\url{http://archive.eso.org/cms.html}}). 

\section*{Affiliations}
\noindent
{\it
$^{11}$ Instituto de Astrof\'{i}sica de Canarias, IAC, V\'{i}a L\'{a}ctea s/n, 38205, La Laguna (S.C. Tenerife), Spain\\
$^{12}$ Departamento de Astrof\'{i}sica, Universidad de La Laguna, 38206, La Laguna (S.C. Tenerife), Spain\\
$^{13}$ School of Physics, University of New South Wales, Sydney, NSW 2052, Australia\\
$^{14}$ Cosmic Dawn Center (DAWN).\\
$^{15}$ Department of Astronomy, School of Physics, Peking University, 5 Yiheyuan Road, Beijing 100871, People’s Republic of China\\
$^{16}$ Kavli Institute for Astronomy and Astrophysics, Peking University, 5 Yiheyuan Road, Beijing 100871, People’s Republic of China\\
$^{17}$ Leiden Observatory, Leiden University, P.O. Box 9513, 2300 RA Leiden, The Netherlands\\
$^{18}$ GSKY, INFN-Sezione di Trieste, via Valerio 2, 34127 Trieste, Italy \\
$^{19}$ SISSA International School for Advanced Studies, Via Bonomea 265, 34136 Trieste, Italy\\
$^{20}$ Department of Physics and Astronomy, University of the Western Cape, Cape Town 7535, South Africa\\
$^{21}$ School of Mathematics and Physics, University of Queensland, Brisbane, QLD 4072, Australia\\
$^{22}$ Centre for Astrophysics and Supercomputing, School of Science, Swinburne University of Technology, Hawthorn, VIC 3122, Australia\\
}



\bibliographystyle{mnras}
\bibliography{example} 

\begin{thebibliography}{}
\makeatletter
\relax
\def\mn@urlcharsother{\let\do\@makeother \do\$\do\&\do\#\do\^\do\_\do\%\do\~}
\def\mn@doi{\begingroup\mn@urlcharsother \@ifnextchar [ {\mn@doi@}
  {\mn@doi@[]}}
\def\mn@doi@[#1]#2{\def\@tempa{#1}\ifx\@tempa\@empty \href
  {http://dx.doi.org/#2} {doi:#2}\else \href {http://dx.doi.org/#2} {#1}\fi
  \endgroup}
\def\mn@eprint#1#2{\mn@eprint@#1:#2::\@nil}
\def\mn@eprint@arXiv#1{\href {http://arxiv.org/abs/#1} {{\tt arXiv:#1}}}
\def\mn@eprint@dblp#1{\href {http://dblp.uni-trier.de/rec/bibtex/#1.xml}
  {dblp:#1}}
\def\mn@eprint@#1:#2:#3:#4\@nil{\def\@tempa {#1}\def\@tempb {#2}\def\@tempc
  {#3}\ifx \@tempc \@empty \let \@tempc \@tempb \let \@tempb \@tempa \fi \ifx
  \@tempb \@empty \def\@tempb {arXiv}\fi \@ifundefined
  {mn@eprint@\@tempb}{\@tempb:\@tempc}{\expandafter \expandafter \csname
  mn@eprint@\@tempb\endcsname \expandafter{\@tempc}}}

\bibitem[\protect\citeauthoryear{{Allen}}{{Allen}}{1998}]{allen_1998_resolving}
{Allen} S.~W.,  1998, \mn@doi [\mnras] {10.1046/j.1365-8711.1998.01358.x},
  \href {https://ui.adsabs.harvard.edu/abs/1998MNRAS.296..392A} {296, 392}

\bibitem[\protect\citeauthoryear{{Auger}, {Treu}, {Bolton}, {Gavazzi},
  {Koopmans}, {Marshall}, {Moustakas}  \& {Burles}}{{Auger}
  et~al.}{2010}]{auger_sloan_2010}
{Auger} M.~W.,  {Treu} T.,  {Bolton} A.~S.,  {Gavazzi} R.,  {Koopmans}
  L.~V.~E.,  {Marshall} P.~J.,  {Moustakas} L.~A.,   {Burles} S.,  2010,
  \mn@doi [\apj] {10.1088/0004-637X/724/1/511}, \href
  {https://ui.adsabs.harvard.edu/abs/2010ApJ...724..511A} {724, 511}

\bibitem[\protect\citeauthoryear{{Barnab{\`e}}, {Czoske}, {Koopmans}, {Treu}
  \& {Bolton}}{{Barnab{\`e}} et~al.}{2011}]{barnabe_two_2011}
{Barnab{\`e}} M.,  {Czoske} O.,  {Koopmans} L. V.~E.,  {Treu} T.,   {Bolton}
  A.~S.,  2011, \mn@doi [\mnras] {10.1111/j.1365-2966.2011.18842.x}, \href
  {https://ui.adsabs.harvard.edu/abs/2011MNRAS.415.2215B} {415, 2215}

\bibitem[\protect\citeauthoryear{Bellstedt et~al.,}{Bellstedt
  et~al.}{2018}]{bellstedt_sluggs_2018}
Bellstedt S.,  et~al., 2018, \mn@doi [MNRAS] {10.1093/mnras/sty456}, 476, 4543

\bibitem[\protect\citeauthoryear{{Bellstedt} et~al.,}{{Bellstedt}
  et~al.}{2020}]{bellstedt_2020_gama}
{Bellstedt} S.,  et~al., 2020, \mn@doi [\mnras] {10.1093/mnras/staa1466}, \href
  {https://ui.adsabs.harvard.edu/abs/2020MNRAS.496.3235B} {496, 3235}

\bibitem[\protect\citeauthoryear{{Bergamini} et~al.,}{{Bergamini}
  et~al.}{2023}]{bergamini_2023_new}
{Bergamini} P.,  et~al., 2023, \mn@doi [\aap] {10.1051/0004-6361/202244575},
  \href {https://ui.adsabs.harvard.edu/abs/2023A&A...670A..60B} {670, A60}

\bibitem[\protect\citeauthoryear{{Bialas}, {Lisker}, {Olczak}, {Spurzem}  \&
  {Kotulla}}{{Bialas} et~al.}{2015}]{Bialas_2015_harassment}
{Bialas} D.,  {Lisker} T.,  {Olczak} C.,  {Spurzem} R.,   {Kotulla} R.,  2015,
  \mn@doi [\aap] {10.1051/0004-6361/201425235}, \href
  {https://ui.adsabs.harvard.edu/abs/2015A&A...576A.103B} {576, A103}

\bibitem[\protect\citeauthoryear{{Bolton}, {Burles}, {Koopmans}, {Treu}  \&
  {Moustakas}}{{Bolton} et~al.}{2006}]{bolton_2006_sloan}
{Bolton} A.~S.,  {Burles} S.,  {Koopmans} L. V.~E.,  {Treu} T.,   {Moustakas}
  L.~A.,  2006, \mn@doi [\apj] {10.1086/498884}, \href
  {https://ui.adsabs.harvard.edu/abs/2006ApJ...638..703B} {638, 703}

\bibitem[\protect\citeauthoryear{{Bolton} et~al.,}{{Bolton}
  et~al.}{2012}]{bolton_boss_2012}
{Bolton} A.~S.,  et~al., 2012, \mn@doi [\apj] {10.1088/0004-637X/757/1/82},
  \href {https://ui.adsabs.harvard.edu/abs/2012ApJ...757...82B} {757, 82}

\bibitem[\protect\citeauthoryear{{Boselli}, {Fossati}  \& {Sun}}{{Boselli}
  et~al.}{2022}]{boselli_2022_ram}
{Boselli} A.,  {Fossati} M.,   {Sun} M.,  2022, \mn@doi [\aapr]
  {10.1007/s00159-022-00140-3}, \href
  {https://ui.adsabs.harvard.edu/abs/2022A&ARv..30....3B} {30, 3}

\bibitem[\protect\citeauthoryear{{Brodie} et~al.,}{{Brodie}
  et~al.}{2014}]{brodie_2014_sluggs}
{Brodie} J.~P.,  et~al., 2014, \mn@doi [\apj] {10.1088/0004-637X/796/1/52},
  \href {https://ui.adsabs.harvard.edu/abs/2014ApJ...796...52B} {796, 52}

\bibitem[\protect\citeauthoryear{{Bundy} et~al.,}{{Bundy}
  et~al.}{2015}]{bundy_2015_manga}
{Bundy} K.,  et~al., 2015, \mn@doi [\apj] {10.1088/0004-637X/798/1/7}, \href
  {https://ui.adsabs.harvard.edu/abs/2015ApJ...798....7B} {798, 7}

\bibitem[\protect\citeauthoryear{{Calvi}, {Poggianti}, {Vulcani}  \&
  {Fasano}}{{Calvi} et~al.}{2013}]{calvi_2013_impact}
{Calvi} R.,  {Poggianti} B.~M.,  {Vulcani} B.,   {Fasano} G.,  2013, \mn@doi
  [\mnras] {10.1093/mnras/stt667}, \href
  {https://ui.adsabs.harvard.edu/abs/2013MNRAS.432.3141C} {432, 3141}

\bibitem[\protect\citeauthoryear{{Cappellari}}{{Cappellari}}{2002}]{MGE}
{Cappellari} M.,  2002, \mn@doi [\mnras] {10.1046/j.1365-8711.2002.05412.x},
  \href {https://ui.adsabs.harvard.edu/abs/2002MNRAS.333..400C} {333, 400}

\bibitem[\protect\citeauthoryear{Cappellari}{Cappellari}{2008}]{cappellari_measuring_2008}
Cappellari M.,  2008, \mn@doi [MNRAS] {10.1111/j.1365-2966.2008.13754.x}, 390,
  71

\bibitem[\protect\citeauthoryear{{Cappellari}}{{Cappellari}}{2017}]{ppxf_2}
{Cappellari} M.,  2017, \mn@doi [\mnras] {10.1093/mnras/stw3020}, \href
  {https://ui.adsabs.harvard.edu/abs/2017MNRAS.466..798C} {466, 798}

\bibitem[\protect\citeauthoryear{{Cappellari} \& {Copin}}{{Cappellari} \&
  {Copin}}{2003}]{cappellari_2003_voronoi}
{Cappellari} M.,  {Copin} Y.,  2003, \mn@doi [\mnras]
  {10.1046/j.1365-8711.2003.06541.x}, \href
  {https://ui.adsabs.harvard.edu/abs/2003MNRAS.342..345C} {342, 345}

\bibitem[\protect\citeauthoryear{{Cappellari} \& {Emsellem}}{{Cappellari} \&
  {Emsellem}}{2004}]{ppxf_1}
{Cappellari} M.,  {Emsellem} E.,  2004, \mn@doi [\pasp] {10.1086/381875}, \href
  {https://ui.adsabs.harvard.edu/abs/2004PASP..116..138C} {116, 138}

\bibitem[\protect\citeauthoryear{{Cappellari} et~al.,}{{Cappellari}
  et~al.}{2007}]{cappellari_2007_sauron}
{Cappellari} M.,  et~al., 2007, \mn@doi [\mnras]
  {10.1111/j.1365-2966.2007.11963.x}, \href
  {https://ui.adsabs.harvard.edu/abs/2007MNRAS.379..418C} {379, 418}

\bibitem[\protect\citeauthoryear{{Cappellari} et~al.,}{{Cappellari}
  et~al.}{2011}]{cappellari_2011_atlas}
{Cappellari} M.,  et~al., 2011, \mn@doi [\mnras]
  {10.1111/j.1365-2966.2010.18174.x}, \href
  {https://ui.adsabs.harvard.edu/abs/2011MNRAS.413..813C} {413, 813}

\bibitem[\protect\citeauthoryear{{Cappellari} et~al.,}{{Cappellari}
  et~al.}{2013a}]{cappellari_2013_benchmark}
{Cappellari} M.,  et~al., 2013a, \mn@doi [\mnras] {10.1093/mnras/stt562}, \href
  {https://ui.adsabs.harvard.edu/abs/2013MNRAS.432.1709C} {432, 1709}

\bibitem[\protect\citeauthoryear{{Cappellari} et~al.,}{{Cappellari}
  et~al.}{2013b}]{cappellari_2013_mass}
{Cappellari} M.,  et~al., 2013b, \mn@doi [\mnras] {10.1093/mnras/stt644}, \href
  {https://ui.adsabs.harvard.edu/abs/2013MNRAS.432.1862C} {432, 1862}

\bibitem[\protect\citeauthoryear{{Cappellari} et~al.,}{{Cappellari}
  et~al.}{2015}]{cappellari_2015_slopes}
{Cappellari} M.,  et~al., 2015, \mn@doi [\apjl] {10.1088/2041-8205/804/1/L21},
  \href {https://ui.adsabs.harvard.edu/abs/2015ApJ...804L..21C} {804, L21}

\bibitem[\protect\citeauthoryear{{Cebri{\'a}n} \& {Trujillo}}{{Cebri{\'a}n} \&
  {Trujillo}}{2014}]{cebrian_2014_effect}
{Cebri{\'a}n} M.,  {Trujillo} I.,  2014, \mn@doi [\mnras]
  {10.1093/mnras/stu1375}, \href
  {https://ui.adsabs.harvard.edu/abs/2014MNRAS.444..682C} {444, 682}

\bibitem[\protect\citeauthoryear{{Chabrier}}{{Chabrier}}{2003}]{chabrier_2002_galactic}
{Chabrier} G.,  2003, \mn@doi [\pasp] {10.1086/376392}, \href
  {https://ui.adsabs.harvard.edu/abs/2003PASP..115..763C} {115, 763}

\bibitem[\protect\citeauthoryear{{Croom} et~al.,}{{Croom}
  et~al.}{2012}]{croom_2012_sydney}
{Croom} S.~M.,  et~al., 2012, \mn@doi [\mnras]
  {10.1111/j.1365-2966.2011.20365.x}, \href
  {https://ui.adsabs.harvard.edu/abs/2012MNRAS.421..872C} {421, 872}

\bibitem[\protect\citeauthoryear{{Deeley} et~al.,}{{Deeley}
  et~al.}{2017}]{deeley_2017_galaxy}
{Deeley} S.,  et~al., 2017, \mn@doi [\mnras] {10.1093/mnras/stx441}, \href
  {https://ui.adsabs.harvard.edu/abs/2017MNRAS.467.3934D} {467, 3934}

\bibitem[\protect\citeauthoryear{{Derkenne}, {McDermid}, {Poci}, {Remus},
  {J{\o}rgensen}  \& {Emsellem}}{{Derkenne} et~al.}{2021}]{derkenne}
{Derkenne} C.,  {McDermid} R.~M.,  {Poci} A.,  {Remus} R.-S.,  {J{\o}rgensen}
  I.,   {Emsellem} E.,  2021, \mn@doi [\mnras] {10.1093/mnras/stab1996}, \href
  {https://ui.adsabs.harvard.edu/abs/2021MNRAS.506.3691D} {506, 3691}

\bibitem[\protect\citeauthoryear{{Dolag}}{{Dolag}}{2015}]{dolag_2015_magneticum}
{Dolag} K.,  2015, in IAU General Assembly. p. 2250156

\bibitem[\protect\citeauthoryear{{Driver} et~al.,}{{Driver}
  et~al.}{2011}]{driver_2011_galaxy}
{Driver} S.~P.,  et~al., 2011, \mn@doi [\mnras]
  {10.1111/j.1365-2966.2010.18188.x}, \href
  {https://ui.adsabs.harvard.edu/abs/2011MNRAS.413..971D} {413, 971}

\bibitem[\protect\citeauthoryear{{Dubois}, {Devriendt}, {Slyz}  \&
  {Teyssier}}{{Dubois} et~al.}{2012}]{dubois_2012_self}
{Dubois} Y.,  {Devriendt} J.,  {Slyz} A.,   {Teyssier} R.,  2012, \mn@doi
  [\mnras] {10.1111/j.1365-2966.2011.20236.x}, \href
  {https://ui.adsabs.harvard.edu/abs/2012MNRAS.420.2662D} {420, 2662}

\bibitem[\protect\citeauthoryear{{Dubois} et~al.,}{{Dubois}
  et~al.}{2014}]{dubois_2014_dancing}
{Dubois} Y.,  et~al., 2014, \mn@doi [\mnras] {10.1093/mnras/stu1227}, \href
  {https://ui.adsabs.harvard.edu/abs/2014MNRAS.444.1453D} {444, 1453}

\bibitem[\protect\citeauthoryear{{Emsellem}, {Monnet}  \& {Bacon}}{{Emsellem}
  et~al.}{1994}]{emsellem_1994_multi}
{Emsellem} E.,  {Monnet} G.,   {Bacon} R.,  1994, \aap, \href
  {https://ui.adsabs.harvard.edu/abs/1994A&A...285..723E} {285, 723}

\bibitem[\protect\citeauthoryear{{Emsellem} et~al.,}{{Emsellem}
  et~al.}{2011}]{emsellem_2011_census}
{Emsellem} E.,  et~al., 2011, \mn@doi [\mnras]
  {10.1111/j.1365-2966.2011.18496.x}, \href
  {https://ui.adsabs.harvard.edu/abs/2011MNRAS.414..888E} {414, 888}

\bibitem[\protect\citeauthoryear{{Fabjan}, {Borgani}, {Tornatore}, {Saro},
  {Murante}  \& {Dolag}}{{Fabjan} et~al.}{2010}]{fabjan_2010_simulating}
{Fabjan} D.,  {Borgani} S.,  {Tornatore} L.,  {Saro} A.,  {Murante} G.,
  {Dolag} K.,  2010, \mn@doi [\mnras] {10.1111/j.1365-2966.2009.15794.x}, \href
  {https://ui.adsabs.harvard.edu/abs/2010MNRAS.401.1670F} {401, 1670}

\bibitem[\protect\citeauthoryear{{Foreman-Mackey}, {Hogg}, {Lang}  \&
  {Goodman}}{{Foreman-Mackey} et~al.}{2013}]{foreman_2013_emcee}
{Foreman-Mackey} D.,  {Hogg} D.~W.,  {Lang} D.,   {Goodman} J.,  2013, \mn@doi
  [\pasp] {10.1086/670067}, \href
  {https://ui.adsabs.harvard.edu/abs/2013PASP..125..306F} {125, 306}

\bibitem[\protect\citeauthoryear{{Foster} et~al.,}{{Foster}
  et~al.}{2021}]{MAGPI}
{Foster} C.,  et~al., 2021, \mn@doi [\pasa] {10.1017/pasa.2021.25}, \href
  {https://ui.adsabs.harvard.edu/abs/2021PASA...38...31F} {38, e031}

\bibitem[\protect\citeauthoryear{{Grillo} et~al.,}{{Grillo}
  et~al.}{2015}]{grillo_2015_clash}
{Grillo} C.,  et~al., 2015, \mn@doi [\apj] {10.1088/0004-637X/800/1/38}, \href
  {https://ui.adsabs.harvard.edu/abs/2015ApJ...800...38G} {800, 38}

\bibitem[\protect\citeauthoryear{{Gruen} et~al.,}{{Gruen}
  et~al.}{2013}]{gruen_2013_weak}
{Gruen} D.,  et~al., 2013, \mn@doi [\mnras] {10.1093/mnras/stt566}, \href
  {https://ui.adsabs.harvard.edu/abs/2013MNRAS.432.1455G} {432, 1455}

\bibitem[\protect\citeauthoryear{{Gr{\"u}tzbauch}, {Conselice}, {Varela},
  {Bundy}, {Cooper}, {Skibba}  \& {Willmer}}{{Gr{\"u}tzbauch}
  et~al.}{2011}]{grutzbauch_2011_how}
{Gr{\"u}tzbauch} R.,  {Conselice} C.~J.,  {Varela} J.,  {Bundy} K.,  {Cooper}
  M.~C.,  {Skibba} R.,   {Willmer} C. N.~A.,  2011, \mn@doi [\mnras]
  {10.1111/j.1365-2966.2010.17727.x}, \href
  {https://ui.adsabs.harvard.edu/abs/2011MNRAS.411..929G} {411, 929}

\bibitem[\protect\citeauthoryear{{Gunn} \& {Gott}}{{Gunn} \&
  {Gott}}{1972}]{gunn_1972_infall}
{Gunn} J.~E.,  {Gott} J.~Richard I.,  1972, \mn@doi [\apj] {10.1086/151605},
  \href {https://ui.adsabs.harvard.edu/abs/1972ApJ...176....1G} {176, 1}

\bibitem[\protect\citeauthoryear{{Hartke}, {Kakkad}, {Reyes}, {Moya-Sierralta},
  {Reyes}, {Kravtsov}, {Kolb}  \& {Selman}}{{Hartke}
  et~al.}{2020}]{hartke_2020_MUSE}
{Hartke} J.,  {Kakkad} D.,  {Reyes} C.,  {Moya-Sierralta} C.,  {Reyes} A.,
  {Kravtsov} T.,  {Kolb} J.,   {Selman} F.,  2020, in {Schreiber} L.,
  {Schmidt} D.,   {Vernet} E.,  eds,  Society of Photo-Optical Instrumentation
  Engineers (SPIE) Conference Series Vol. 11448, Adaptive Optics Systems VII.
  p. 114480V, \mn@doi{10.1117/12.2560793}

\bibitem[\protect\citeauthoryear{{Hilz}, {Naab}  \& {Ostriker}}{{Hilz}
  et~al.}{2013}]{hilz_2013_minor}
{Hilz} M.,  {Naab} T.,   {Ostriker} J.~P.,  2013, \mn@doi [\mnras]
  {10.1093/mnras/sts501}, \href
  {https://ui.adsabs.harvard.edu/abs/2013MNRAS.429.2924H} {429, 2924}

\bibitem[\protect\citeauthoryear{{Hirschmann}, {Dolag}, {Saro}, {Bachmann},
  {Borgani}  \& {Burkert}}{{Hirschmann}
  et~al.}{2014}]{hirschmann_2014_cosmological}
{Hirschmann} M.,  {Dolag} K.,  {Saro} A.,  {Bachmann} L.,  {Borgani} S.,
  {Burkert} A.,  2014, \mn@doi [\mnras] {10.1093/mnras/stu1023}, \href
  {https://ui.adsabs.harvard.edu/abs/2014MNRAS.442.2304H} {442, 2304}

\bibitem[\protect\citeauthoryear{{Ho}, {Ntampaka}, {Rau}, {Chen}, {Lansberry},
  {Ruehle}  \& {Trac}}{{Ho} et~al.}{2022}]{ho_2022_dynamical}
{Ho} M.,  {Ntampaka} M.,  {Rau} M.~M.,  {Chen} M.,  {Lansberry} A.,  {Ruehle}
  F.,   {Trac} H.,  2022, \mn@doi [Nature Astronomy]
  {10.1038/s41550-022-01711-1}, \href
  {https://ui.adsabs.harvard.edu/abs/2022NatAs.tmp..148H} {}

\bibitem[\protect\citeauthoryear{{Hogg} \& {Foreman-Mackey}}{{Hogg} \&
  {Foreman-Mackey}}{2018}]{hogg_2018_emcee}
{Hogg} D.~W.,  {Foreman-Mackey} D.,  2018, \mn@doi [\apjs]
  {10.3847/1538-4365/aab76e}, \href
  {https://ui.adsabs.harvard.edu/abs/2018ApJS..236...11H} {236, 11}

\bibitem[\protect\citeauthoryear{{Jauzac} et~al.,}{{Jauzac}
  et~al.}{2016}]{jauzac_2016_extraordinary}
{Jauzac} M.,  et~al., 2016, \mn@doi [\mnras] {10.1093/mnras/stw2251}, \href
  {https://ui.adsabs.harvard.edu/abs/2016MNRAS.463.3876J} {463, 3876}

\bibitem[\protect\citeauthoryear{{Jeans}}{{Jeans}}{1922}]{jeans_1922_motions}
{Jeans} J.~H.,  1922, \mn@doi [\mnras] {10.1093/mnras/82.3.122}, \href
  {https://ui.adsabs.harvard.edu/abs/1922MNRAS..82..122J} {82, 122}

\bibitem[\protect\citeauthoryear{{Jian}, {Lin}  \& {Chiueh}}{{Jian}
  et~al.}{2012}]{Jian_2012_environmental}
{Jian} H.-Y.,  {Lin} L.,   {Chiueh} T.,  2012, \mn@doi [\apj]
  {10.1088/0004-637X/754/1/26}, \href
  {https://ui.adsabs.harvard.edu/abs/2012ApJ...754...26J} {754, 26}

\bibitem[\protect\citeauthoryear{{Jones} \& {Forman}}{{Jones} \&
  {Forman}}{1984}]{jones_1984_structure}
{Jones} C.,  {Forman} W.,  1984, \mn@doi [\apj] {10.1086/161591}, \href
  {https://ui.adsabs.harvard.edu/abs/1984ApJ...276...38J} {276, 38}

\bibitem[\protect\citeauthoryear{{Kimmig}, {Remus}, {Dolag}  \&
  {Biffi}}{{Kimmig} et~al.}{2022}]{kimmig_2022_hateful}
{Kimmig} L.~C.,  {Remus} R.-S.,  {Dolag} K.,   {Biffi} V.,  2022, arXiv
  e-prints, \href {https://ui.adsabs.harvard.edu/abs/2022arXiv220909916K} {p.
  arXiv:2209.09916}

\bibitem[\protect\citeauthoryear{{Lagattuta} et~al.,}{{Lagattuta}
  et~al.}{2022}]{lagattuta_2022_pilot}
{Lagattuta} D.~J.,  et~al., 2022, \mn@doi [\mnras] {10.1093/mnras/stac418},
  \href {https://ui.adsabs.harvard.edu/abs/2022MNRAS.514..497L} {514, 497}

\bibitem[\protect\citeauthoryear{{Lagos}, {Schaye}, {Bah{\'e}}, {van de Sande},
  {Kay}, {Barnes}, {Davis}  \& {Dalla Vecchia}}{{Lagos}
  et~al.}{2018}]{lagos_2018_connection}
{Lagos} C. d.~P.,  {Schaye} J.,  {Bah{\'e}} Y.,  {van de Sande} J.,  {Kay}
  S.~T.,  {Barnes} D.,  {Davis} T.~A.,   {Dalla Vecchia} C.,  2018, \mn@doi
  [\mnras] {10.1093/mnras/sty489}, \href
  {https://ui.adsabs.harvard.edu/abs/2018MNRAS.476.4327L} {476, 4327}

\bibitem[\protect\citeauthoryear{{Lauer} et~al.,}{{Lauer}
  et~al.}{1995}]{lauer_1995_nuker}
{Lauer} T.~R.,  et~al., 1995, \mn@doi [\aj] {10.1086/117719}, \href
  {https://ui.adsabs.harvard.edu/abs/1995AJ....110.2622L} {110, 2622}

\bibitem[\protect\citeauthoryear{{Leitherer} et~al.,}{{Leitherer}
  et~al.}{1999}]{leitherer_1999_starburst}
{Leitherer} C.,  et~al., 1999, \mn@doi [\apjs] {10.1086/313233}, \href
  {https://ui.adsabs.harvard.edu/abs/1999ApJS..123....3L} {123, 3}

\bibitem[\protect\citeauthoryear{{Li}, {Shu}  \& {Wang}}{{Li}
  et~al.}{2018}]{rui_2018_strong}
{Li} R.,  {Shu} Y.,   {Wang} J.,  2018, \mn@doi [\mnras]
  {10.1093/mnras/sty1813}, \href
  {https://ui.adsabs.harvard.edu/abs/2018MNRAS.480..431L} {480, 431}

\bibitem[\protect\citeauthoryear{{Li} et~al.,}{{Li}
  et~al.}{2019}]{li_manga_2019}
{Li} R.,  et~al., 2019, \mn@doi [\mnras] {10.1093/mnras/stz2565}, \href
  {https://ui.adsabs.harvard.edu/abs/2019MNRAS.490.2124L} {490, 2124}

\bibitem[\protect\citeauthoryear{{Lilly} et~al.,}{{Lilly}
  et~al.}{2007}]{lilly_2007_zcosmos}
{Lilly} S.~J.,  et~al., 2007, \mn@doi [\apjs] {10.1086/516589}, \href
  {https://ui.adsabs.harvard.edu/abs/2007ApJS..172...70L} {172, 70}

\bibitem[\protect\citeauthoryear{{Limousin}, {Kneib}, {Bardeau}, {Natarajan},
  {Czoske}, {Smail}, {Ebeling}  \& {Smith}}{{Limousin}
  et~al.}{2007}]{limousin_2007_truncation}
{Limousin} M.,  {Kneib} J.~P.,  {Bardeau} S.,  {Natarajan} P.,  {Czoske} O.,
  {Smail} I.,  {Ebeling} H.,   {Smith} G.~P.,  2007, \mn@doi [\aap]
  {10.1051/0004-6361:20065543}, \href
  {https://ui.adsabs.harvard.edu/abs/2007A&A...461..881L} {461, 881}

\bibitem[\protect\citeauthoryear{{Limousin}, {Sommer-Larsen}, {Natarajan}  \&
  {Milvang-Jensen}}{{Limousin} et~al.}{2009}]{limousin_2009_probing}
{Limousin} M.,  {Sommer-Larsen} J.,  {Natarajan} P.,   {Milvang-Jensen} B.,
  2009, \mn@doi [\apj] {10.1088/0004-637X/696/2/1771}, \href
  {https://ui.adsabs.harvard.edu/abs/2009ApJ...696.1771L} {696, 1771}

\bibitem[\protect\citeauthoryear{{L{\'o}pez-Sanjuan}
  et~al.,}{{L{\'o}pez-Sanjuan} et~al.}{2015}]{lopez_2015_alhambra}
{L{\'o}pez-Sanjuan} C.,  et~al., 2015, \mn@doi [\aap]
  {10.1051/0004-6361/201424913}, \href
  {https://ui.adsabs.harvard.edu/abs/2015A&A...576A..53L} {576, A53}

\bibitem[\protect\citeauthoryear{{Lotz}, {Jonsson}, {Cox}, {Croton}, {Primack},
  {Somerville}  \& {Stewart}}{{Lotz} et~al.}{2011}]{lotz_2011_major}
{Lotz} J.~M.,  {Jonsson} P.,  {Cox} T.~J.,  {Croton} D.,  {Primack} J.~R.,
  {Somerville} R.~S.,   {Stewart} K.,  2011, \mn@doi [\apj]
  {10.1088/0004-637X/742/2/103}, \href
  {https://ui.adsabs.harvard.edu/abs/2011ApJ...742..103L} {742, 103}

\bibitem[\protect\citeauthoryear{{Lotz} et~al.,}{{Lotz}
  et~al.}{2017}]{lotz_2017_frontier}
{Lotz} J.~M.,  et~al., 2017, \mn@doi [\apj] {10.3847/1538-4357/837/1/97}, \href
  {https://ui.adsabs.harvard.edu/abs/2017ApJ...837...97L} {837, 97}

\bibitem[\protect\citeauthoryear{{Maltby} et~al.,}{{Maltby}
  et~al.}{2010}]{maltby_2010_environmental}
{Maltby} D.~T.,  et~al., 2010, \mn@doi [\mnras]
  {10.1111/j.1365-2966.2009.15953.x}, \href
  {https://ui.adsabs.harvard.edu/abs/2010MNRAS.402..282M} {402, 282}

\bibitem[\protect\citeauthoryear{{Mann} \& {Ebeling}}{{Mann} \&
  {Ebeling}}{2012}]{mann_2012_xray}
{Mann} A.~W.,  {Ebeling} H.,  2012, \mn@doi [\mnras]
  {10.1111/j.1365-2966.2011.20170.x}, \href
  {https://ui.adsabs.harvard.edu/abs/2012MNRAS.420.2120M} {420, 2120}

\bibitem[\protect\citeauthoryear{{Maulbetsch}, {Avila-Reese}, {Col{\'\i}n},
  {Gottl{\"o}ber}, {Khalatyan}  \& {Steinmetz}}{{Maulbetsch}
  et~al.}{2007}]{maulbetsch_2007_dependence}
{Maulbetsch} C.,  {Avila-Reese} V.,  {Col{\'\i}n} P.,  {Gottl{\"o}ber} S.,
  {Khalatyan} A.,   {Steinmetz} M.,  2007, \mn@doi [\apj] {10.1086/509706},
  \href {https://ui.adsabs.harvard.edu/abs/2007ApJ...654...53M} {654, 53}

\bibitem[\protect\citeauthoryear{{Mentz} et~al.,}{{Mentz}
  et~al.}{2016}]{mentz_2016_abundance}
{Mentz} J.~J.,  et~al., 2016, \mn@doi [\mnras] {10.1093/mnras/stw2129}, \href
  {https://ui.adsabs.harvard.edu/abs/2016MNRAS.463.2819M} {463, 2819}

\bibitem[\protect\citeauthoryear{{Merten} et~al.,}{{Merten}
  et~al.}{2011}]{merten_2011_creation}
{Merten} J.,  et~al., 2011, \mn@doi [\mnras]
  {10.1111/j.1365-2966.2011.19266.x}, \href
  {https://ui.adsabs.harvard.edu/abs/2011MNRAS.417..333M} {417, 333}

\bibitem[\protect\citeauthoryear{{Mihos}}{{Mihos}}{2003}]{mihos_2003_interactions}
{Mihos} C.,  2003, arXiv e-prints, \href
  {https://ui.adsabs.harvard.edu/abs/2003astro.ph..5512M} {pp
  astro--ph/0305512}

\bibitem[\protect\citeauthoryear{{Moore}, {Katz}, {Lake}, {Dressler}  \&
  {Oemler}}{{Moore} et~al.}{1996}]{moore_1996_harassment}
{Moore} B.,  {Katz} N.,  {Lake} G.,  {Dressler} A.,   {Oemler} A.,  1996,
  \mn@doi [\nat] {10.1038/379613a0}, \href
  {https://ui.adsabs.harvard.edu/abs/1996Natur.379..613M} {379, 613}

\bibitem[\protect\citeauthoryear{{Naab}, {Johansson}  \& {Ostriker}}{{Naab}
  et~al.}{2009}]{naab_2009_minor}
{Naab} T.,  {Johansson} P.~H.,   {Ostriker} J.~P.,  2009, \mn@doi [\apjl]
  {10.1088/0004-637X/699/2/L178}, \href
  {https://ui.adsabs.harvard.edu/abs/2009ApJ...699L.178N} {699, L178}

\bibitem[\protect\citeauthoryear{{Navarro}, {Frenk}  \& {White}}{{Navarro}
  et~al.}{1997}]{navarro_1997_nfw}
{Navarro} J.~F.,  {Frenk} C.~S.,   {White} S. D.~M.,  1997, \mn@doi [\apj]
  {10.1086/304888}, \href
  {https://ui.adsabs.harvard.edu/abs/1997ApJ...490..493N} {490, 493}

\bibitem[\protect\citeauthoryear{{Nedkova} et~al.,}{{Nedkova}
  et~al.}{2021}]{nedkova_2021_extending}
{Nedkova} K.~V.,  et~al., 2021, \mn@doi [\mnras] {10.1093/mnras/stab1744},
  \href {https://ui.adsabs.harvard.edu/abs/2021MNRAS.506..928N} {506, 928}

\bibitem[\protect\citeauthoryear{{Oldham} \& {Auger}}{{Oldham} \&
  {Auger}}{2018}]{oldham_2018_dark}
{Oldham} L.~J.,  {Auger} M.~W.,  2018, \mn@doi [\mnras] {10.1093/mnras/sty065},
  \href {https://ui.adsabs.harvard.edu/abs/2018MNRAS.476..133O} {476, 133}

\bibitem[\protect\citeauthoryear{{Owers}, {Couch}, {Nulsen}  \&
  {Randall}}{{Owers} et~al.}{2012}]{owers_2012_shocking}
{Owers} M.~S.,  {Couch} W.~J.,  {Nulsen} P. E.~J.,   {Randall} S.~W.,  2012,
  \mn@doi [\apjl] {10.1088/2041-8205/750/1/L23}, \href
  {https://ui.adsabs.harvard.edu/abs/2012ApJ...750L..23O} {750, L23}

\bibitem[\protect\citeauthoryear{{Papovich} et~al.,}{{Papovich}
  et~al.}{2012}]{papovich_2012_candels}
{Papovich} C.,  et~al., 2012, \mn@doi [\apj] {10.1088/0004-637X/750/2/93},
  \href {https://ui.adsabs.harvard.edu/abs/2012ApJ...750...93P} {750, 93}

\bibitem[\protect\citeauthoryear{{Pillepich} et~al.,}{{Pillepich}
  et~al.}{2018}]{pillepich_2018_simulating}
{Pillepich} A.,  et~al., 2018, \mn@doi [\mnras] {10.1093/mnras/stx2656}, \href
  {https://ui.adsabs.harvard.edu/abs/2018MNRAS.473.4077P} {473, 4077}

\bibitem[\protect\citeauthoryear{{Poci}, {Cappellari}  \& {McDermid}}{{Poci}
  et~al.}{2017}]{poci}
{Poci} A.,  {Cappellari} M.,   {McDermid} R.~M.,  2017, \mn@doi [\mnras]
  {10.1093/mnras/stx101}, \href
  {https://ui.adsabs.harvard.edu/abs/2017MNRAS.467.1397P} {467, 1397}

\bibitem[\protect\citeauthoryear{{Poggianti} et~al.,}{{Poggianti}
  et~al.}{2017}]{poggianti_2017_gasp}
{Poggianti} B.~M.,  et~al., 2017, \mn@doi [\apj] {10.3847/1538-4357/aa78ed},
  \href {https://ui.adsabs.harvard.edu/abs/2017ApJ...844...48P} {844, 48}

\bibitem[\protect\citeauthoryear{{Reiprich} \& {B{\"o}hringer}}{{Reiprich} \&
  {B{\"o}hringer}}{2002}]{reiprich_2002_mass}
{Reiprich} T.~H.,  {B{\"o}hringer} H.,  2002, \mn@doi [\apj] {10.1086/338753},
  \href {https://ui.adsabs.harvard.edu/abs/2002ApJ...567..716R} {567, 716}

\bibitem[\protect\citeauthoryear{{Remus}, {Burkert}, {Dolag}, {Johansson},
  {Naab}, {Oser}  \& {Thomas}}{{Remus} et~al.}{2013}]{remus_2013_dark}
{Remus} R.-S.,  {Burkert} A.,  {Dolag} K.,  {Johansson} P.~H.,  {Naab} T.,
  {Oser} L.,   {Thomas} J.,  2013, \mn@doi [\apj] {10.1088/0004-637X/766/2/71},
  \href {https://ui.adsabs.harvard.edu/abs/2013ApJ...766...71R} {766, 71}

\bibitem[\protect\citeauthoryear{Remus, Dolag, Naab, Burkert, Hirschmann,
  Hoffmann  \& Johansson}{Remus et~al.}{2017}]{remus_co-evolution_2017}
Remus R.-S.,  Dolag K.,  Naab T.,  Burkert A.,  Hirschmann M.,  Hoffmann T.~L.,
    Johansson P.~H.,  2017, \mn@doi [MNRAS] {10.1093/mnras/stw2594}, 464, 3742

\bibitem[\protect\citeauthoryear{{Richard}, {Kneib}, {Limousin}, {Edge}  \&
  {Jullo}}{{Richard} et~al.}{2010}]{richard_2010_abell}
{Richard} J.,  {Kneib} J.~P.,  {Limousin} M.,  {Edge} A.,   {Jullo} E.,  2010,
  \mn@doi [\mnras] {10.1111/j.1745-3933.2009.00796.x}, \href
  {https://ui.adsabs.harvard.edu/abs/2010MNRAS.402L..44R} {402, L44}

\bibitem[\protect\citeauthoryear{{Robotham}, {Davies}, {Driver}, {Koushan},
  {Taranu}, {Casura}  \& {Liske}}{{Robotham}
  et~al.}{2018}]{robotham_2018_profound}
{Robotham} A.~S.~G.,  {Davies} L.~J.~M.,  {Driver} S.~P.,  {Koushan} S.,
  {Taranu} D.~S.,  {Casura} S.,   {Liske} J.,  2018, \mn@doi [\mnras]
  {10.1093/mnras/sty440}, \href
  {https://ui.adsabs.harvard.edu/abs/2018MNRAS.476.3137R} {476, 3137}

\bibitem[\protect\citeauthoryear{{Ruff}, {Gavazzi}, {Marshall}, {Treu}, {Auger}
   \& {Brault}}{{Ruff} et~al.}{2011}]{ruff_sl2s_2011}
{Ruff} A.~J.,  {Gavazzi} R.,  {Marshall} P.~J.,  {Treu} T.,  {Auger} M.~W.,
  {Brault} F.,  2011, \mn@doi [\apj] {10.1088/0004-637X/727/2/96}, \href
  {https://ui.adsabs.harvard.edu/abs/2011ApJ...727...96R} {727, 96}

\bibitem[\protect\citeauthoryear{{Salpeter}}{{Salpeter}}{1955}]{saltpeter_1955_luminosity}
{Salpeter} E.~E.,  1955, \mn@doi [\apj] {10.1086/145971}, \href
  {https://ui.adsabs.harvard.edu/abs/1955ApJ...121..161S} {121, 161}

\bibitem[\protect\citeauthoryear{{Schwarzschild}}{{Schwarzschild}}{1979}]{schwarzschild_1979_numerical}
{Schwarzschild} M.,  1979, \mn@doi [\apj] {10.1086/157282}, \href
  {https://ui.adsabs.harvard.edu/abs/1979ApJ...232..236S} {232, 236}

\bibitem[\protect\citeauthoryear{{Schwinn}, {Jauzac}, {Baugh}, {Bartelmann},
  {Eckert}, {Harvey}, {Natarajan}  \& {Massey}}{{Schwinn}
  et~al.}{2017}]{schwinn_2017_abell}
{Schwinn} J.,  {Jauzac} M.,  {Baugh} C.~M.,  {Bartelmann} M.,  {Eckert} D.,
  {Harvey} D.,  {Natarajan} P.,   {Massey} R.,  2017, \mn@doi [\mnras]
  {10.1093/mnras/stx277}, \href
  {https://ui.adsabs.harvard.edu/abs/2017MNRAS.467.2913S} {467, 2913}

\bibitem[\protect\citeauthoryear{{Scott} et~al.,}{{Scott}
  et~al.}{2013}]{scott_2013_atlas}
{Scott} N.,  et~al., 2013, \mn@doi [\mnras] {10.1093/mnras/sts422}, \href
  {https://ui.adsabs.harvard.edu/abs/2013MNRAS.432.1894S} {432, 1894}

\bibitem[\protect\citeauthoryear{{Scott}, {Brinks}, {Cortese}, {Boselli}  \&
  {Bravo-Alfaro}}{{Scott} et~al.}{2018}]{scott_2018_abell}
{Scott} T.~C.,  {Brinks} E.,  {Cortese} L.,  {Boselli} A.,   {Bravo-Alfaro} H.,
   2018, \mn@doi [\mnras] {10.1093/mnras/sty063}, \href
  {https://ui.adsabs.harvard.edu/abs/2018MNRAS.475.4648S} {475, 4648}

\bibitem[\protect\citeauthoryear{{Serra}, {Oosterloo}, {Cappellari}, {den
  Heijer}  \& {J{\'o}zsa}}{{Serra} et~al.}{2016}]{serra_2016_linear}
{Serra} P.,  {Oosterloo} T.,  {Cappellari} M.,  {den Heijer} M.,   {J{\'o}zsa}
  G. I.~G.,  2016, \mn@doi [\mnras] {10.1093/mnras/stw1010}, \href
  {https://ui.adsabs.harvard.edu/abs/2016MNRAS.460.1382S} {460, 1382}

\bibitem[\protect\citeauthoryear{{Shankar}, {Marulli}, {Bernardi}, {Mei},
  {Meert}  \& {Vikram}}{{Shankar} et~al.}{2013}]{shankar_2013_size}
{Shankar} F.,  {Marulli} F.,  {Bernardi} M.,  {Mei} S.,  {Meert} A.,   {Vikram}
  V.,  2013, \mn@doi [\mnras] {10.1093/mnras/sts001}, \href
  {https://ui.adsabs.harvard.edu/abs/2013MNRAS.428..109S} {428, 109}

\bibitem[\protect\citeauthoryear{{Sharma}, {Salucci}  \& {van de Ven}}{{Sharma}
  et~al.}{2022}]{sharma_2022_observational}
{Sharma} G.,  {Salucci} P.,   {van de Ven} G.,  2022, \mn@doi [\aap]
  {10.1051/0004-6361/202141822}, \href
  {https://ui.adsabs.harvard.edu/abs/2022A&A...659A..40S} {659, A40}

\bibitem[\protect\citeauthoryear{{Simionescu}, {Werner}, {Mantz}, {Allen}  \&
  {Urban}}{{Simionescu} et~al.}{2017}]{simionescu_2017_witnessing}
{Simionescu} A.,  {Werner} N.,  {Mantz} A.,  {Allen} S.~W.,   {Urban} O.,
  2017, \mn@doi [\mnras] {10.1093/mnras/stx919}, \href
  {https://ui.adsabs.harvard.edu/abs/2017MNRAS.469.1476S} {469, 1476}

\bibitem[\protect\citeauthoryear{{Smith}, {Choi}, {Lee}, {Rhee},
  {Sanchez-Janssen}  \& {Yi}}{{Smith} et~al.}{2016}]{smith_2016_preferential}
{Smith} R.,  {Choi} H.,  {Lee} J.,  {Rhee} J.,  {Sanchez-Janssen} R.,   {Yi}
  S.~K.,  2016, \mn@doi [\apj] {10.3847/1538-4357/833/1/109}, \href
  {https://ui.adsabs.harvard.edu/abs/2016ApJ...833..109S} {833, 109}

\bibitem[\protect\citeauthoryear{{Sonnenfeld}, {Treu}, {Gavazzi}, {Suyu},
  {Marshall}, {Auger}  \& {Nipoti}}{{Sonnenfeld} et~al.}{2013}]{sonn_sl2s_2013}
{Sonnenfeld} A.,  {Treu} T.,  {Gavazzi} R.,  {Suyu} S.~H.,  {Marshall} P.~J.,
  {Auger} M.~W.,   {Nipoti} C.,  2013, \mn@doi [\apj]
  {10.1088/0004-637X/777/2/98}, \href
  {https://ui.adsabs.harvard.edu/abs/2013ApJ...777...98S} {777, 98}

\bibitem[\protect\citeauthoryear{{Soto}, {Lilly}, {Bacon}, {Richard}  \&
  {Conseil}}{{Soto} et~al.}{2016}]{soto_2016_zap}
{Soto} K.~T.,  {Lilly} S.~J.,  {Bacon} R.,  {Richard} J.,   {Conseil} S.,
  2016, \mn@doi [\mnras] {10.1093/mnras/stw474}, \href
  {https://ui.adsabs.harvard.edu/abs/2016MNRAS.458.3210S} {458, 3210}

\bibitem[\protect\citeauthoryear{{Springel}}{{Springel}}{2010}]{springel_2010_galilean}
{Springel} V.,  2010, \mn@doi [\mnras] {10.1111/j.1365-2966.2009.15715.x},
  \href {https://ui.adsabs.harvard.edu/abs/2010MNRAS.401..791S} {401, 791}

\bibitem[\protect\citeauthoryear{{Springel} \& {Hernquist}}{{Springel} \&
  {Hernquist}}{2003}]{springel_2003_cosmogloical}
{Springel} V.,  {Hernquist} L.,  2003, \mn@doi [\mnras]
  {10.1046/j.1365-8711.2003.06206.x}, \href
  {https://ui.adsabs.harvard.edu/abs/2003MNRAS.339..289S} {339, 289}

\bibitem[\protect\citeauthoryear{{Springel}, {Di Matteo}  \&
  {Hernquist}}{{Springel} et~al.}{2005a}]{springel_2005_modelling}
{Springel} V.,  {Di Matteo} T.,   {Hernquist} L.,  2005a, \mn@doi [\mnras]
  {10.1111/j.1365-2966.2005.09238.x}, \href
  {https://ui.adsabs.harvard.edu/abs/2005MNRAS.361..776S} {361, 776}

\bibitem[\protect\citeauthoryear{{Springel} et~al.,}{{Springel}
  et~al.}{2005b}]{springel_2005_simulations}
{Springel} V.,  et~al., 2005b, \mn@doi [\nat] {10.1038/nature03597}, \href
  {https://ui.adsabs.harvard.edu/abs/2005Natur.435..629S} {435, 629}

\bibitem[\protect\citeauthoryear{{Stone}}{{Stone}}{1996}]{stone_1996_spectrophotometry}
{Stone} R. P.~S.,  1996, \mn@doi [\apjs] {10.1086/192369}, \href
  {https://ui.adsabs.harvard.edu/abs/1996ApJS..107..423S} {107, 423}

\bibitem[\protect\citeauthoryear{{Teyssier}}{{Teyssier}}{2002}]{teyssier_2002_cosmological}
{Teyssier} R.,  2002, \mn@doi [\aap] {10.1051/0004-6361:20011817}, \href
  {https://ui.adsabs.harvard.edu/abs/2002A&A...385..337T} {385, 337}

\bibitem[\protect\citeauthoryear{{Thomas} et~al.,}{{Thomas}
  et~al.}{2011}]{thomas_2011_dynamical}
{Thomas} J.,  et~al., 2011, \mn@doi [\mnras]
  {10.1111/j.1365-2966.2011.18725.x}, \href
  {https://ui.adsabs.harvard.edu/abs/2011MNRAS.415..545T} {415, 545}

\bibitem[\protect\citeauthoryear{{Valdes}, {Gupta}, {Rose}, {Singh}  \&
  {Bell}}{{Valdes} et~al.}{2004}]{valdes_2005_indous}
{Valdes} F.,  {Gupta} R.,  {Rose} J.~A.,  {Singh} H.~P.,   {Bell} D.~J.,  2004,
  \mn@doi [\apjs] {10.1086/386343}, \href
  {https://ui.adsabs.harvard.edu/abs/2004ApJS..152..251V} {152, 251}

\bibitem[\protect\citeauthoryear{{Vazdekis}, {S{\'a}nchez-Bl{\'a}zquez},
  {Falc{\'o}n-Barroso}, {Cenarro}, {Beasley}, {Cardiel}, {Gorgas}  \&
  {Peletier}}{{Vazdekis} et~al.}{2010}]{vazdekis_2010_evolutionary}
{Vazdekis} A.,  {S{\'a}nchez-Bl{\'a}zquez} P.,  {Falc{\'o}n-Barroso} J.,
  {Cenarro} A.~J.,  {Beasley} M.~A.,  {Cardiel} N.,  {Gorgas} J.,   {Peletier}
  R.~F.,  2010, \mn@doi [\mnras] {10.1111/j.1365-2966.2010.16407.x}, \href
  {https://ui.adsabs.harvard.edu/abs/2010MNRAS.404.1639V} {404, 1639}

\bibitem[\protect\citeauthoryear{{Wang} et~al.,}{{Wang}
  et~al.}{2019}]{wang_early_2019}
{Wang} Y.,  et~al., 2019, \mn@doi [\mnras] {10.1093/mnras/stz2907}, \href
  {https://ui.adsabs.harvard.edu/abs/2019MNRAS.490.5722W} {490, 5722}

\bibitem[\protect\citeauthoryear{{Wang} et~al.,}{{Wang}
  et~al.}{2020}]{wang_2020_early}
{Wang} Y.,  et~al., 2020, \mn@doi [\mnras] {10.1093/mnras/stz3348}, \href
  {https://ui.adsabs.harvard.edu/abs/2020MNRAS.491.5188W} {491, 5188}

\bibitem[\protect\citeauthoryear{{Watson} et~al.,}{{Watson}
  et~al.}{2019}]{watson_2019_galaxy}
{Watson} C.,  et~al., 2019, \mn@doi [\apj] {10.3847/1538-4357/ab06ef}, \href
  {https://ui.adsabs.harvard.edu/abs/2019ApJ...874...63W} {874, 63}

\bibitem[\protect\citeauthoryear{{Weilbacher} et~al.,}{{Weilbacher}
  et~al.}{2020}]{weilbacher_2020_data}
{Weilbacher} P.~M.,  et~al., 2020, \mn@doi [\aap]
  {10.1051/0004-6361/202037855}, \href
  {https://ui.adsabs.harvard.edu/abs/2020A&A...641A..28W} {641, A28}

\bibitem[\protect\citeauthoryear{{Weinberger} et~al.,}{{Weinberger}
  et~al.}{2018}]{weinberger_2018_supermassive}
{Weinberger} R.,  et~al., 2018, \mn@doi [\mnras] {10.1093/mnras/sty1733}, \href
  {https://ui.adsabs.harvard.edu/abs/2018MNRAS.479.4056W} {479, 4056}

\bibitem[\protect\citeauthoryear{{Williamson} et~al.,}{{Williamson}
  et~al.}{2011}]{Williamson_2011_sunyaev}
{Williamson} R.,  et~al., 2011, \mn@doi [\apj] {10.1088/0004-637X/738/2/139},
  \href {https://ui.adsabs.harvard.edu/abs/2011ApJ...738..139W} {738, 139}

\bibitem[\protect\citeauthoryear{{Zheng} et~al.,}{{Zheng}
  et~al.}{2012}]{zheng_2012_magnified}
{Zheng} W.,  et~al., 2012, \mn@doi [\nat] {10.1038/nature11446}, \href
  {https://ui.adsabs.harvard.edu/abs/2012Natur.489..406Z} {489, 406}

\bibitem[\protect\citeauthoryear{{Zitrin} \& {Broadhurst}}{{Zitrin} \&
  {Broadhurst}}{2009}]{zitrin_2009_discovery}
{Zitrin} A.,  {Broadhurst} T.,  2009, \mn@doi [\apjl]
  {10.1088/0004-637X/703/2/L132}, \href
  {https://ui.adsabs.harvard.edu/abs/2009ApJ...703L.132Z} {703, L132}

\bibitem[\protect\citeauthoryear{{de Ravel} et~al.,}{{de Ravel}
  et~al.}{2011}]{ravel_2011_zcomos}
{de Ravel} L.,  et~al., 2011, arXiv e-prints, \href
  {https://ui.adsabs.harvard.edu/abs/2011arXiv1104.5470D} {p. arXiv:1104.5470}

\makeatother
\end{thebibliography}



\appendix

\section{Validation of MUSE-based MGEs}
\label{sec:simulations}

The JAM method has previously assumed the availability of high-resolution photometric data from which stellar MGEs are obtained. The MAGPI survey currently has no space-based imaging for its targets (aside from the two archival Frontier Fields clusters), and so the highest resolution imaging available is actually synthetic images constructed from the IFU data themselves. Even with GLAO, the PSF FWHM of the MAGPI fields is around 10 times larger than that of, e.g., \textit{HST} data. The pixel scale is also around 7 times larger for MUSE data than \textit{HST}, comparing the 0.2 arcsecond per pixel size of MUSE for the MAGPI fields to the 0.03 arcsecond pixel size of the \textit{HST} data of the supplementary MAGPI fields, Abell 2744 and Abell 370. In this section we investigate how this difference in resolution and pixel scale affects the resulting MGEs and derived inner density slopes.

To explore this issue, we completed end-to-end simulations of the JAM modelling process with mock galaxies ``observed" with a MUSE pixel scale, MAGPI-like noise for the photometry and kinematics, and a PSF FWHM of 0.6 arcseconds. A MGE measured from a galaxy in Abell 2744 from the Frontier Fields sample with \textit{HST} photometry was used as the underlying MGE from which all others were created. The suite of mock galaxies were made by scaling the Gaussian sigmas to change the effective radius, and scaling the Gaussian heights to change the surface density. The MGEs were scaled so that all mock galaxies fall on the \textit{HST}-CANDELS mass-size relation for redshifts $z = 0.2 - 0.5 $ \citep{nedkova_2021_extending}.

Two parameters were used to investigate whether a MUSE-based MGE could recover the inner density slope ($\gamma'$) without bias. We choose to focus on the recovery of the inner density slope as it closely correlates to the total density slope, aside from the radial bounds used as described in Section \ref{sec:calculation}. The  parameters we investigated are: 1) The effective radius as a fraction of the PSF FWHM, and 2) the radial extent of the kinematic data $\mathrm{R}_{\mathrm{max}}$, expressed as a fraction of the effective radius. A suite of 288 mock galaxies were created, spanning $0.5 - 3.95$ in \re/PSF FWHM and $0.5 - 3.25$ in $\mathrm{R_{max}}$/\re. Each of the 288 mock galaxies is an independent `observation' due to the random noise added to both the photometry before the MGEs were measured and the \vf fields before JAM modelling.

Each mock galaxy image was convolved with the adopted PSF, as if it were observed. Adding noise to the photometry and kinematics was informed by MAGPI data, ensuring a realistic treatment. A relationship between noise (from the MAGPI data) and surface brightness (from MAGPI galaxies) was created. The value of noise at each surface brightness was treated as the sigma of a Gaussian distribution, so that MAGPI-like noise could be randomly drawn from that Gaussian distribution and added onto the model galaxies before MGEs were measured. 

For the \vf fields a similar process was used. The underlying scaled MGE (without noise or PSF convolution) was used to create a \vf field, which was then convolved with the adopted PSF.  Measured uncertainties in velocity and velocity dispersion, as described in Section \ref{sec:kinematics}, were used to create a relationship between surface brightness of MAGPI galaxies and \vf noise in \kms. This relationship was used to again add noise to the model input \vf fields based on the surface brightness of the modelled galaxy. 

Models of the second velocity moments were then constructed using these mock luminous MGEs and mock \vf fields as if they were observed data, following exactly the method outlined in Section \ref{sec:models}. As with actual MAGPI data, the constrained parameters are from PSF-deconvolved models. These simulations show in what circumstances the PSF-deconvolved MGE surface brightness model is not an accurate description of the underlying galaxy, and therefore constrains a biased total density profile.

\begin{figure}

	\includegraphics[width=\columnwidth]{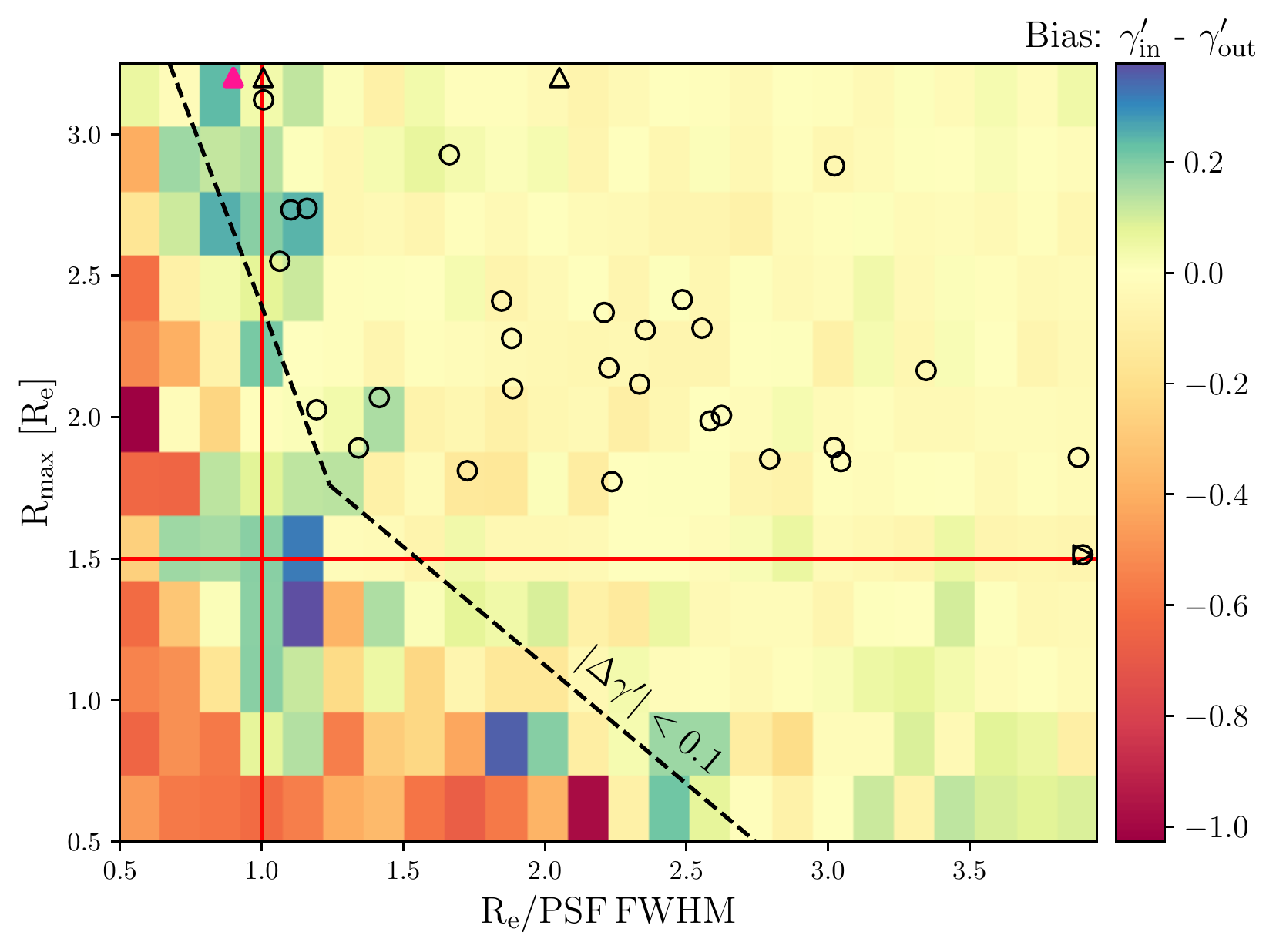}
    \caption{The results of the end-to-end simulations for the parameter $\gamma'$, the inner density slope of the total potential. The y-axis shows the extent of the kinematic coverage expressed as a fraction of the effective radii. The x-axis shows the effective radius as a fraction of the PSF FWHM of the data used to measure the MGEs. The yellow region is where the input inner density slope is well-recovered. The red lines indicate the cuts used to define the sample in this work (upper right quadrant, with a median shift of 1 per cent toward shallower slopes). The black dashed line shows where the absolute bias is less than 0.1. The sample of 30 MAGPI galaxies are shown as black circles, with arrow markers indicating points off the plot region (3 objects). The pink triangle shows the test data case, object 4423 from Abell 2744.}
    \label{fig:muse_image_simulations_bias}
\end{figure}

\begin{figure}

	\includegraphics[width=\columnwidth]{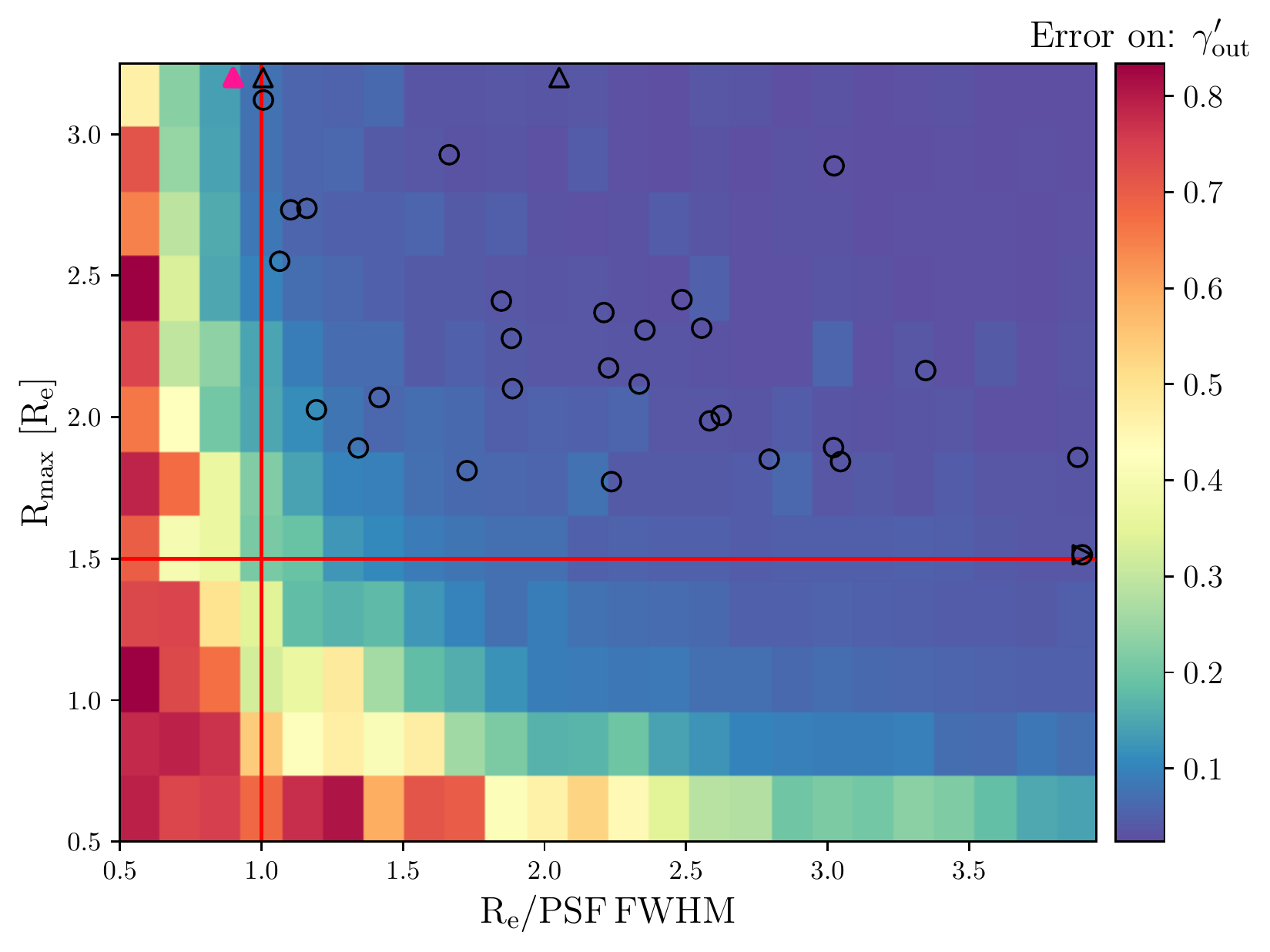}
    \caption{The same as Figure \ref{fig:muse_image_simulations_bias}, however instead showing the standard deviation of the {\sc{emcee}} posterior distribution for the inner density slope. The blue region indicates the parameter space for which the inner density slope is well constrained. The upper right quadrant as defined by the red lines has a median uncertainty of $\pm 0.03$ on $\gamma'$. The sample of 30 MAGPI galaxies are shown as black circles, with arrow markers indicating points off the plot region (3 objects). The pink triangle shows the test data case, object 4423 from Abell 2744.}
    \label{fig:muse_image_simulations_error}
\end{figure}

The difference between the known, input inner density slopes and the recovered inner density slopes is shown in Figure \ref{fig:muse_image_simulations_bias}. The results of the simulation show that there is a compromise between having kinematic data across a large radial extent and having a well-resolved effective radius compared to the PSF FWHM of observation. Poorly resolved galaxies with 1 \re comparable to the PSF FWHM, but with multiple effective radii of kinematic coverage, return the inner density slope with little bias (upper left regions in yellow of Figure \ref{fig:muse_image_simulations_bias}). However, the opposite is also apparent; galaxies that are \textit{well} resolved but with small radial coverage for the kinematics do not recover the input inner density slope well (e.g., bottom row of Figure \ref{fig:muse_image_simulations_bias}). The errors expressed as a standard deviation of the {\sc{emcee}} posterior distribution are shown in Figure \ref{fig:muse_image_simulations_error}.

Based on the results of these end-to-end simulations, we make simple cuts in the parameter space as a data-quality requisite for MAGPI galaxies included in the final sample. These cuts are shown as red lines in both Figure \ref{fig:muse_image_simulations_bias} and Figure \ref{fig:muse_image_simulations_error}. On Figure \ref{fig:muse_image_simulations_bias} we also show a dashed-line to indicate where the input and recovered inner density slope have an absolute difference of less than $0.1$, after smoothing. To make the final sample, a MAGPI object must have \re/PSF FWHM $> 1$ and $\mathrm{R}_{\mathrm{max}}$/\re $> 1.5$. Although a curve or some other more complex function encapsulates the unbiased and well-constrained region of the simulations more accurately, it is not necessary given the distribution of MAGPI objects in this parameter space. The chosen cuts result in a median difference between $\gamma'_{\mathrm{in}}$ and $\gamma'_{\mathrm{out}}$ of 1 per cent toward shallower (less negative) slopes, and a median uncertainty of $\pm 0.03$ on $\gamma'_{\mathrm{out}}$. Here we show only the results for $\gamma'$, however for the chosen cuts the other parameters used in the JAM models ($\rho$, $\beta_z$, and $i$) are also recovered with comparable accuracy. Making a more stringent data cut of \re/PSF FWHM $> 1.5$ does not change the results claimed in this work, resulting in a sample of 21 galaxies and median $\gamma = -2.22 \pm 0.05$, or 14 galaxies when also excluding spiral galaxies with median $\gamma = -2.22 \pm 0.04$.

The cuts imply a certain level of bias in the sample selection of MAGPI objects used in this work. Well-resolved objects with kinematic data that extends less than $1.5$ \re imply low-mass, faint galaxies are excluded from the sample. This is not an issue when comparing the MAGPI and Frontier Fields samples, as we only compare against Frontier Fields objects with the same radial coverage of kinematic data, as described in Section \ref{sec:calculation}. For the other extreme, compact objects that are poorly resolved are also preferentially removed from the selected sample based on the cuts made here. As compact objects tend to have steeper (more negative) total density slopes, we do not expect this cut to influence our conclusion of an observed difference in the distributions of total density slopes between the MAGPI and Frontier Fields galaxies.

\begin{figure*}

	\includegraphics[width=2\columnwidth]{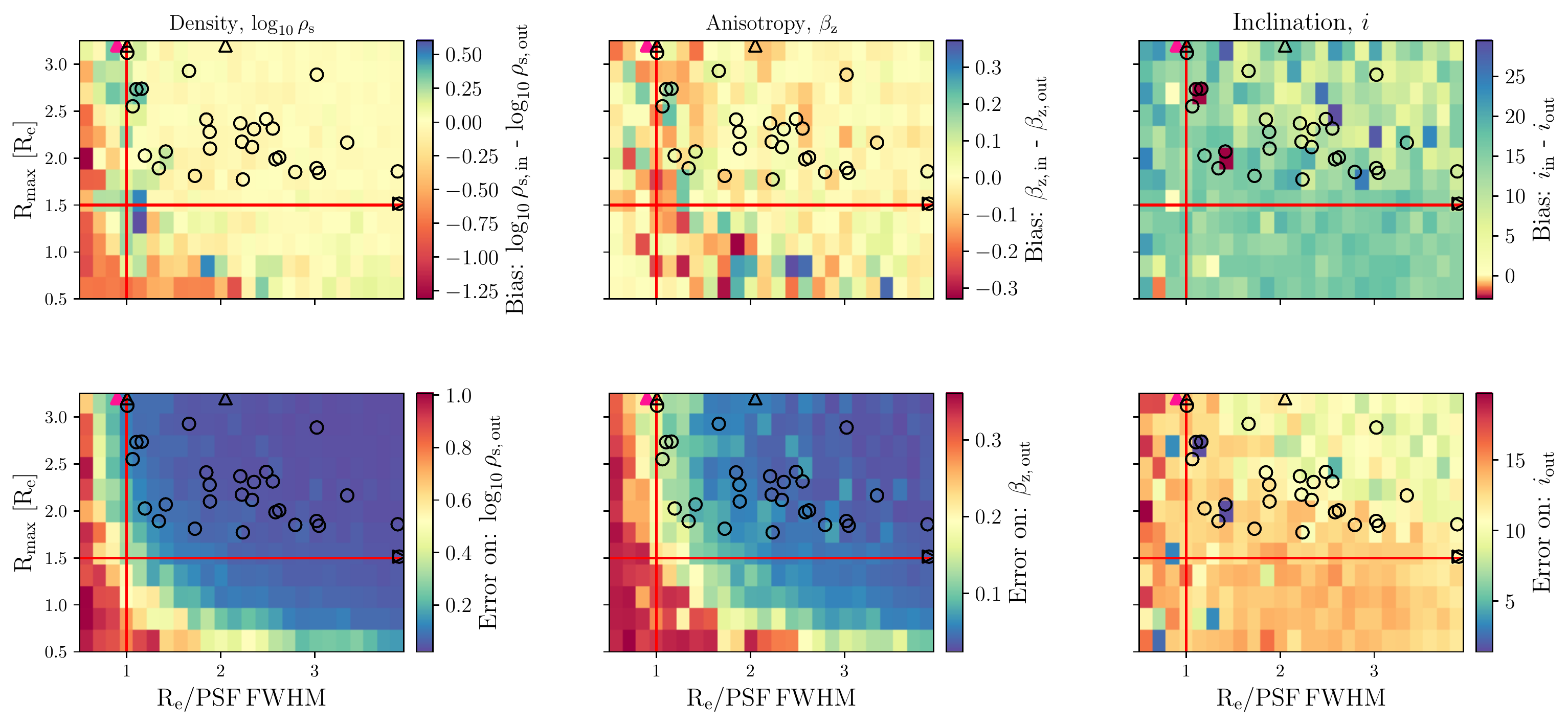}
    \caption{The same as Figures \ref{fig:muse_image_simulations_bias} and \ref{fig:muse_image_simulations_error}, however for the other three parameters used when constructing dynamical models in this work: The density at the break radius, $\log_{10} \rho_{\mathrm{s}}$, orbital anisotropy, $\beta_{\mathrm{z}}$, and inclination, $i$. The red lines indicate the parameter cuts used to select the final MAGPI sample. The top row shows the bias estimated when recovering each parameter, and the bottom row shows the standard deviation of the {\sc{emcee}} posterior distribution. The pink triangle shows the test data case, object 4423 from Abell 2744.}
    \label{fig:muse_image_simulations_all}
\end{figure*}

The entire modelling process was repeated with a different underlying MGE (with a more elliptical morphology) and with different input parameters (e.g., a different input galaxy inclination, inner density slope, density at the break radius, and anisotropy). The simulation results shown in Figure  \ref{fig:muse_image_simulations_bias} and \ref{fig:muse_image_simulations_error} are indicative of those for these different underlying MGEs and parameter sets. 
In Figure \ref{fig:muse_image_simulations_all} we show the simulation results for the remaining parameters. Inclination is recovered the worst, although this parameter is not used for the density slope calculation.

\begin{figure*}
	\includegraphics[width=2\columnwidth]{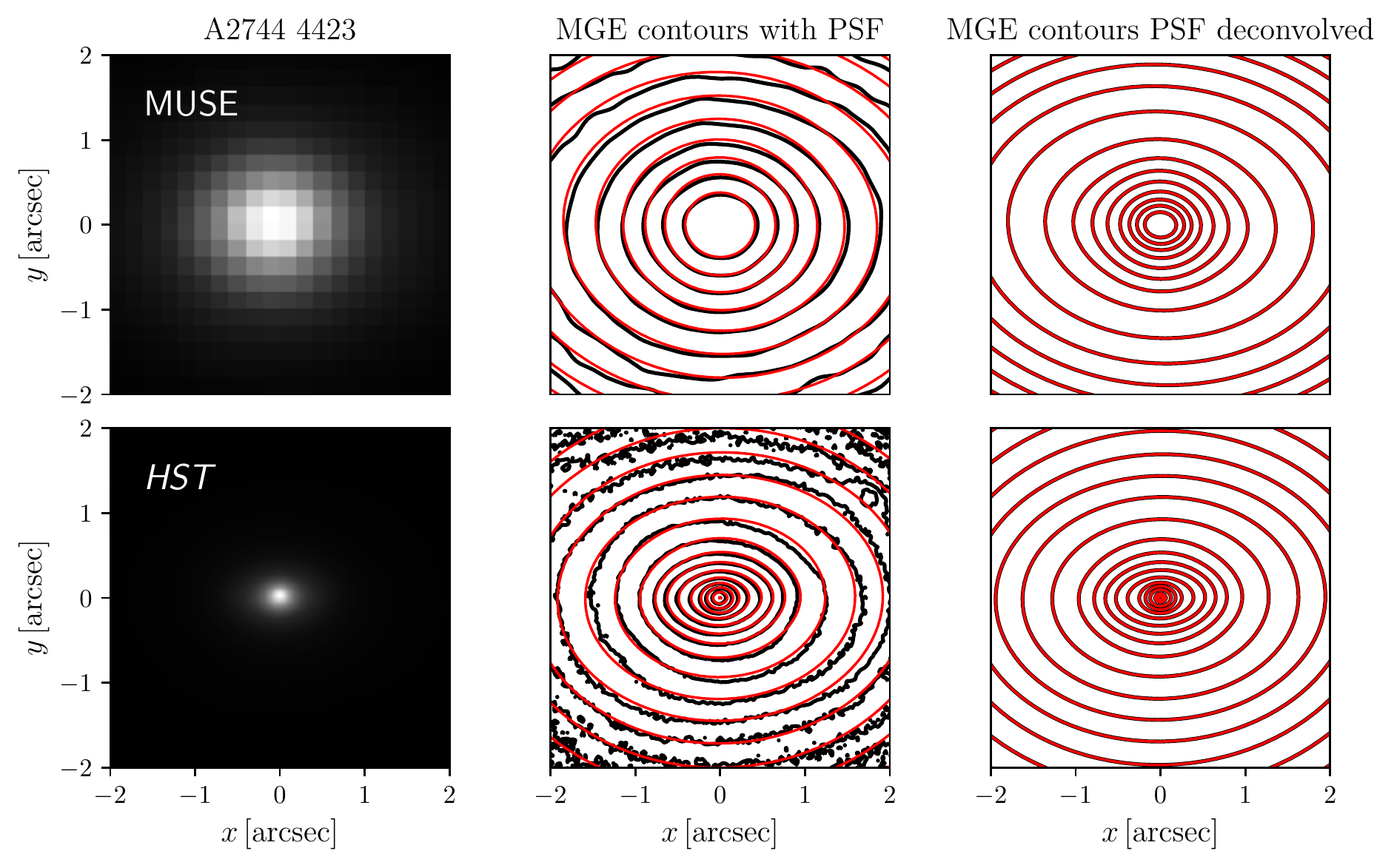}
	\caption{Top row: The SDSS \textit{r}-band `image' made from the MUSE data cube for galaxy A2744 4423, and the associated MGE fit contours (red) overlaid on the galaxy isophotes (black). The third column shows the  MUSE PSF deconvolved galaxy surface brightness model. Bottom row: The \textit{F814W} filter \textit{HST} thumbnail of A2744 4423, and associated MGE fit (red) overlaid on the galaxy isophotes (black), in steps of 0.5 magnitudes per square arcsecond. The third column shows the \textit{HST} PSF deconvolved galaxy surface brightness model.}
    \label{fig:mge_comparison}
\end{figure*}

\begin{figure*}
\centering
	\includegraphics[width=2\columnwidth]{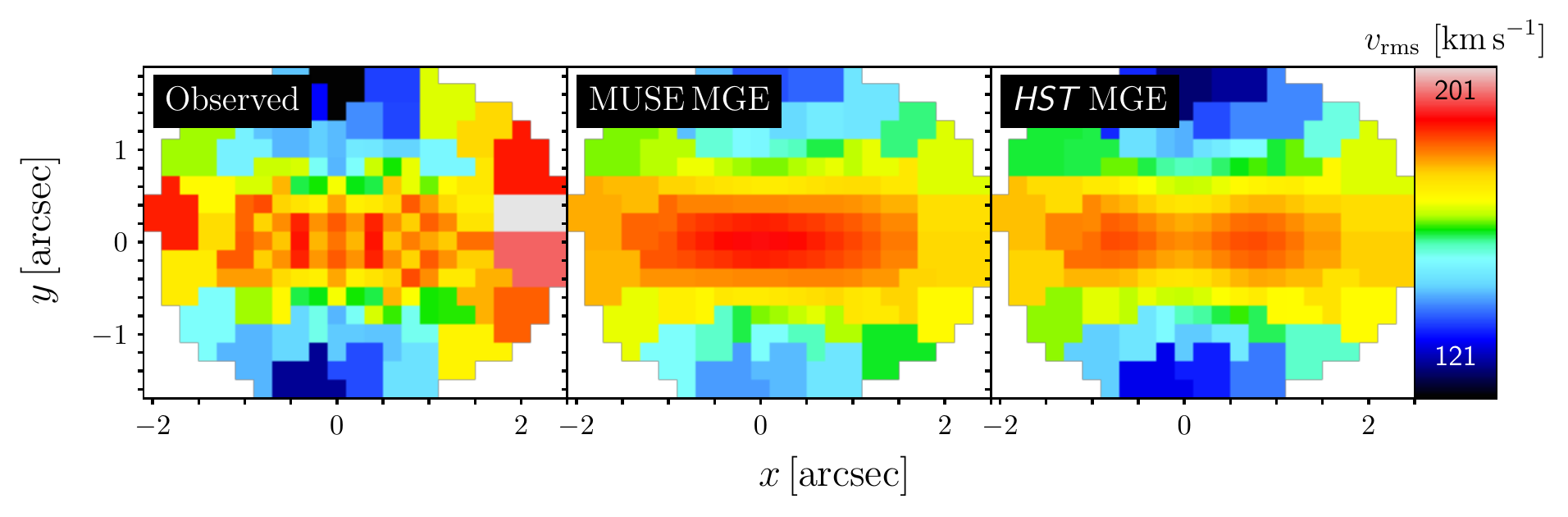}
	\caption{The left panel shows the observed \vf field for galaxy Abell 2744 4423. The middle panel shows the \vf field generated from the median of the posterior distribution using JAM and a MUSE-based MGE. The right panel shows the same, but for a modelled \vf field using JAM and a \textit{HST}-based MGE. The \textit{HST}-based model recovered $\gamma' = -2.087 \pm 0.02$  and the MUSE-based model recovered $\gamma' = -2.089 \pm 0.04$.}
    \label{fig:hst_muse_compare_models}
\end{figure*}

Given some of the Frontier Fields already have existing \textit{HST} and MUSE data, and an effective radius that approximates the MUSE PSF size, it is possible to test the accuracy of this simulation process. We show a comparison between the inner regions of a MUSE-based and \textit{HST}-based MGE for the Abell 2744 galaxy used to create all the mock galaxies in Figure \ref{fig:mge_comparison}, which was modelled in \citet{derkenne}. The \textit{HST}-based MGE model clearly has more structure visible than the MUSE-based MGE model, the inner regions of which are noticeably blurred by the PSF compared to the \textit{HST} model on the same scale. Although the PSF deconvolved models using \textit{HST} and MUSE data are similar they are not identical, which is what motivated the end-to-end simulation process using mock galaxies. The observed and modelled \vf fields using the MUSE-based and \textit{HST}-based MGEs are shown in Figure \ref{fig:hst_muse_compare_models}. 

\begin{figure}
	\includegraphics[width=1\columnwidth]{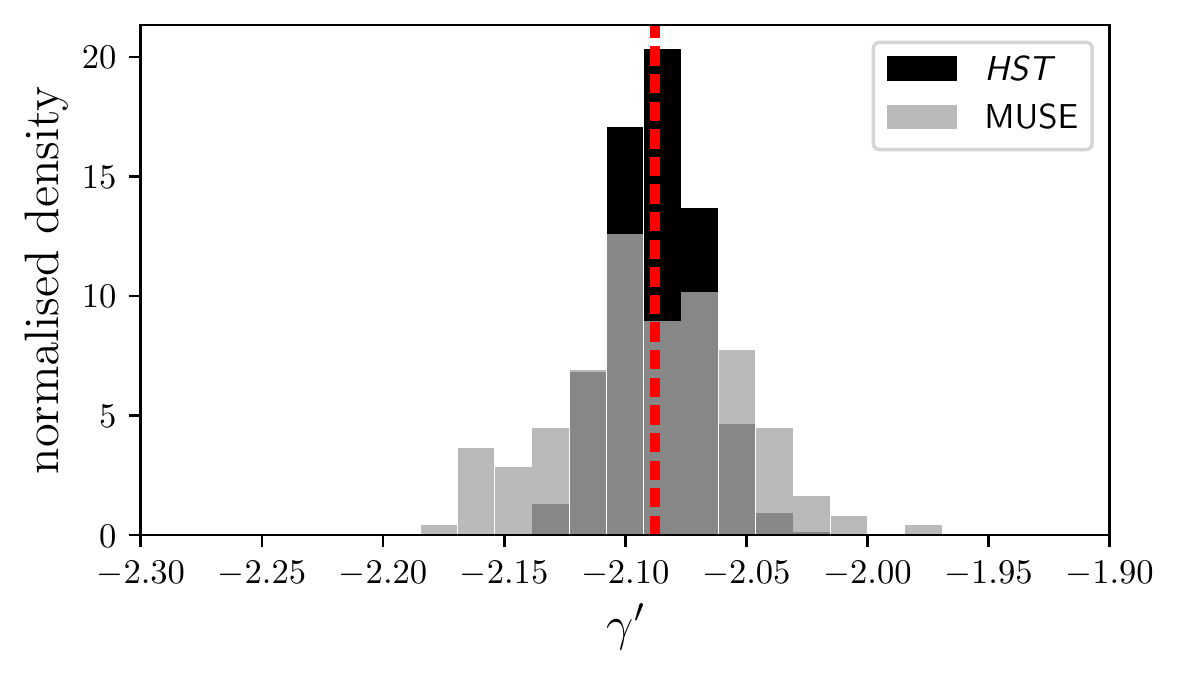}
	\caption{The normalised posterior distribution from  {\sc{emcee}} for the inner density slope using MUSE and \textit{HST}-based MGEs. A red dashed line indicates the median of each distribution, but overlaps as the recovered inner density slopes are near identical.}
    \label{fig:muse_hst_emcee_chain}
\end{figure}

As with the MGEs, the JAM model using the \textit{HST} stellar MGE and MUSE kinematics exhibits more structure than the MUSE-based model. However, the MUSE-based model of this galaxy is able to return an almost identical parameter set to the \textit{HST}-based model. The recovered inner density slope from both models are shown in Figure \ref{fig:muse_hst_emcee_chain}, with a median of $\gamma' = -2.087 \pm 0.02$ using \textit{HST} data and $\gamma' = -2.089 \pm 0.04$ using MUSE data. The other three parameters were recovered for MUSE (\textit{HST}) data as $\log \rho_{\mathrm{s}} = -2.76 \pm 0.03$ ($-2.80 \pm 0.02$), $\beta_z = = -0.47\pm 0.04$ ($-0.45 \pm 0.04$), and $i = 71 \pm 9$ ($79\pm 9$) degrees.  The galaxy A2744 4423 has \re as a fraction of the MUSE PSF FWHM of $\sim ~ 0.9$ and $\mathrm{R}_{\mathrm{max}}$/\re of $\sim 3.7$, which demonstrates well the compromise between resolution and kinematic coverage on recovering the inner density slope (and the other parameters used to create the JAM models) discussed above.

\section{Impact of radial bounds and break radius on the total slope}
\label{sec:app_break_radius}
To construct the dynamical models in this work we have assumed a break radius (the radius at which the total density profile transitions from a free inner slope to a fixed outer slope of -3) is $20$ kpc. This choice was motivated to be consistent with other dynamical modelling works \citep{cappellari_2015_slopes,poci,bellstedt_sluggs_2018,derkenne}, as it is expected the derived total slope will vary systematically with the break radius.

\begin{figure}
	\includegraphics[width=1\columnwidth]{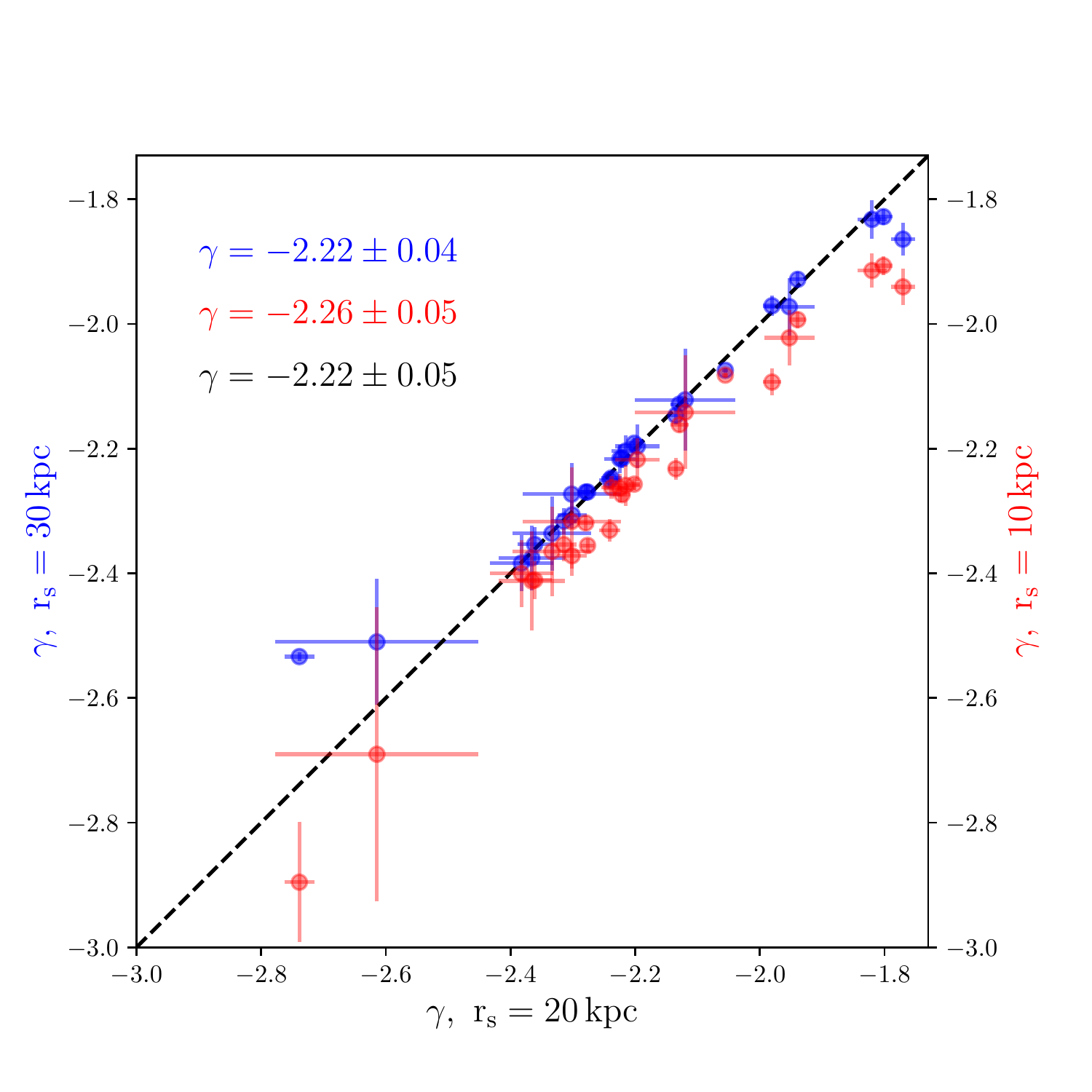}
	\caption{The total density slope, $\gamma$, derived with different choices of break radius, $\mathrm{r}_{\mathrm{s}}$: 10, 20, and 30 kpc. The median gamma values with standard error on the median are inset. There is no difference in the derived median density slope for break radii values of 20 and 30 kpc, however the derived density slopes for a break radius of 10 kpc are on average steeper than those derived with a break radius of 20 kpc.}
    \label{fig:rs_tests}
\end{figure}

We re-ran the MAGPI JAM models using two other choices of break radius to explore the impact of this parameter on the derived total density slopes; 10 kpc and 30 kpc. The results of this test are shown in Figure \ref{fig:rs_tests}.

The total density slopes derived when using a break radius of 20 kpc or 30 kpc are generally consistent within errors, with median derived total density slope $\gamma = -2.22 \pm 0.05$ and $\gamma = -2.22 \pm 0.04$, respectively. A break radius of 10 kpc results in steeper slopes on average, as the transition between a free slope and a fixed outer slope of $-3$ physically occurs at smaller radii for each galaxy. The median with the break radius set as 10 kpc is $\gamma = -2.26 \pm 0.05$. We conclude that setting the break radius to smaller than 20 kpc systematically drives the slopes to steeper values, and setting the break radius to values larger than 20 kpc has no significant effect.

Another potential source of systematic bias in this work is the choice of radial bounds when computing the total density slope. We have computed all MAGPI, Frontier Fields and \atlas slopes on the radial range of $r = [\re/10, 2\,\re]$. As a double power-law is used to describe the total density profile, the exact choice of radial bounds will necessarily have an effect on the derived median slope.

To test the impact of the radial bounds when computing total density slopes we re-calculated all the slopes for the MAGPI, Frontier Fields, and \atlas samples on three other radial ranges in addition to the ones reported in the main text. In total the four radial ranges tested are: $r = [\re/10, 1\,\re]$, $r = [\re/10, 1.5\,\re]$ , $r = [\re/5, 1.5\,\re]$, and $r = [\re/10, 2\,\re]$.

\begin{figure}
	\includegraphics[width=1\columnwidth]{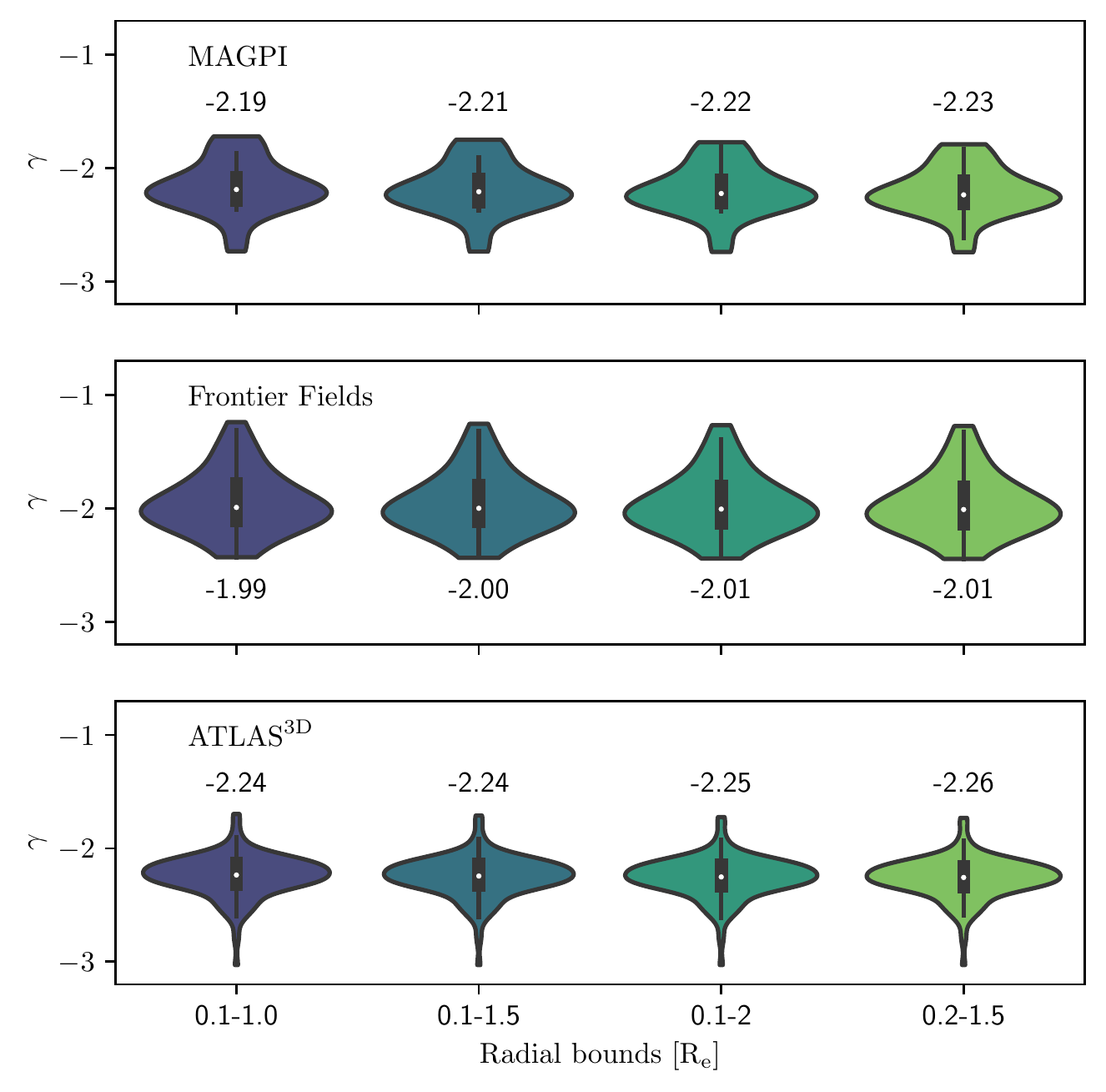}
	\caption{A `violin' plot showing the slope distributions of the MAGPI, Frontier Fields, and \atlas samples, for four different choices of radial bounds: $r = [\re/10, 1\,\re]$, $r = [\re/10, 1.5\,\re]$ , $r = [\re/5, 1.5\,\re]$, and $r = [\re/10, 2\,\re]$. The choice of radial bounds $r = [\re/10, 2\,\re]$ is the one on which we base our conclusions in the main text. The violin plots shows a kernel density estimate of the underlying points. A box plot is inset  with whiskers set to 1.5 the interquartile range. The median total density slope is shown as a white dot, and is also annotated above each data set.}
    \label{fig:radial_bounds}
\end{figure}

We show the results of this test as a `violin' plot in Figure \ref{fig:radial_bounds}. As the outer radial bound gets larger the derived median density slope becomes marginally steeper. As discussed above, this is because the outer fixed slope of $-3$ influences the total derived slope at physical radii around 20 kpc. However, the difference between the derived median density slopes for these choices of radial bounds are minimal, and the relative trends between them are maintained. That is, the Frontier Fields slopes remain significantly more shallow than either the MAGPI or \atlas samples for all choices of radial bounds. The shape of the distribution for different radial bounds is also constant. Finally, the change in the median slope for these different bound is consistent with the quoted uncertainty on the total density slope ($\pm 0.05$ for MAGPI).

We note that the radial range of $r = [\re/10, 1\,\re]$ (first column on Figure \ref{fig:radial_bounds}) is most similar to the radial ranges used in the majority of gravitational lensing works (see Table \ref{tab:comparison_gamma} in Appendix \ref{sec:results_app} for a summary), as the Einstein radius generally approximates an effective radius. 

\section{Tabulated total slopes}
\label{sec:results_app}

We present the tabulated results for the MAGPI sample in Table \ref{tab:results_table}, including their IDs, coordinates, redshift, central velocity dispersion, total mass (defined as twice the mass integrated within a sphere of 1 \re), the circularised effective radius, total density slope and associated uncertainties. 

Table \ref{tab:comparison_gamma} summaries the median total density slopes are measured by various works, and includes all studies as shown on Figure \ref{fig:main_results}. In particular, we give the redshift, sample size, total density slope, and radial range across which the slope was measured or constrained. 

\begin{table*}
	\centering
	\caption{Columns 1-4: The MAGPI ID, right-ascension and declination, and redshift of each object used in this work. The Field of each object corresponds to the first 4 numbers of the MAGPI ID. Column 5: The measured 1 \re aperture velocity dispersion from the kinematics.  Column 6: Twice the mass integrated within a sphere of 1 \re using the median total potential parametrisation from the {\sc{emcee}} posterior distribution, as described in Section \ref{sec:calculation}. Column 7: The circularised effective radius containing half the galaxy light. Column 8: The average density slope measured between 0.1 \re and 2 \re for each galaxy. Columns 9 and 10: The 16\textsuperscript{th} and 84\textsuperscript{th} percentiles offset from the 50\textsuperscript{th} percentile of the {\textsc{emcee}} posterior distribution. These are Monte Carlo errors as described in Section \ref{sec:calculation}. No uncertainties are given for the central dispersion or effective radius as these measurements are dominated by systematic effects.}
	\label{tab:results_table}
	\begin{tabular}{lccccccccc} 
		\hline
		MAGPI ID & R.A (J2000) & Decl. (J2000) & $z$ &$\sigma_\mathrm{e}$ [\kms] & $\log_{10}\left(\mathrm{M}/\mathrm{M}_{\odot}\right)$ &$\mathrm{R}_{\mathrm{e}}$  [kpc] &  $\gamma$ & +err & -err\\
	
(1) & (2) & (3) & (4) & (5) & (6) & (7) & (8) &(9) & (10) \\
		\hline
1202197197 & 175.3388 &  -1.5825 & 0.2920 & 148 & $11.4 \pm 0.01 $ & 7.68 & -1.82 & 0.02 &0.03 \\
1203040085 & 175.3562 &  0.6272 & 0.3132 & 271 & $11.4 \pm 0.04 $ & 3.34 & -2.33 & 0.06 &0.06 \\
1203060081 & 175.3549 &  0.6268 & 0.2799 & 203 & $11.0 \pm 0.01 $ & 2.49 & -2.12 & 0.08 &0.09 \\
1203070184 & 175.3549 &  0.6323 & 0.3089 & 223 & $11.5 \pm 0.02 $ & 6.02 & -2.22 & 0.02 &0.02 \\
1203087201 & 175.3535 &  0.6335 & 0.3094 & 108 & $10.6 \pm 0.03 $ & 3.16 & -2.30 & 0.08 &0.08 \\
1203196196 & 175.3474 &  0.6333 & 0.3103 & 299 & $12.0 \pm 0.01 $ & 9.23 & -1.94 & 0.01 &0.01 \\
1203230310 & 175.3456 &  0.6396 & 0.3120 & 162 & $10.9 \pm 0.01 $ & 2.99 & -2.20 & 0.04 &0.03 \\
1203305151 & 175.3413 &  0.6308 & 0.3152 & 241 & $11.2 \pm 0.01 $ & 2.61 & -2.38 & 0.05 &0.05 \\
1204141177 & 175.6646 &  -0.7958 & 0.1070 & 177 & $10.9 \pm 0.00 $ & 2.22 & -2.06 & 0.00 &0.01 \\
1204198199 & 175.6615 &  -0.7946 & 0.3164 & 163 & $11.4 \pm 0.00 $ & 7.94 & -1.80 & 0.01 &0.02 \\
1205093221 & 178.0856 &  -0.8259 & 0.2918 & 189 & $11.2 \pm 0.01 $ & 5.01 & -2.30 & 0.02 &0.03 \\
1205196165 & 178.0798 &  -0.8290 & 0.2925 & 81 & $10.4 \pm 0.03 $ & 3.28 & -2.17 & 0.14 &0.17 \\
1205197197 & 178.0798 &  -0.8272 & 0.2919 & 97 & $10.9 \pm 0.02 $ & 7.43 & -2.28 & 0.02 &0.03 \\
1206110186 & 180.1670 &  -1.4530 & 0.2677 & 264 & $11.7 \pm 0.00 $ & 6.73 & -2.20 & 0.01 &0.01 \\
1206196198 & 180.1622 &  -1.4524 & 0.3271 & 255 & $11.7 \pm 0.00 $ & 8.81 & -2.28 & 0.01 &0.01 \\
1206276211 & 180.1578 &  -1.4516 & 0.2689 & 206 & $11.3 \pm 0.01 $ & 5.10 & -2.22 & 0.03 &0.03 \\
1207128248 & 182.0037 &  -2.4805 & 0.3215 & 170 & $11.2 \pm 0.01 $ & 4.91 & -2.37 & 0.05 &0.06 \\
1207197197 & 181.9999 &  -2.4834 & 0.3212 & 103 & $10.8 \pm 0.01 $ & 6.63 & -2.36 & 0.03 &0.05 \\
1208197197 & 184.9718 &  -2.4811 & 0.3008 & 208 & $11.4 \pm 0.01 $ & 5.03 & -2.13 & 0.01 &0.01 \\
1209131247 & 185.6022 &  -1.3774 & 0.2960 & 155 & $11.0 \pm 0.01 $ & 3.50 & -1.95 & 0.04 &0.05 \\
1209197197 & 185.5985 &  -1.3801 & 0.2960 & 159 & $11.2 \pm 0.09 $ & 5.86 & -2.13 & 0.01 &0.01 \\
1209206324 & 185.5981 &  -1.3731 & 0.5710 & 177 & $11.0 \pm 0.04 $ & 4.58 & -2.61 & 0.16 &0.19 \\
1501196198 & 212.3056 &  1.7839 & 0.3100 & 242 & $11.9 \pm 0.03 $ & 10.54 & -1.98 & 0.01 &0.02 \\
1501224275 & 212.3040 &  1.7882 & 0.3118 & 158 & $10.7 \pm 0.01 $ & 2.91 & -2.74 & 0.02 &0.04 \\
1507196198 & 215.6209 &  0.4080 & 0.3152 & 213 & $11.2 \pm 0.00 $ & 4.16 & -2.31 & 0.02 &0.02 \\
1508197198 & 215.8838 &  2.7133 & 0.3164 & 177 & $11.2 \pm 0.01 $ & 5.31 & -1.77 & 0.02 &0.02 \\
1523197197 & 219.5414 &  -1.0993 & 0.2813 & 184 & $11.3 \pm 0.01 $ & 5.75 & -2.28 & 0.01 &0.01 \\
1525170222 & 219.6684 &  0.3349 & 0.3210 & 327 & $11.7 \pm 0.04 $ & 5.52 & -2.24 & 0.02 &0.01 \\
1525196197 & 219.6669 &  0.3335 & 0.3187 & 228 & $11.5 \pm 0.01 $ & 6.15 & -2.22 & 0.01 &0.01 \\
1530197196 & 222.1438 &  2.9410 & 0.3108 & 288 & $11.3 \pm 0.04 $ & 2.49 & -2.24 & 0.02 &0.01 \\
		\hline
	\end{tabular}
\end{table*}

\begin{table*}
	\centering
	\caption{A summary of the literature total density slopes values shown in Figure \ref{fig:main_results}. All values are calculated as a median unless except for \citet{bolton_boss_2012} where the fitted relation is given, and \citet{rui_2018_strong} where the mean is given. Column 1 gives the sample and study from which the slopes are drawn. Note that slopes from \citet{derkenne}, \citet{poci}, \citet{bellstedt_sluggs_2018}, and \citet{thomas_2011_dynamical} were all re-analysed to be on the same radial range as MAGPI total density slopes. Column 2: The redshift range or value. Column 3: The sample size of galaxies. Column 4: The radial range across which the total density slope was measured.  Column 5: The median total density slope with the standard error on the median. Column 6: The standard deviation of the sample total density slopes.}
	\label{tab:comparison_gamma}
	\begin{tabular}{lccccc} 
		\hline
		Sample & $z$ & N & Radial range & Median $\gamma$ & $\sigma_{\gamma}$ \\
		(1) & (2) & (3) & (4) & (5) & (6) \\
		\hline
Dynamics \\\hline
MAGPI (this work) & 0.31 & 28 &$0.1\, \re - 2\, \re$& $-2.22 \pm 0.05$ & 0.22 \\

Frontier Fields \citep{derkenne} & $0.29 < z < 0.55$ & 64 &$0.1\, \re - 2\, \re$& $-2.01 \pm 0.04$ & 0.26\\

\atlas \citep{poci} & $\sim 0$ & 150 &$0.1\, \re - 2\, \re$ & $-2.25 \pm 0.02$ & 0.17 \\

Coma \citep{thomas_2011_dynamical} & 0.0231 & 17 &$0.1\, \re - 2\, \re$& $-2.00 \pm 0.06$ & 0.20\\

SLUGGS \citep{bellstedt_sluggs_2018} & $\sim 0$ & 22 &$0.1\, \re  -2\, \re$& $-2.06 \pm 0.04$ &0.13\\

MaNGA \citep{li_manga_2019} & $\sim 0$ & 2110 &$ < \re$ & $-2.22 \pm 0.006$ & 0.22\\
\\
Simulations \\\hline
Magneticum \citep{remus_co-evolution_2017} & 0.34 & 93 &$0.4\,\rmass - 4\,\rmass$ &$- 2.11 \pm 0.02$ & 0.16 \\

Magneticum \citep{remus_co-evolution_2017} & 0.066 & 96 & $0.4\,\rmass - 4\,\rmass$ & $-2.05 \pm 0.016$ & 0.13 \\

IllustrisTNG \citep{wang_early_2019} & $\sim 0 $ & 559 & $0.4\,\rmass - 4\,\rmass$ & $-2.00 \pm 0.0004$ & 0.17\\

IllustrisTNG \citep{wang_early_2019} & 0.3 & 731 & $0.4\,\rmass - 4\,\rmass$ & $-1.98 \pm 0.0086$ & 0.19\\

Horizon-AGN &  0.018 & 2659 &$0.1\, \re - 2\, \re$ & $-1.99 \pm 0.004$ & 0.17 \\

Horizon-AGN & 0.305 & 2170 &$0.1\, \re - 2\, \re$ &$-1.93 \pm 0.004$ & 0.15 \\
\\
Lensing \\\hline
SLACS \citep{auger_sloan_2010} & $0.063 < z < 0.358$& 58  & $< \mathrm{R}_{\mathrm{Einstein}}$  & $-2.09 \pm 0.04$ & 0.22 \\

SLACS \citep{barnabe_two_2011} & $0.0808 < z< 0.3475$ & 16 & $< \re$ &$-2.079 \pm 0.05$ & 0.16\\

SL2S \citep{ruff_sl2s_2011} & $0.238 < z < 0.65$ & 11 & $< \mathrm{R}_{\mathrm{Einstein}}$& $-2.09 \pm  0.11$ & 0.29\\

SLACS and BELLS \citep{bolton_boss_2012} & $0.1 < z < 0.6$ & 85 &$< \mathrm{R}_{\mathrm{Einstein}}$ & $\gamma(z) = (-2.11 \pm 0.02) + z(0.6 \pm 0.15)$ &  -  \\

SL2S \citep{sonn_sl2s_2013}& $0.238 < z <0.78$ & 25 & $< \mathrm{R}_{\mathrm{Einstein}}$ &$-2.01 \pm 0.05$& 0.21\\

BELLS, BELLS GALLERY, and SL2S \citep{rui_2018_strong} &$\sim 0.5$ &63 & $ < 3\,\re$&$\langle{\gamma}\rangle = -2.00 \pm 0.03$ & 0.18\\
		\hline
	\end{tabular}
\end{table*}


\bsp	
\label{lastpage}
\end{document}